\documentclass[12pt]{article}
\usepackage{bbm}
\usepackage{mathtools}
\usepackage{algorithm}
\usepackage{accents}
\usepackage{algpseudocode}
\usepackage{enumerate}
\usepackage{xcolor}
\usepackage{pifont}
\usepackage{makecell}
\usepackage{natbib}
\usepackage{hyperref}
\usepackage{mathdots}
\RequirePackage{epsfig}
\RequirePackage{amssymb}
\RequirePackage{natbib}
\RequirePackage{graphicx}

\if@usehyper
\RequirePackage[colorlinks=true,allbordercolors={1 1 1}]{hyperref}
\fi

\if@abbrvbib
\else
\fi

\bibpunct{(}{)}{;}{a}{,}{,}

\makeatletter
\newcommand*{\addFileDependency}[1]{
  \typeout{(#1)}
  \@addtofilelist{#1}
  \IfFileExists{#1}{}{\typeout{No file #1.}}
}
\makeatother

\newcommand{\blind}{0}

\DeclareMathAlphabet\mathbfcal{OMS}{cmsy}{b}{n}
\algnewcommand\INPUT{\item[\textbf{Input:}]}
\algnewcommand\OUTPUT{\item[\textbf{Output:}]}

\newcommand{\tenX}{\mathbfcal{X}}
\newcommand{\tenY}{\mathbfcal{Y}}
\newcommand{\tenG}{\mathbfcal{G}}
\newcommand{\htenG}{\widehat{\mathbfcal{G}}}

\newcommand{\tenA}{\mathbfcal{A}}
\newcommand{\tenB}{\mathbfcal{B}}
\newcommand{\tenR}{\mathbfcal{R}}

\newcommand{\matX}{\mathbf{X}}
\newcommand{\matG}{\mathbf{G}}
\newcommand{\matA}{\mathbf{A}}
\newcommand{\matB}{\mathbf{B}}
\newcommand{\matC}{\mathbf{C}}
\newcommand{\matD}{\mathbf{D}}
\newcommand{\hmatA}{\widehat{\mathbf{A}}}
\newcommand{\hmatB}{\widehat{\mathbf{B}}}
\newcommand{\hmatC}{\widehat{\mathbf{C}}}
\newcommand{\matE}{\mathbf{E}}

\newcommand{\matK}{\mathbf{K}}
\newcommand{\matI}{\mathbf{I}}
\newcommand{\matJ}{\mathbf{J}}
\newcommand{\matL}{\mathbf{L}}
\newcommand{\matO}{\mathbf{O}}
\newcommand{\matR}{\mathbf{R}}
\newcommand{\matH}{\mathbf{H}}
\newcommand{\matW}{\mathbf{W}}
\newcommand{\matS}{\mathbf{S}}
\newcommand{\matU}{\mathbf{U}}
\newcommand{\matV}{\mathbf{V}}
\newcommand{\matY}{\mathbf{Y}}
\newcommand{\matKR}{\mathbf{K}_{R}}
\newcommand{\wmatX}{\widetilde{\mathbf{X}}}
\newcommand{\bSigma}{\boldsymbol{\Sigma}}
\newcommand{\bSigmar}{\boldsymbol{\Sigma}_{r}}
\newcommand{\bSigmac}{\boldsymbol{\Sigma}_{c}}
\newcommand{\matgamma}{\boldsymbol{\Gamma}}
\newcommand{\matTheta}{\boldsymbol{\Theta}}
\newcommand{\matLambda}{\boldsymbol{\Lambda}}
\newcommand{\matOmega}{\boldsymbol{\Omega}}
\newcommand{\matPhi}{\boldsymbol{\Phi}}
\newcommand{\matPsi}{\boldsymbol{\Psi}}
\newcommand{\matXi}{\boldsymbol{\Xi}}

\newcommand{\vecx}{\mathbf{x}}
\newcommand{\vecy}{\mathbf{y}}
\newcommand{\vecz}{\mathbf{z}}
\newcommand{\vece}{\mathbf{e}}
\newcommand{\vecr}{\mathbf{r}}
\newcommand{\vecu}{\mathbf{u}}
\newcommand{\vecep}{\mathbfcal{E}}
\newcommand{\vecalpha}{\boldsymbol{\alpha}}
\newcommand{\vecbeta}{\boldsymbol{\beta}}
\newcommand{\vecgamma}{\boldsymbol{\gamma}}
\newcommand{\vectheta}{\boldsymbol{\theta}}
\newcommand{\vecnu}{\boldsymbol{\nu}}
\newcommand{\veceta}{\boldsymbol{\eta}}
\newcommand{\veczeta}{\boldsymbol{\zeta}}
\newcommand{\vecphi}{\boldsymbol{\phi}}
\newcommand{\vecxi}{\boldsymbol{\xi}}
\newcommand{\wvecz}{\widetilde{\mathbf{z}}}

\newcommand{\Qtilde}{\widetilde{Q}}
\newcommand{\qtilde}{\widetilde{q}}

\newcommand{\setS}{\mathbb{S}}

\newcommand{\setF}{\mathbb{F}}
\newcommand{\RKHS}{\mathbb{H}_{k}}

\newcommand{\inner}[2]{\langle #1, #2\rangle}
\newcommand{\binner}[2]{\left\langle #1, #2\right\rangle}
\newcommand{\twonorm}[1]{\|#1\|_{\mathrm{F}}}
\newcommand{\specnorm}[1]{\|#1\|_{s}}
\newcommand{\tvprod}{\bar{\times}}
\newcommand{\vect}[1]{\mathrm{\mathbf{vec}}\left(#1\right)}
\newcommand{\widesim}[2][1.5]{
  \mathrel{\overset{#2}{\scalebox{#1}[1]{$\sim$}}}
}
\newcommand{\tr}[1]{\text{tr}\left(#1\right)}
\newcommand{\ubar}[1]{\underaccent{\bar}{#1}}
\newcommand{\maxeigen}[1]{\bar{\rho}(#1)}
\newcommand{\mineigen}[1]{\underaccent{\bar}{\rho}(#1)}

\DeclareMathOperator*{\argmin}{arg\,min}

\newcommand{\BlackBox}{\rule{1.5ex}{1.5ex}}  
\ifdefined\proof
    \renewenvironment{proof}{\par\noindent{\bf Proof\ }}{\hfill\BlackBox\\[2mm]}
\else
    \newenvironment{proof}{\par\noindent{\bf Proof\ }}{\hfill\BlackBox\\[2mm]}
\fi
 
\newtheorem{theorem}{Theorem}
\newtheorem{lemma}[theorem]{Lemma} 
\newtheorem{proposition}[theorem]{Proposition} 
\newtheorem{remark}[theorem]{Remark}
\newtheorem{corollary}[theorem]{Corollary}

\newtheorem{assumption}[theorem]{Assumption}

\hypersetup{
     colorlinks=true,
     citecolor=blue,
     linkcolor=red
}

\begin{document}
\def\spacingset#1{\renewcommand{\baselinestretch}%
{#1}\small\normalsize} \spacingset{1}

\if0\blind
{
  \title{\bf Matrix Autoregressive Model with Vector Time Series Covariates for Spatio-Temporal Data}
  \author{Hu Sun\\
    Department of Statistics, University of Michigan, Ann Arbor\\
    Zuofeng Shang \\
    Department of Mathematical Sciences, New Jersey Institute of Technology \\
    and \\
    Yang Chen \\
    Department of Statistics, University of Michigan, Ann Arbor}
  \maketitle
} \fi

\if1\blind
{
  \title{\bf Matrix Autoregressive Model with Vector Time Series Covariates for Spatio-Temporal Data}
  \author{Anonymous Authors}
  \maketitle
} \fi

\bigskip
\begin{abstract}
We develop a new methodology for forecasting matrix-valued time series with historical matrix data and auxiliary vector time series data. We focus on a time series of matrices defined on a static 2-D spatial grid and an auxiliary time series of non-spatial vectors. The proposed model, Matrix AutoRegression with Auxiliary Covariates (MARAC), contains an autoregressive component for the historical matrix predictors and an additive component that maps the auxiliary vector predictors to a matrix response via tensor-vector product. The autoregressive component adopts a bi-linear transformation framework following~\citet{chen2021autoregressive}, significantly reducing the number of parameters. The auxiliary component posits that the tensor coefficient, which maps non-spatial predictors to a spatial response, contains slices of spatially smooth matrix coefficients that are discrete evaluations of smooth functions on a spatial grid from a Reproducing Kernel Hilbert Space (RKHS). We propose to estimate the model parameters under a penalized maximum likelihood estimation framework coupled with an alternating minimization algorithm. We establish the joint asymptotics of the autoregressive and tensor parameters under fixed and high-dimensional regimes. Extensive simulations and a geophysical application for forecasting the global Total Electron Content (TEC) are conducted to validate the performance of MARAC.
\end{abstract}

\noindent%
\vfill

\newpage
\spacingset{1.5}
\section{Introduction}\label{sec:intro}
\begin{figure}[!htb]
    \centering
    \includegraphics[width=0.98\textwidth]{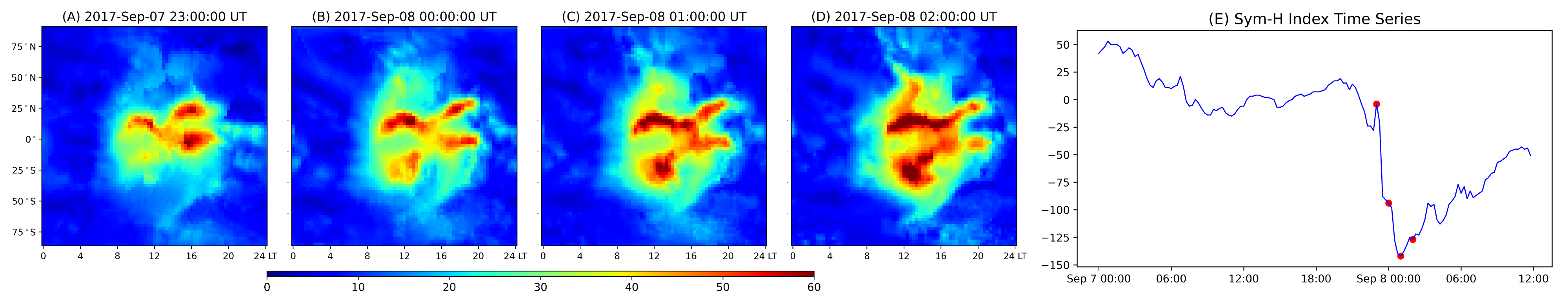}
    \caption{An example of matrix time series with auxiliary vector time series. Panels (A)-(D) show the global Total Electron Content (TEC) distribution at four timestamps, separated by 1 hour, on the latitude-local time grid (source: the IGS TEC database~\citep{hernandez2009igs}). Panel (E) plots the auxiliary Sym-H index time series, which measures the impact of solar eruptions on Earth. We highlight the time of panels (A)-(D) with dots.}
    \label{fig:TEC-Example}
\end{figure}

Matrix-valued time series data have received increasing attention in multiple scientific fields, such as economics~\citep{wang2019factor}, geophysics~\citep{sun2022matrix}, and environmental science~\citep{dong2020envsci}, where scientists are interested in modeling the joint dynamics of data observed on a 2-D grid over time. This paper focuses on the matrix-valued data defined on a 2-D spatial grid that contains the geographical information of the individual observations. As a concrete example, we visualize the global Total Electron Content (TEC) distribution in Figure \ref{fig:TEC-Example}. TEC is the density of electrons in the Earth's ionosphere along the vertical pathway connecting a radio transmitter and a ground-based receiver. An accurate prediction of global TEC is critical as it can predict the impact of space weather on positioning, navigation, and timing (PNT) services~\citep{wang2021,Younas2022}. Every image in panel (A)-(D) is a $71\times 73$ matrix, distributed on a spatial grid with $2.5^{\circ}$-latitude-by-$5^\circ$-longitude resolution and is exactly 1 hour apart in time.

The matrix-valued time series, such as the TEC time series, is often associated with auxiliary vector time series that measure the same object, such as the Earth's ionosphere, from a different data modality. In panel (E) of Figure \ref{fig:TEC-Example}, we plot the global SYM-H index, which measures the geomagnetic activity caused by solar eruptions that can finally affect the global TEC distribution. These non-spatial auxiliary covariates carry additional information related to the matrix time series dynamics, as one can tell from the sudden decrease of the Sym-H index around 00:00 UT on September 8, 2017, and the associated intensification of the global TEC near the equatorial regions.


This paper investigates the problem of forecasting future matrix data jointly with the historical matrices and the vector time series covariates. There are two significant challenges in this modeling context. In order to build a matrix-variate regression model, we need to integrate the information of predictors with non-uniform modes, namely, both matrices and vectors. Adding the auxiliary vector covariates benefits the prediction and enables domain scientists to understand the interplay between different data modalities, but complicates the modeling and the subsequent theoretical analysis. From the perspective of spatio-temporal analysis~\citep{cressie2015statistics}, we need to properly leverage the spatial information of the data and transform the classical spatial statistics framework to accommodate the grid geometry of matrix-valued data. In the remainder of this section, we briefly review the related literature that can shed light on these challenges and then summarize our contributions.

A naive but straightforward prediction model is to vectorize the matrices as vectors and make predictions via Vector Autoregression (VAR)~\citep{stock2001vector}. The auxiliary vector covariates can be incorporated once concatenated with the vectorized matrix predictors. However, vectorizing matrix data leads to the loss of spatial information and also requires a significant amount of parameters, given the high dimensionality of the data. To avoid vectorizing the matrix data, scalar-on-tensor regression~\citep{zhou2013tensor,guhaniyogi2017bayesian,li2018tucker,papadogeorgou2021soft} tackles the problem by using matrix predictors directly. However, these models are built for \textit{scalar} responses while in our setting we are dealing with \textit{matrix} responses. Dividing the matrix into individual scalars and fitting scalar-on-tensor regressions still requires a significant number of parameters and, more importantly, it fails to take the spatial information of the response into account.

The statistical framework that can incorporate matrices as both predictors and response is the tensor-on-tensor regression~\citep{lock2018tensor,liu2020low,luo2022tensor} and, more specifically, for time series data, the matrix/tensor autoregression~\citep{chen2021autoregressive,li2021multi,hsu2021matrix,wang2024high}. The matrix/tensor predictors are mapped to matrix/tensor responses via multi-linear transformations that significantly reduce the number of parameters. Our work builds on this framework and incorporates the non-spatial vector predictors under a unified framework.

To incorporate the vector predictor in the same model, we need to map vector predictors to matrix responses. Tensor-on-scalar regression~\citep{rabusseau2016low,sun2017store,li2017parsimonious,guha2021bayesian} illustrates a way of mapping low-order scalar/vector predictors to high-order matrix/tensor responses via taking the tensor-vector product between the vector predictor and a high-order tensor coefficient. Similarly, we introduce a 3-D tensor coefficient for the vector predictors such that our model can take predictors with non-uniform modes, which is a key distinction of our model compared to existing works.

The other distinction of our model is that our model utilizes the spatial information of the matrix response. In our model, a key assumption is that the vector predictor has similar predictive effects on neighboring locations in the matrix response. This is equivalent to saying that the tensor coefficient is spatially smooth and is typically done via adding a total-variation (TV) penalty~\citep{wang2017generalized,shen2022smooth,sun2023tensorGP} to the unknown tensor. The TV penalty leads to piecewise smooth estimators with sharp edges and enables feature selections. However, the estimation with the TV penalty requires solving non-convex optimization problems, making the subsequent theoretical analysis difficult. Our model uses a simpler approach by assuming that the tensor coefficients are discrete evaluations of functional parameters from a Reproducing Kernel Hilbert Space (RKHS). Such a kernel method has been widely used in scalar-on-image regressions~\citep{kang2018scalar} where the regression coefficients of the image predictor are constrained to be spatially smooth.

We facilitate the estimation of the unknown functional parameters with the functional norm penalty. Functional norm penalties have been widely used for estimating smooth functions in classic semi/non-parametric learning in which data variables are either scalar/vector-valued \citep[see][]{hastie2009, gu2013, yuan2010, cai2012, shang2013aos, shang2015aos, shang2015aos:b, shang:colt:2020}. To the best of our knowledge, the present article is the first to consider a functional norm penalty for tensor coefficient estimation in a matrix autoregressive setting. 

To summarize, our paper has two major contributions. Firstly, we build a unified matrix autoregression framework for spatio-temporal data that incorporates lower-order scalar/vector time series covariates. Such a framework has strong application motivation where domain scientists are curious about integrating spatial and non-spatial data information for predictions and inference. The framework also bridges regression methodologies with tensor predictors and responses of non-uniform modes, making the theoretical investigation itself an interesting topic. Secondly, we propose to estimate coefficients of the auxiliary covariates, together with the autoregressive coefficients, in a single penalized maximum likelihood estimation (MLE) framework with the RKHS functional norm penalty. The RKHS framework builds spatial continuity into the regression coefficients. We establish the joint asymptotics of the autoregressive coefficients and the functional parameters under fixed/high matrix dimensionality regimes and propose an efficient alternating minimization algorithm for estimation and validate it with extensive simulations and real applications.

The remainder of the paper is organized as follows. We introduce our model formally in Section~\ref{sec:model} and provide model interpretations and comparisons in sufficient detail. Section~\ref{sec:algorithm} introduces the penalized MLE framework and describes an alternating minimization framework for estimation. Large sample properties of the estimators under fixed and high matrix dimensionality are established in Section~\ref{sec:theory}. Section~\ref{sec:simulation} provides extensive simulation studies for validating the consistency of the estimators, demonstrating BIC-based model selection results, and comparing our method with various competitors. We apply our method to the global TEC data in Section~\ref{sec:application} and draw conclusions in Section~\ref{sec:summary}. Technical proofs and additional details of the algorithm and simulations are deferred to supplemental materials.

\section{Model}\label{sec:model}
\subsection{Notation}
We adopt the following notations throughout the article. We use calligraphic bold-face letters (e.g., $\tenX,\tenG$) for tensors with at least three modes, uppercase bold-face letters (e.g., $\matX,\matG$) for matrices, and lowercase bold-face letters (e.g., $\vecx, \vecz$) for vectors and blackboard bold-faced letters for sets (e.g., $\mathbb{R}, \setS$). To subscript any tensor/matrix/vector, we use square brackets with subscripts such as $[\tenG]_{ijd},[\vecz_{t}]_{d},[\matX_{t}]_{ij}$, and we keep the subscript $t$ inside the square bracket to index time. Any fibers and slices of tensor are subscripted with colons, such as $[\tenG]_{ij:}$, $[\tenG]_{::d}$, and thus any row and column of a matrix is denoted as $[\matX_{t}]_{i:}$ and $[\matX_{t}]_{:j}$. If the slices of tensor/matrix are based on the last mode such as $[\tenG]_{::d}$ and $[\matX_{t}]_{:j}$, we will often omit the colons and write as $[\tenG]_{d}$ and $[\matX_{t}]_{j}$ for brevity. For any tensor $\tenX$, we use $\vect{\tenX}$ to denote the vectorized tensor. For any two tensors $\tenX, \mathbfcal{Y}$ with identical size, we define their inner product as: $\inner{\tenX}{\mathbfcal{Y}} = \vect{\tenX}^{\top}\vect{\tenY}$, and we use $\twonorm{\tenX}$ to denote the Frobenius norm of a tensor and one has $\twonorm{\tenX} = \sqrt{\inner{\tenX}{\tenX}}$. 

Following \citet{li2017parsimonious}, the \textit{tensor-vector product} between a tensor $\tenG$ of size $d_{1}\times\cdots\times d_{K+1}$ and a vector $\vecz\in\mathbb{R}^{d_{K+1}}$, denoted as $\tenG \tvprod_{(K+1)} \vecz$, or simply $\tenG\tvprod\vecz$, is a tensor of size $d_{1}\times\cdots\times d_{K}$ with $[\tenG \tvprod \vecz]_{i_{1}\ldots i_{K}} = \sum_{i_{K+1}} [\tenG]_{i_{1}\ldots i_{K}i_{K+1}}\cdot [\vecz]_{i_{K+1}}$. For tensor $\tenX\in\mathbb{R}^{d_{1}\times\cdots\times d_{K}}$, we use $\matX_{(k)}\in\mathbb{R}^{d_{k}\times \prod_{m\neq k}d_{m}}$ to denote its $k$-mode matricization. The Kronecker product between matrices is denoted via $\matA\otimes\matB$ and the trace of a square matrix $\matA$ is denoted as $\tr{\matA}$. We use $\maxeigen{\cdot},\mineigen{\cdot},\rho_i(\cdot)$ to denote the maximum, minimum and $i^{\rm{th}}$ largest eigenvalue of a matrix. We use $diag(\matC_1,\ldots,\matC_d)$ to denote a block diagonal matrix with $\matC_1,\ldots,\matC_d$ along the diagonal. More on tensor notations can be found in \citet{kolda2009tensor}.

For the matrix time series $\matX_t\in\mathbb{R}^{M\times N}$ in our modeling context, we assume that all $S=MN$ grid locations are points on an $M\times N$ grid within the domain $\bar{\setS} = [0,1]^2$. The collection of all the spatial locations is denoted as $\setS$ and any particular element of $\setS$ corresponding to the $(i,j)^{\rm{th}}$ entry of the matrix is denoted as $s_{ij}$. We will often index the $(i,j)^{\rm{th}}$ entry of the matrix $\matX_t$ with a single index $u=i+(j-1)M$ and thus $s_{ij}$ will be denoted as $s_u$. We use $[N]$ to denote index set, i.e., $[N] = \{1,2,\ldots,N\}$. Finally, we use $k(\cdot,\cdot): \bar{\setS}\times \bar{\setS} \mapsto \mathbb{R}$ to represent a spatial kernel function and $\RKHS$ to denote the corresponding Reproducing Kernel Hilbert Space (RKHS).

\subsection{Matrix AutoRegression with Auxiliary Covariates (MARAC)}
Let $\{\matX_{t},\vecz_{t}\}_{t=1}^{T}$ be a joint observation of the matrix and the auxiliary vector time series with $\matX_{t} \in\mathbb{R}^{M\times N},\vecz_{t}\in\mathbb{R}^{D}$. To forecast $\matX_{t}$, we propose our Matrix AutoRegression with Auxiliary Covariates, or MARAC, as:
\begin{equation}\label{eq:MARAC-model}
    \matX_{t} = \sum_{p=1}^{P} \matA_p \matX_{t-p} \matB_{p}^{\top} + \sum_{q=1}^{Q} \tenG_{q} \tvprod \vecz_{t-q} + \matE_{t},
\end{equation}
where $\matA_{p}\in\mathbb{R}^{M\times M},\matB_{p}\in\mathbb{R}^{N\times N}$ are the autoregressive coefficients for the lag-$p$ matrix predictor and $\tenG_{q}\in\mathbb{R}^{M\times N\times D}$ is the tensor coefficient for the lag-$q$ vector predictor, and $\matE_{t}$ is a noise term whose distribution will be specified later. The lag parameters $P,Q$ are hyperparameters of the model, and we often refer to the model~\eqref{eq:MARAC-model} as MARAC$(P,Q)$.

Based on model~\eqref{eq:MARAC-model}, for the $(i,j)^{\rm{th}}$ element of $\matX_t$, the MARAC$(P,Q)$ specifies the following model:
\begin{equation}\label{eq:MARAC-element-model}
    [\matX_t]_{ij} = \sum_{p=1}^{P} \binner{[\matA_{p}]_{i:}^{\top}[\matB_{p}]_{j:}}{\matX_{t-p}} + \sum_{q=1}^{Q} [\tenG_{q}]_{ij:}^{\top}\vecz_{t-q} + [\matE_{t}]_{ij},
\end{equation}
where each autoregressive term is associated with a rank-$1$ coefficient matrix determined by the specific rows from $\matA_p,\matB_p$, and each non-spatial auxiliary covariate is associated with a coefficient vector that is location-specific, i.e., $[\tenG_{q}]_{ij:}$. It now becomes more evident from~\eqref{eq:MARAC-element-model} that the auxiliary vector covariates enter the model via an elementwise linear model. The autoregressive term utilizes $\matA_p,\matB_p$ to transform each lag-$p$ predictor in a bi-linear form. Using such a bi-linear transformation greatly reduces the total number of parameters of the autoregressive term from $O(M^{2}N^{2})$ to $O((M^{2}+N^{2}))$. When the spatial dimensions $M, N$ are high, one can further reduce the dimensionality by assuming $\matA_p,\matB_p$ are low-rank matrices~\citep{xiao2022reduced}, or the tensors with frontal slices being $\matA_p,\matB_p$ are low-rank~\citep{wang2022high}. Apart from low-rank structure, one could also constrain the autoregressive coefficients to generate smooth predictions via restricting $\matA_p,\matB_p$ to be matrices with a banded structure~\citep{guo2016banded,hsu2021matrix}. However, these configurations would significantly complicate the modeling and theoretical analysis and result in additional model selection problems. Furthermore, in our satellite imaging data, the dependency structure can be arbitrary (among spatial locations). Thus, we do not want to be constrained by any assumptions. In this paper, we keep a more straightforward setup and put no constraints on $\matA_p,\matB_p$ while acknowledging that additional constraints can benefit the computational efficiency under high-dimensional settings. Additionally, we consider a setting where $D$, the dimension of the auxiliary covariates $\vecz_t$, is fixed instead of growing with $M$ and $N$. This setup greatly simplifies the theoretical analysis and reflects the application scenario where one has a fixed set of auxiliary predictors but a matrix-valued data with growing spatial resolution.

For the tensor coefficient $\tenG_q$, we assume that it is spatially smooth. More specifically, we assume that $[\tenG_{q}]_{ijd}$ and $[\tenG_{q}]_{uvd}$ are similar if $s_{ij},s_{uv}$ are spatially close. Formally, we assume that each $[\tenG_{q}]_{d}$, i.e. the coefficient matrix for the $d^{\rm{th}}$ covariate at lag-$q$, is a discrete evaluation of a function $g_{q,d}(\cdot): [0,1]^{2}\mapsto\mathbb{R}$ on $\setS$. Furthermore, each $g_{q,d}(\cdot)$ comes from an RKHS $\RKHS$ endowed with the spatial kernel function $k(\cdot,\cdot)$. The spatial kernel function specifies the spatial smoothness of the functional parameters $g_{q,d}(\cdot)$ and thus the tensor coefficient $\tenG_{q}$. 

An alternative formulation for $\tenG_{q}$ would be a low-rank form~\citep{li2017parsimonious}. Similar low-rank assumptions can be found in matrix time series factor model~\citep{chen2023statistical,gao2023denoising,gao2023two}, where our vector predictors $\{\vecz_t\}_{t=-\infty}^{\infty}$ become a matrix-valued, unknown factor time series. Typically, low-rank representations could significantly reduce the dimensionality of parameters in contexts with limited data~\citep{zhou2013tensor}. However, even with a low-rank structure over $\tenG_{q}$, the number of parameters is still at the order of $O(D(M^2+N^2))$, bounded by the autoregressive parameters, and thus cannot benefit from the low-rank form but complicates the theoretical analysis. Also, we are motivated by applications that forecast time series of spatial data with vector predictors, and we want to model the spatial continuity of the regression coefficients explicitly. Therefore, we choose an RKHS framework over the low-rank framework.


Finally, for the additive noise term $\matE_t$, we assume that it is i.i.d. from a multivariate normal distribution with a separable Kronecker-product covariance:
\begin{equation}\label{eq:Et_Covariance}
    \vect{\matE_{t}} \widesim{\mathrm{i.i.d.}} \mathcal{N}\left(\mathbf{0}, \bSigmac \otimes \bSigmar\right), \quad t\in[T]
\end{equation}
where $\bSigmar\in\mathbb{R}^{M\times M}, \bSigmac\in\mathbb{R}^{N\times N}$ are the row/column covariance components. Such a Kronecker-product covariance is commonly seen in the covariance models for multi-way data \citep{hoff2011separable,tsiligkaridis2013convergence,fosdick2014separable,zhou2014gemini,lyu2019tensor,li2021multi} with the merit of reducing the number of parameters significantly.

Compared to existing models that only deal with either matrix or vector predictors, our model~\eqref{eq:MARAC-model} can incorporate predictors with non-uniform modes. If one redefines $\matE_t$ in our model as $\sum_{q=1}^{Q} \tenG_{q} \tvprod \vecz_{t-q} + \matE_{t}$, i.e., all terms except the autoregressive term, then our model ends up specifying:
\begin{align*}
    \mathrm{Cov}(\vect{\matE_t},\vect{\matE_{t^{\prime}}}) & = \mathbbm{1}_{\{t=t^{\prime}\}}\cdot \bSigmac\otimes\bSigmar + \mathbf{F}\mathbf{M}\mathbf{F}^{\top} \\
    \mathbf{F} = [\left(\tenG_1\right)_{(3)}^{\top}:\cdots:\left(\tenG_Q\right)_{(3)}^{\top}], & \quad \mathbf{M} = \left[\mathrm{Cov}(\vecz_{t-q_1},\vecz_{t^{\prime}-q_2})\right]_{q_1,q_2\in[Q]}
\end{align*}
where $(\tenG_q)_{(3)}$ is the mode-3 matricization of $\tenG_{q}$ and we will use $\matG_{q}$ to denote it for the rest of the paper. This new formulation reveals how our model differs from other autoregression models with matrix predictors. The covariance of $\matE_t, \matE_{t^\prime}$ in our model contains a separable covariance matrix $\bSigmac\otimes\bSigmar$ that is based on the matrix grid geometry, a locally smooth coefficient matrix $\mathbf{F}$ that captures the local spatial dependency, and an auto-covariance matrix $\mathbf{M}$ that captures the temporal dependency. Consequently, entries of $\matE_t$ are more correlated if they are either spatially/temporally close or share the same row/column index and are thus more flexible for spatial data distributed on a matrix grid.

As a comparison, in the kriging framework~\citep{cressie1986kriging}, the covariance of $\matE_t, \matE_{t^\prime}$ is characterized by a spatio-temporal kernel that captures the dependencies among spatial and temporal neighbors. Such a kernel method can account for the local dependency but not the spatial dependency based on the matrix grid geometry. In the matrix autoregression model~\citep{chen2021autoregressive}, the authors do not consider the local spatial dependencies among entries of $\matE_t$ nor the temporal dependency across different $t$. In~\citet{hsu2021matrix}, the matrix autoregression model is generalized to adapt to spatial data via fixed-rank co-kriging (FRC)~\citep{cressie2008fixed} with $\mathrm{Cov}(\vect{\matE_t},\vect{\matE_{t^{\prime}}}) = \mathbbm{1}_{\{t=t^{\prime}\}}\cdot\bSigmac\otimes\bSigmar + \mathbf{F}\mathbf{M}\mathbf{F}^{\top}$, where $\mathbf{M}$ is a $k\times k$ coefficient matrix and $\mathbf{F}$ is a pre-specified $MN\times k$ spatial basis matrix. Such a co-kriging framework does not account for the temporal dependency of the noises, nor does it consider the auxiliary covariates. Our model generalizes these previous works to allow for temporally dependent noise with both local and grid spatial dependency.

\begin{figure}[h]
    \centering
    \includegraphics[width=0.98\linewidth]{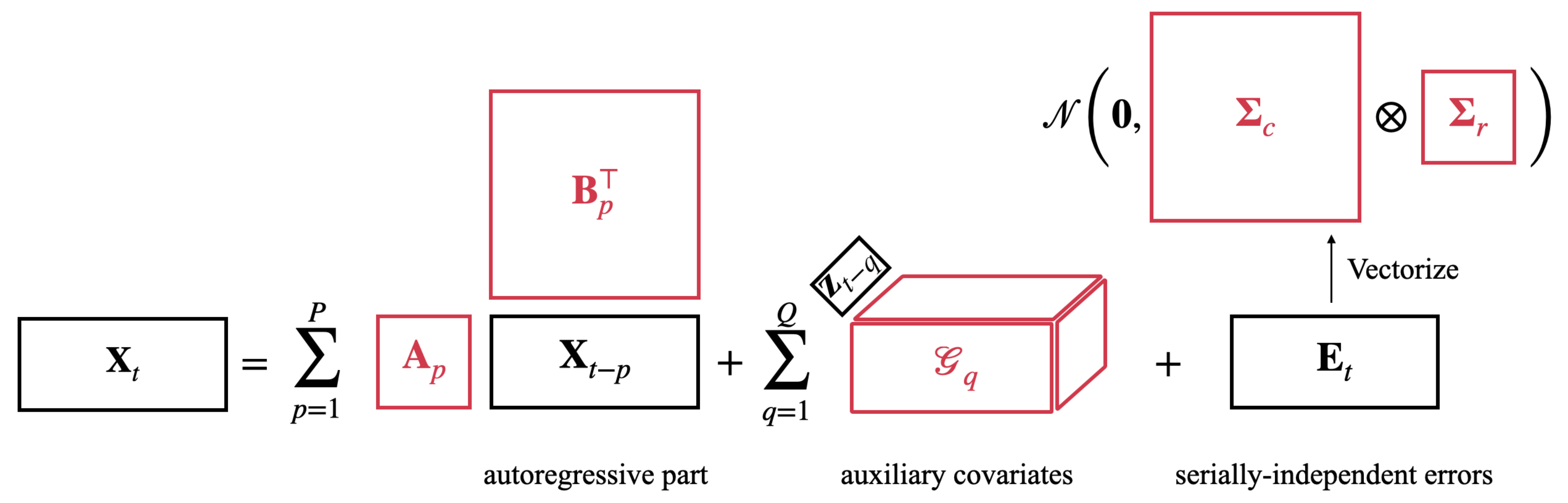}
    \caption{A schematic illustration of the MARAC model in~\eqref{eq:MARAC-model} and~\eqref{eq:Et_Covariance}. All parameters are highlighted in red.}
    \label{fig:visualmodel}
\end{figure}

The combination of~\eqref{eq:MARAC-model} and~\eqref{eq:Et_Covariance} specifies the complete MARAC$(P,Q)$ model. We visualize our MARAC model in Figure~\ref{fig:visualmodel}.
Vectorizing both sides of~\eqref{eq:MARAC-model} yields the vectorized MARAC$(P,Q)$ model:
\begin{equation}\label{eq:MARAC-model-vec}
    \vecx_{t} = \sum_{p=1}^{P} \left(\matB_{p}\otimes \matA_{p}\right) \vecx_{t-p} + \sum_{q=1}^{Q} \matG_{q}^{\top}\vecz_{t-q} + \vece_{t}, \quad \vece_{t}\widesim{\mathrm{i.i.d.}} \mathcal{N}\left(\mathbf{0}, \bSigmac \otimes \bSigmar\right)
\end{equation}
where $\vecx_{t} = \vect{\matX_{t}}, \vece_{t} = \vect{\matE_{t}}$, and recall that $\matG_{q} = (\tenG_q)_{(3)}$. It is now more evident that the Kronecker-product structure over the autoregressive coefficient matrix and the noise covariance matrix greatly reduces the number of parameters, making the regression estimation feasible given finite samples. The spatially smooth structure of $\matG_{q}$ leverages the spatial information of the spatial data. In the next section, we will discuss the estimating algorithm of the model parameters of MARAC.

\section{Estimating Algorithm}\label{sec:algorithm}
This section discusses the parameter estimation for the MARAC$(P,Q)$ model in~\eqref{eq:MARAC-model}. We propose a penalized maximum likelihood estimator (MLE) in Section~\ref{subsec:PMLE} for exact parameter estimation. Then in Section~\ref{subsec:model_selection}, we outline the model selection criterion for selecting the lag hyperparameters whose consistency will be validated empirically in Section~\ref{sec:simulation}. 

\subsection{Penalized Maximum Likelihood Estimation (MLE)}\label{subsec:PMLE}
To estimate the parameters of the MARAC$(P,Q)$ model, which we denote collectively as $\matTheta$, we propose a penalized maximum likelihood estimation (MLE) approach. Following the distribution assumption on $\matE_t$ in~\eqref{eq:Et_Covariance}, we can write the negative log-likelihood of $\{\matX_t\}_{t=1}^{T}$ with a squared RKHS functional norm penalty, after dropping the constants, as:
\begin{equation}\label{eq:MARAC-PMLE-Likelihood}
    \mathfrak{L}_{\lambda}(\matTheta) = -\frac{1}{T}\sum_{t\in[T]} \ell\left(\matX_{t}|\matX_{t-1:P},\vecz_{t-1:Q};\matTheta\right) + \frac{\lambda}{2} \sum_{q\in[Q]}\sum_{d\in[D]} \|g_{q,d}\|_{\RKHS}^{2},
\end{equation}
where $\ell(\cdot)$ is the conditional log-likelihood of $\matX_{t}$:
\begin{equation}\label{eq:log-likelihood}
    \ell\left(\matX_{t}|\matX_{t-1:P},\vecz_{t-1:Q};\matTheta\right) = -\frac12 \log|\bSigmac\otimes\bSigmar| - \frac12 \vecr_{t}^{\top}\left(\bSigmac^{-1}\otimes\bSigmar^{-1}\right)\vecr_{t},
\end{equation}
and $\vecr_{t} = \vecx_{t} - \sum_{p} (\matB_p\otimes\matA_p)\vecx_{t-p} - \sum_{q} \matG_{q}^{\top}\vecz_{t-q}$ is the vectorized residual at $t$. To estimate the parameters, one needs to solve a constrained minimization problem:
\begin{equation}\label{eq:MARAC-PMLE-Problem}
    \min_{\matTheta} \mathfrak{L}_{\lambda}(\matTheta), \quad \text{s.t. } g_{q,d}(s_{ij}) = [\tenG_{q}]_{ijd}, \text{ for all }s_{ij}\in\setS.
\end{equation}

We now explicitly define the functional norm penalty in~\eqref{eq:MARAC-PMLE-Likelihood} and derive a \textit{finite-dimensional equivalent} of the optimization problem above. We assume that the spatial kernel function $k(\cdot,\cdot)$ is continuous and square-integrable. Thus, it has an eigen-decomposition following Mercer's Theorem \citep{williams2006gaussian}:
\begin{equation}\label{eq:Kernel-Decomposition}
k(s_{ij},s_{uv}) = \sum_{r=1}^{\infty}\lambda_{r}\psi_{r}(s_{ij})\psi_{r}(s_{uv}), \quad s_{ij}, s_{uv} \in [0,1]^2,
\end{equation}
where $\lambda_{1}\ge \lambda_{2}\ge \ldots$ is a sequence of non-negative eigenvalues and $\psi_{1},\psi_{2},\ldots$ is a set of orthonormal eigen-functions on $[0,1]^{2}$. The functional norm of function $g$ from the RKHS $\RKHS$ endowed with kernel $k(\cdot,\cdot)$ is defined as:
\begin{equation}\label{eq:func_norm_penalty}
    \|g\|_{\RKHS} = \sqrt{\sum_{r=1}^{\infty}\frac{\beta_{r}^{2}}{\lambda_r}}, \qquad \text{where } g(\cdot) = \sum_{r=1}^{\infty} \beta_{r}\psi_r(\cdot),
\end{equation}
following~\citet{van2008reproducing}.

Given any $\lambda > 0$ in~\eqref{eq:MARAC-PMLE-Likelihood}, the generalized representer theorem~\citep{scholkopf2001generalized} suggests that the solution of the functional parameters, denoted as $\{\widetilde{g}_{q,d}\}_{q=1,d=1}^{Q,D}$, of the minimization problem~\eqref{eq:MARAC-PMLE-Problem}, with all other parameters held fixed, is a linear combination of the representers $\{k(\cdot,s)\}_{s\in \setS}$ plus a linear combination of the basis functions $\{\phi_{1},\ldots,\phi_{J}\}$ of the null space of $\RKHS$, i.e.,
\begin{equation}\label{eq:Representer-Theorem-Result}
    \widetilde{g}_{q,d}(\cdot) = \sum_{s\in\setS} \gamma_{s}k(\cdot,s) + \sum_{j=1}^{J} \alpha_{j}\phi_{j}(\cdot), \quad \|\phi_{j}\|_{\RKHS} = 0,
\end{equation}
where we omit the subscript $(q,d)$ for the coefficient $\gamma_s,\alpha_{j}$ for brevity but they are essentially different for each $(q,d)$. We assume that the null space of $\RKHS$ contains only the zero function for the remainder of the paper. As a consequence of~\eqref{eq:Representer-Theorem-Result}, the minimization problem in~\eqref{eq:MARAC-PMLE-Problem} can be reduced to a finite-dimensional Kernel Ridge Regression (KRR) problem. We summarize the discussion above in Proposition~\ref{thm:KRR-equivalence}:
\begin{proposition}\label{thm:KRR-equivalence}
    If $\lambda > 0$, the constrained minimization problem in~\eqref{eq:MARAC-PMLE-Problem} is equivalent to the following unconstrained kernel ridge regression problem:
    \begin{equation}\label{eq:MARAC-PMLE-KRR}
        \min_{\matTheta} \left\{\frac12\log|\bSigmac\otimes\bSigmar| + \frac{1}{2T}\sum_{t\in[T]} \vecr_{t}^{\top}\left(\bSigmac^{-1}\otimes\bSigmar^{-1}\right)\vecr_{t} + \frac{\lambda}{2}\sum_{q\in[Q]} \tr{\matgamma_{q}^{\top}\matK\matgamma_{q}}\right\},
    \end{equation}
    where $\vecr_{t} = \vecx_{t} - \sum_{p} (\matB_p\otimes\matA_p)\vecx_{t-p} - \sum_{q} \matK\matgamma_{q}\vecz_{t-q}$ is the vectorized residual, $\matK\in\mathbb{R}^{MN\times MN}$ is the kernel Gram matrix with $[\matK]_{u_1u_2} = k(s_{i_1j_1},s_{i_2j_2}), s_{i_lj_l}\in\setS, u_l=i_l+(j_l-1)M,l=1,2$ and $\matgamma_{q}\in\mathbb{R}^{MN\times D}$ contains the coefficients of the representers with $[\matgamma_{q}]_{ud}$ being the coefficient for the $u^{\rm{th}}$ representer $k(\cdot,s_{u})$ and the $d^{\rm{th}}$ auxiliary covariate at lag $q$. 
\end{proposition}

We provide proof in the supplemental material. After introducing the functional norm penalty, the original tensor coefficient is now converted to a linear combination of the representer functions with the relationship that $[\tenG_{q}]_{ijd} = \inner{[\matK]_{u:}^{\top}}{[\matgamma_q]_{:d}}$ where $u=i+(j-1)M$. For more efficient computation, one can use a set of basis functions based on the spectral decomposition of the selected kernel as an approximation: $[\tenG_{q}]_{ijd}\approx\sum_{r\in R} [\vectheta_{q,d}]_r\psi_r(s_{ij})$. The choice of the number of basis functions can be determined via cross-validation, and generally, we observe better results with more basis functions, given enough data. We discuss this approach in Section~\ref{subsec:truncation} of the supplemental material.

The choice of the kernel function $k(\cdot, \cdot)$ depends on the application context. For spatial data in Euclidean space, common choices include the radial-basis function (RBF) kernel or the Mat\'ern kernel~\citep[Sec.~4.2]{williams2006gaussian}. For data distributed on a sphere, which is the context of the TEC data used in this paper, one could consider the Lebedev kernel~\citep{kennedy2013classification} or the von Mises-Fisher kernel~\citep{banerjee2005clustering}. 

We attempt to solve the minimization problem in \eqref{eq:MARAC-PMLE-KRR} with an alternating minimization algorithm~\citep{attouch2013convergence} where we update one block of parameters at a time, keeping the others fixed, following the order of: 
$\matA_{1}\rightarrow\matB_{1}\rightarrow\cdots\rightarrow\matA_{P}\rightarrow\matB_{P}\rightarrow\matgamma_{1}\rightarrow\cdots\rightarrow\matgamma_{Q}\rightarrow\bSigmar\rightarrow\bSigmac\rightarrow\matA_{1}\rightarrow\cdots$. We choose the alternating minimization algorithm for its simplicity and efficiency. Each step of the algorithm conducts exact minimization over one block of the parameters, leading to a non-increasing sequence of the objective function, which guarantees the convergence of the algorithm towards a local stationary point. 
We abstract away the exact updating formula for each parameter here and include them in Section~\ref{sec:algorithm_detail} of the supplemental material. We conclude this session with a remark on identifiability.

\begin{remark}{(Identifiability Constraint)}
    The MARAC$(P,Q)$ model specified in~\eqref{eq:MARAC-model} is scale-unidentifiable in that one can re-scale each pair of $(\matA_{p},\matB_{p})$ by a non-zero constant $c$ and obtain $(c\cdot\matA_{p},c^{-1}\cdot\matB_{p})$ without changing their Kronecker product. To enforce scale identifiability, we re-scale the algorithm output for each pair of $(\matA_{p},\matB_{p})$ such that $\twonorm{\matA_p} = 1$, $\text{sign}(\tr{\matA_{p}}) = 1$. The identifiability constraint is enforced before outputting the estimators.
\end{remark}

\subsection{Lag Selection}\label{subsec:model_selection}
The MARAC$(P,Q)$ model~\eqref{eq:MARAC-model} has three hyperparameters: the autoregressive lag $P$, the auxiliary covariate lag $Q$, and the RKHS norm penalty weight $\lambda$. In practice, $\lambda$ can be chosen by cross-validation, while choosing $P$ and $Q$ requires a more formal model selection criterion. We propose to select $P$ and $Q$ by using information criterion, including the Akaike Information Criterion (AIC)~\citep{akaike1998information} and the Bayesian Information Criterion (BIC)~\citep{schwarz1978estimating}. Here, we formally define the AIC and BIC for the MARAC$(P,Q)$ model and empirically validate their consistency via simulation experiments in Section~\ref{sec:simulation}.

Let $\widehat{\matTheta}$ be the set of the estimated parameters of the MARAC$(P,Q)$ model, and $\mathbf{df}_{P,Q,\lambda}$ be the \textit{effective degrees of freedom} of the model. We can then define the AIC and the BIC as follows:
\begin{align}
    \mathrm{AIC}(P,Q,\lambda) & = -2 \sum_{t\in[T]} \ell(\matX_t|\matX_{t-1:P},\vecz_{t-1:Q},\widehat{\matTheta}) + 2\cdot\mathbf{df}_{P,Q,\lambda} , \label{eq:Akaike-IC} \\
    \mathrm{BIC}(P,Q,\lambda) & = -2 \sum_{t\in[T]} \ell(\matX_t|\matX_{t-1:P},\vecz_{t-1:Q},\widehat{\matTheta}) + \log(T)\cdot\mathbf{df}_{P,Q,\lambda} \label{eq:Bayesian-IC}.
\end{align}
 
To calculate $\mathbf{df}_{P,Q,\lambda}$, we decompose it into the sum of three components: 1) for each pair of the autoregressive coefficient $\widehat{\matA}_p,\widehat{\matB}_p$, the model has $(M^2 + N^2 - 1)$ degrees of freedom; 2) for the noise covariance $\widehat{\bSigma}_r,\widehat{\bSigma}_c$, the model has $(M^2 + N^2)$ degrees of freedom; and 3) for the auxiliary covariate functional parameters $\widehat{g}_{q,1},\ldots,\widehat{g}_{q,D}$, inspired by the kernel ridge regression estimator in~\eqref{eq:Update-Gammaq}, we define the sum of their degrees of freedom as:
\begin{equation*}
    \mathbf{df}_{q}(\widehat{g}) = \text{tr}\left\{\left[\widetilde{\matK} + \lambda\left(\matI_{D}\otimes\widehat{\bSigma}_{c}\otimes\widehat{\bSigma}_r\right)\right]^{-1}\widetilde{\matK}\right\},
\end{equation*}
where $\widetilde{\matK} = \left(T^{-1}\sum_{t\in[T]} \vecz_{t-q}\vecz_{t-q}^{\top}\right)\otimes \matK$. 
As $\lambda\rightarrow 0$, we have $\mathbf{df}_{q}(\widehat{g})\rightarrow MND$; namely, each covariate has $MN$ free parameters, which then reduces to the element-wise linear regression model. Empirically, we find that the BIC is a consistent lag selection criterion for our model.

\section{Theoretical Analysis}\label{sec:theory}
This section presents the theoretical analyses of the MARAC model. We first formulate the condition under which the matrix and vector time series are \textit{jointly stationary}. Under this condition, we then establish the consistency and asymptotic normality of the penalized MLE under \textit{fixed} matrix dimensionality as $T\rightarrow\infty$. Finally, we consider the case where the matrix size goes to infinity as $T\rightarrow\infty$ and derive the convergence rate of the penalized MLE estimator and the optimal order of the functional norm penalty tuning parameter $\lambda$. Without loss of generality, we assume that the matrix and vector time series have zero means, and we use $S=MN$ to denote the spatial dimensionality of the matrix data. All proofs are deferred to the supplemental material.

\subsection{Stationarity Condition}\label{subsec:stationarity}
To facilitate the theoretical analysis, we make another assumption for the vector time series $\vecz_{t}$, which significantly simplifies the presentation of our theoretical analysis.
\begin{assumption}\label{assump:vector_VAR}
    The $D$-dimensional auxiliary vector time series $\{\vecz_t\}_{t=-\infty}^{\infty}$ follows a stationary VAR$(\Qtilde)$ process:
    \begin{equation}\label{eq:Z-VAR-Assumption}
        \vecz_t = \sum_{\qtilde=1}^{\Qtilde} \matC_{\qtilde} \vecz_{t-\qtilde} + \vecnu_t,
    \end{equation}
    where $\matC_{\qtilde} \in \mathbb{R}^{D\times D}$ is the lag-$\qtilde$ transition matrix and $\vecnu_t$ has independent sub-Gaussian entries and is independent of $\matE_t$.
\end{assumption}
Given Assumption~\ref{assump:vector_VAR}, we now derive the condition for $\vecx_t$ and $\vecz_t$ to be \textit{jointly stationary}:
\begin{theorem}[MARAC Stationarity Condition]\label{thm:Joint-Stationarity}
    Assume that Assumption~\ref{assump:vector_VAR} holds for the auxiliary time series $\{\vecz_t\}_{t=-\infty}^{\infty}$, and that the matrix time series $\{\matX_t\}_{t=-\infty}^{\infty}$ is generated by the MARAC$(P,Q)$ model in~\eqref{eq:MARAC-model}, then $\{\matX_t,\vecz_t\}_{t=-\infty}^{\infty}$ are jointly stationary if and only if for any $y\in\mathbb{C}$ in the complex plane such that $|y|\le 1$, we have
    \begin{equation}\label{eq:characteristic_poly}
        \mathrm{det}\left[\matI_{S} - \sum_{p=1}^{P} \left(\matB_p\otimes\matA_p\right)y^{p}\right] \neq 0, \quad \mathrm{det}\left[\matI_D - \sum_{\qtilde=1}^{\Qtilde}\matC_{\qtilde}y^{\qtilde}\right]\neq 0.
    \end{equation}
\end{theorem}
As a special case where $P=\Qtilde=1$, the stationarity condition in~\eqref{eq:characteristic_poly} is equivalent to $\bar{\rho}(\matA_1)\cdot\bar{\rho}(\matB_1) < 1$ and $\bar{\rho}(\matC_1) < 1$, where $\bar{\rho}(\cdot)$ is the spectral radius of a square matrix. Based on Theorem~\ref{thm:Joint-Stationarity}, the stationarity of the matrix and vector time series relies on the stationarity of the autoregressive coefficients of the MARAC$(P,Q)$ and VAR$(\Qtilde)$ models. The tensor coefficients $\tenG_1,\ldots,\tenG_Q$ do not affect the stationarity.

We can relax Assumption~\ref{assump:vector_VAR} to $\{\vecz_t\}_{t=-\infty}^{\infty}$ being covariance-stationary and independent of $\{\matE_t\}_{t=-\infty}^{\infty}$ without affecting most of the theory below, just as the VARX model~\citep{hamilton2020time}. But we decide to keep this assumption for the rest of the paper since having a joint autoregressive process for $\{\matX_t\}_{t=-\infty}^{\infty}$ and $\{\vecz_t\}_{t=-\infty}^{\infty}$ greatly simplifies the analysis, especially under the high-dimensional regime in Section~\ref{subsec:high_dim_asymp}.

\subsection{Finite Spatial Dimension Asymptotics}\label{subsec:fixed_dim_asymp}
In this subsection, we establish the consistency and asymptotic normality of the MARAC model estimators under the scenario that $M,N$ are \textit{fixed}. Given a fixed matrix dimensionality, the functional parameters $g_{q,d}\in\RKHS$ can only be estimated at $S=MN$ fixed locations. Thus, the asymptotic normality result is established for the corresponding tensor coefficient $\widehat{\tenG}_q$. In Section~\ref{subsec:high_dim_asymp}, we will discuss the \textit{double} asymptotics when both $S, T\rightarrow\infty$. For the remainder of the paper, we denote the true model coefficient with an asterisk superscript, such as $\matA_1^{*},\matB_1^{*},\tenG_1^{*}$ and $\bSigma^{*}$. 

To start with, we make an assumption on the Gram matrix $\matK$:
\begin{assumption}\label{assump:Gq-Gammaq}
    The minimum eigenvalue of $\matK$ is bounded below, i.e. $\mineigen{\matK} = \ubar{c} > 0$.
\end{assumption}
As a result of Assumption~\ref{assump:Gq-Gammaq}, every $\tenG_q^{*}$ has a unique kernel decomposition: $\vect{\tenG_{q}^{*}} = (\matI_D\otimes\matK)\vecgamma_{q}^{*}$. Now we are ready to establish the consistency of the covariance matrix estimator $\widehat{\bSigma} = \widehat{\bSigma}_c\otimes\widehat{\bSigma}_r$, which we summarize in Proposition~\ref{thm:error-cov-consistency}.
\begin{proposition}[Covariance Consistency]\label{thm:error-cov-consistency}
    Assume that $\lambda\rightarrow 0$ as $T\rightarrow\infty$ and $S$ is fixed, and Assumption~\ref{assump:vector_VAR},~\ref{assump:Gq-Gammaq} and the stationarity condition in Theorem~\ref{thm:Joint-Stationarity} hold, 
    , then $\widehat{\bSigma}\overset{p.}{\rightarrow}\bSigma^{*}$.
\end{proposition}
We can further establish the asymptotic normality of the other model estimators:
\begin{theorem}[Asymptotic Normality]\label{thm:PMLE-Asymp-Normal}
    Assume that the matrix time series $\{\matX_{t}\}_{t=-\infty}^{\infty}$ follows the MARAC$(P,Q)$ model~\eqref{eq:MARAC-model} with i.i.d. noise $\{\matE_t\}_{t=-\infty}^{\infty}$ following~\eqref{eq:Et_Covariance} and Assumption~\ref{assump:vector_VAR},~\ref{assump:Gq-Gammaq} and the stationarity condition in Theorem~\ref{thm:Joint-Stationarity} hold and $\lambda=o(T^{-1/2})$. Additionally, assume that $\mineigen{\mathrm{Var}([\vect{\matX_t}^{\top}, \vecz_{t}^{\top}]^{\top})} = \ubar{c}^{'} > 0$. Then suppose $M,N$ are fixed and $P,Q$ are known and denote $\vect{\matA_p}, \vect{\matB_p^{\top}}$ as $\vecalpha_p$ and $\vecbeta_p$ for any $p\in[P]$, the penalized MLE of the MARAC$(P,Q)$ model is asymptotically normal:
    \begin{equation}\label{eq:PMLE-Asymp-Dist}
        \sqrt{T}\begin{bmatrix}
            \widehat{\vecbeta}_1\otimes\widehat{\vecalpha}_1 - \vecbeta_1^*\otimes\vecalpha_1^* \\
            \vdots \\
            \widehat{\vecbeta}_P\otimes\widehat{\vecalpha}_P - \vecbeta_P^*\otimes\vecalpha_P^* \\
            \vect{\widehat{\tenG}_1 - \tenG^*_1} \\
            \vdots \\
            \vect{\widehat{\tenG}_Q - \tenG^*_Q}
        \end{bmatrix} \overset{d.}{\longrightarrow} \mathcal{N}\left(\mathbf{0}, \matV\matXi\matV^{\top}\right),
    \end{equation}
    where $\matV$ is: 
    \begin{equation*}
        \matV = \begin{bmatrix}
        \text{diag}(\matV_1,\ldots,\matV_P) & \matO \\
        \matO & \matI_{QD}\otimes\matK
    \end{bmatrix}, \quad \matV_p = [\vecbeta^*_p\otimes \matI_{M^2},\matI_{N^2}\otimes\vecalpha_p^*],
    \end{equation*}
    and $\matXi = \matH^{-1}\mathrm{E}\left[\matW_t^{\top}(\bSigma^{*})^{-1}\matW_t\right]\matH^{-1}$, and $\matW_t$ is defined as:
    \begin{equation*}
        \matW_t = \left[\matW_{0,t}\otimes\matI_{M},
        \matI_N\otimes\matW_{1,t},[\vecz_{t-1}^{\top},\ldots,\vecz_{t-Q}^{\top}]\otimes\matK\right],
    \end{equation*}
    where $\matW_{0,t}=[\matB_1^*\matX_{t-1}^{\top},\ldots,\matB_P^*\matX_{t-P}^{\top}]$, $\matW_{1,t}=[\matA_1^*\matX_{t-1},\ldots,\matA_P^*\matX_{t-P}]$, and:
    \begin{equation*}
        \matH = \mathrm{E}\left[\matW_t^{\top}(\bSigma^*)^{-1}\matW_t\right]+\veczeta\veczeta^{\top}, \quad \veczeta^{\top} = \left[(\vecalpha_1^*)^{\top},\cdots,(\vecalpha_P^*)^{\top},\mathbf{0}^{\top}\right].
    \end{equation*}
\end{theorem}
The asymptotic distribution~\eqref{eq:PMLE-Asymp-Dist} indicates that all parameters are $\sqrt{T}$-consistent under fixed matrix dimensionality. Given this result, we have a corollary on testing the existence of the auxiliary covariates in the model:
\begin{corollary}[Specification Test]\label{cor:SpecTest}
Given the same assumption as Theorem~\ref{thm:PMLE-Asymp-Normal}, we have:
\begin{equation}\label{eq:asymp-dist-G}
T \cdot \left(\widehat{\mathbf{g}}-\mathbf{g}^*\right)^\top \mathbf{\Psi}^{\dagger}\left(\widehat{\mathbf{g}}-\mathbf{g}^*\right) \overset{d.}{\longrightarrow} \chi^2_{r},
\end{equation}
where $\mathbf{g}^* = [\vect{\tenG_1^*}^\top, \ldots, \vect{\tenG_Q^*}^\top]^\top$ and $\widehat{\mathbf{g}}$ is its estimator and $\mathbf{\Psi}^\dagger$ is the Moore-Penrose pseudo-inverse of $\mathbf{\Psi}\coloneqq [\matO:\matI_{QD}\otimes\matK]\matXi[\matO:\matI_{QD}\otimes\matK]^\top$. Furthermore, we can prove that $r \ge MNQD-1$. To test the hypothesis at significance level $\alpha$:
\begin{equation*}
    H_0: \mathbf{g}^* = \mathbf{0}, \quad \text{vs. } H_1: \mathbf{g}^* \neq \mathbf{0},
\end{equation*}
we have a test statistics $T\cdot \widehat{\mathbf{g}}^\top\mathbf{\Psi}^{\dagger}\widehat{\mathbf{g}}$ with a rejection region $\{\widehat{\mathbf{g}}|T\cdot \widehat{\mathbf{g}}^\top\mathbf{\Psi}^{\dagger}\widehat{\mathbf{g}} \ge \chi^2_{r, 1-\alpha}\}$.
\end{corollary}
In practice, we will use plug-in estimators to estimate $\mathbf{\Psi}$ and use $\chi^2_{MNQD-1,1-\alpha}$ as the critical value. The test statistics can take a longer time to compute under large hyperparameters $M, N, P, Q, D$. We will discuss simulation results of this test in Section~\ref{subsec:estimator-consistency} for relatively smaller model hyperparameters, and show a real use case for data application in Section~\ref{sec:application}. We leave the problem of scaling this test to higher-dimensional contexts for future work.

\subsection{High Spatial Dimension Asymptotics}\label{subsec:high_dim_asymp}
The previous section presents the asymptotic normality of the MARAC estimators under a \textit{fixed} matrix dimensionality $S$. In this section, we relax this assumption and establish the convergence rate of the MARAC estimators when $S,T\rightarrow\infty$. For technical reasons, we assume that the covariance of $\vect{\matE_t}$ is known but allows for an arbitrary covariance $\bSigma$.
To establish the convergence rate,
we make several additional assumptions.
\begin{assumption}\label{assump:kernel-eigen-decay}
    The spatial kernel function $k(\cdot,\cdot)$ can be decomposed into the product of a row kernel $k_1(\cdot,\cdot)$ and a column kernel $k_2(\cdot,\cdot)$ that satisfies $k((u,v),(s,t)) = k_1(u,s)k_2(v,t)$. Both $k_1,k_2$ have their eigenvalues decaying at a polynomial rate: $\lambda_{j}(k_1)\asymp j^{-r_0}, \lambda_{j}(k_2)\asymp j^{-r_0}$ with $r_0 \in (1/2,2)$.
\end{assumption}

\begin{assumption}\label{assump:location-uniform-sample}
    The spatial locations of the rows and columns of $\matX_t$ are sampled independently from a uniform distribution on $[0,1]$. 
\end{assumption}

Assumption~\ref{assump:kernel-eigen-decay} elicits a simple eigen-spectrum characterization of the spatial kernel $k(\cdot,\cdot)$, whose eigenvalue can be written as $\lambda_{i}(k_1)\lambda_{j}(k_2)$. Also, the Gram matrix $\matK$ is separable, i.e. $\matK=\matK_2\otimes\matK_1$ and all eigenvalues of $\matK$ have the form of $\rho_i(\matK_1)\rho_j(\matK_2)$, where $\matK_1\in\mathbb{R}^{M\times M},\matK_2\in\mathbb{R}^{N\times N}$ are the Gram matrix for the kernel $k_1,k_2$, respectively. The separability of the kernel can accommodate the grid structure of the spatial locations. 

Under Assumption~\ref{assump:location-uniform-sample}, we further have $\rho_{i}(\matK_1) \rightarrow M\lambda_i(k_1)$ and $\rho_{j}(\matK_2)\rightarrow N\lambda_j(k_2)$, as $M,N\rightarrow\infty$. 
We refer our readers to~\citet{koltchinskii2000random,braun2006accurate} for more references about the eigen-analysis of the kernel Gram matrix. One can generalize Assumption~\ref{assump:location-uniform-sample} to non-uniform sampling, but here, we stick to this more straightforward setting. 
With these assumptions, we are ready to present the main result in Theorem~\ref{thm:high_dimensional_MARAC}.

\begin{theorem}[Asymptotics for High-Dimensional MARAC]\label{thm:high_dimensional_MARAC}
    Assume that Assumptions~\ref{assump:vector_VAR}, ~\ref{assump:kernel-eigen-decay} and~\ref{assump:location-uniform-sample} hold and $\matX_t$ is generated by the MARAC$(P,Q)$ model~\eqref{eq:MARAC-model} with $\vect{\matE_t}\overset{i.i.d}{\sim} \mathcal{N}(\mathbf{0}, \bSigma)$ and $\bSigma$ is known. With $S, T\rightarrow\infty$ ($D$ is fixed) and $S\log S/T\rightarrow 0$, and assume that:
    \begin{enumerate}
        \item $M=O(\sqrt{S}), N=O(\sqrt{S})$;
        \item $\gamma_S \coloneqq \lambda/S\rightarrow 0$ and $\gamma_S\cdot S^{r_0}\rightarrow C_1$ as $S\rightarrow\infty$, with $0 < C_1 \le \infty$;
        \item $\mineigen{\bSigma_{\vecx,\vecx}^*-(\bSigma_{\vecz,\vecx}^*)^{\top}(\bSigma_{\vecz,\vecz}^*)^{-1}\bSigma_{\vecz,\vecx}} = c_{0,S} > 0$ as $S, T\rightarrow\infty$, where $\bSigma_{\vecx,\vecx}^*,\bSigma_{\vecz,\vecz}^*,\bSigma_{\vecz,\vecx}^*$ are $\mathrm{Var}(\vecx_t)$, $\mathrm{Var}(\vecz_t)$ and $\mathrm{Cov}(\vecz_t,\vecx_t)$, respectively. $c_{0,S}$ is a constant that only relates to $S$;
        \item For any $S$, we have $0 < \mineigen{\matK} < \maxeigen{\matK} \le C_0$, where $C_0$ is a finite constant;
        \item $\maxeigen{\bSigma}/\mineigen{\bSigma} \le C_1 < \infty$, where $C_1$ is a constant, and $\maxeigen{\bSigma}=c_{1,S}$,
    \end{enumerate}
    Then we have:
    \begin{equation}\label{eq:high-dim-autoregressive-error}
        \frac{1}{\sqrt{P}S}\sqrt{\sum_{p=1}^{P}\left\|\widehat{\matB}_p\otimes\widehat{\matA}_p- \matB^*_p\otimes\matA_p^*\right\|_{\mathrm{F}}^2} \lesssim O_P\left(\sqrt{\frac{C_g\cdot\gamma_S}{c_{0,S}\cdot S}}\right) + O_P\left(\sqrt{\frac{c_{1,S}\cdot D}{c_{0,S}\cdot TS}}\right),
    \end{equation}
    where $C_g = \sum_{q=1}^{Q}\sum_{d=1}^{D}\|g_{q,d}\|_{\RKHS}^2$. Furthermore, we also have:
    \begin{align}
    & \sqrt{(TS)^{-1}\sum_{t=1}^{T}\left\|\sum_{q=1}^{Q}\left(\widehat{\tenG}_q-\tenG_q^*\right)\tvprod\vecz_{t-q}\right\|_{\mathrm{F}}^2} \notag\\
    & \lesssim O_P\left(\frac{\sqrt{c_{1,S}\cdot\gamma_S^{-1/{2r_0}}}}{\sqrt{T}\sqrt[4]{S}}\right) + O_P(\sqrt{\gamma_S}) + O_P\left(\frac{1}{\sqrt{S}}\right) + O_P\left(\sqrt{\frac{c_{1,S}}{T}}\right) + O_P\left(\frac{\sqrt{c_{1,S}\cdot\gamma_S^{-1}}}{\sqrt{TS}}\right). \label{eq:high-dim-nonparametric-error}
    \end{align}
\end{theorem}
In Theorem~\ref{thm:high_dimensional_MARAC}, \eqref{eq:high-dim-autoregressive-error} gives the error bound of the autoregressive coefficients and~\eqref{eq:high-dim-nonparametric-error} gives the error bound of the prediction made by the auxiliary time series, which contains the functional parameter estimators. As a special case of~\eqref{eq:high-dim-autoregressive-error} where $\gamma_S=0$ and $S$ is fixed, the convergence rate for the autoregressive coefficients is $O_P(T^{-1/2})$, which reproduces the result in Theorem~\ref{thm:PMLE-Asymp-Normal}. For the discussion below, we use $\text{AR}_{err}$ and $\text{AC}_{err}$ as acronyms for the quantity on the left-hand side of~\eqref{eq:high-dim-autoregressive-error} and~\eqref{eq:high-dim-nonparametric-error}.


\begin{remark}[Optimal Choice of $\lambda$ and Phase Transition]\label{remark:optimal-lambda}
    According to our proof, the error bound~\eqref{eq:high-dim-nonparametric-error} can be decomposed into the sum of:
    \begin{itemize}
        \item nonparametric error: $O_{P}\left(\frac{\sqrt{c_{1,S}\cdot\gamma_S^{-1/{2r_0}}}}{\sqrt{T}\sqrt[4]{S}}\right) + O_P(\sqrt{\gamma_S})$,
        \item autoregressive error: $O_P\left(\sqrt{\gamma_S}\right) + O_P\left(S^{-1/2}\right) + O_P\left(\sqrt{\frac{c_{1,S}}{T}}\right) + O_P\left(\frac{\sqrt{c_{1,S}\cdot\gamma_S^{-1}}}{\sqrt{TS}}\right)$,
    \end{itemize}
    where the autoregressive error stems from the estimation error of $\widehat{\matB}_p\otimes\widehat{\matA}_p$. In our model, if there is no autoregressive error, the optimal tuning parameter satisfies $\gamma_S\asymp (TS^{1/2}c_{1,S}^{-1})^{-2r_0/(2r_0+1)}$. Compared to the classical semi-parametric regression result~\citep{cui2018rkhs}, our optimal rate does not exactly scale with the number of data points $TS$, but scales with $T\sqrt{S}$.
    This is a special result for matrix-shaped data. Also, the optimal rate depends upon the correlations among the errors, which is a special result for spatial data. Notably, under $S\log S/T\rightarrow 0$, the autoregressive error dominates the nonparametric error.

    To simplify the discussion of the optimal order of $\gamma_S$, we assume that $S = T^{c}$, where $c<1$ is a constant. Under this condition, when $P, Q\ge 1$, the optimal tuning parameter $\gamma_{S}=\lambda/S$ shows an interesting phase transition phenomenon under different spatial smoothness $r_0$ and matrix dimensionality $c = \log_T S$, which we summarize in Table~\ref{tab:conv-rate-summary}.
    \begin{table}[!htb]
    \centering
    \begin{tabular}{|c|c|c|c|}
    \hline
    $r_0$ & $\log_TS$ & Optimal $\gamma_S$ & Estimator Error \\ \hline
    $[1,2)$ & $[\frac{1}{2r_0-1},1)$ & $O((TS)^{-\frac12})$ & \makecell{$\mathrm{AR}_{err} = O_P(T^{-\frac14}S^{-\frac34})$ \\ $\mathrm{AC}_{err} = O_P(S^{-\frac12})$} \\\hline
    $[1,2)$ & $(0,\frac{1}{2r_0-1})$ & $O(S^{-r_0})$ & \makecell{$\mathrm{AR}_{err} = O_P(S^{-\frac{r_0+1}{2}})$ \\ $\mathrm{AC}_{err} = O_P(S^{-\frac12})$} \\\hline
    $(\frac12,1)$ & $[2r_0-1,1)$ & $O(S^{-r_0(2r_0-1)})$ & \makecell{$\mathrm{AR}_{err} = O_P(S^{-\frac{r_0(2r_0-1)+1}{2}})$ \\ $\mathrm{AC}_{err} = O_P(S^{-\frac12})$} \\\hline
    $(\frac12,1)$ & $(0,2r_0-1)$ & $O((T\sqrt{S})^{-\frac{2r_0}{2r_0+1}})$ & \makecell{$\mathrm{AR}_{err} = O_P((TS)^{-\frac12})+ O_P((T\sqrt{S})^{-\frac{r_0}{2r_0+1}}S^{-\frac12})$ \\ $\mathrm{AC}_{err} = O_P(S^{-\frac12})+ O_P((T\sqrt{S})^{-\frac{r_0}{2r_0+1}})$}\\
    \hline
    \end{tabular}
    \caption{Summary of optimal tuning parameter $\gamma_S$ and estimators error following~\eqref{eq:high-dim-autoregressive-error} and~\eqref{eq:high-dim-nonparametric-error}, under the assumption that $c_{0,S} \ge c_{0} > 0$, $c_{1,S} \le c_1 < \infty$, for all $S$ and $S = T^{c}$ for some constant $0 < c < 1$ such that $S\log S/T\rightarrow 0$. $\mathrm{AR}_{err}$ and $\mathrm{AC}_{err}$ are the quantity on the left-hand side of~\eqref{eq:high-dim-autoregressive-error} and~\eqref{eq:high-dim-nonparametric-error}.}
    \label{tab:conv-rate-summary}
    \end{table}
    Based on the results in Table~\ref{tab:conv-rate-summary}, the faster $S$ grows with respect to $T$, the smaller the optimal tuning parameter $\gamma_S$ is. This is an intuitive result since when one has more spatial locations, the observations are denser, and thus less smoothing is needed. Furthermore, we achieve an optimal tuning order of $\gamma_S$ that is close to the classic nonparametric optimal rate at $(TS)^{-2r_0/(2r_0+1)}$ only under the regime where $1/2 < r_0 < 1$ and $\log_TS < 2r_0-1$. This regime specifies the scenario where the functional parameter is relatively unsmooth, and the spatial dimensionality grows slowly with respect to $T$. Only under this regime will the discrepancy between the nonparametric and autoregressive errors remain small, leading to an optimal tuning parameter close to the result of nonparametric regression.
\end{remark}




In~\eqref{eq:high-dim-autoregressive-error}, the constant $c_{0,S}$ appears in the error bound of the autoregressive term. This constant characterizes the spatial correlation of the matrix time series $\matX_t$, conditioning on the auxiliary vector time series $\vecz_t$, and can vary across different assumptions made on the covariances of $\matE_t$ and $\vecnu_t$. In Table~\ref{tab:conv-rate-summary}, we assume that $c_{0,S} \ge c_{0} > 0$ for some universal constant $c_0$. Unfortunately, in practice, it is common to have $c_{0,S}\rightarrow 0$ as $S\rightarrow\infty$, which makes the autoregressive coefficient converge at a slower rate but does not affect the functional parameter convergence. We leave the constant $c_{0,S}$ here in~\eqref{eq:high-dim-autoregressive-error} to give a general result and leave the characterization of $c_{0,S}$ under specific assumptions for future works.

\section{Simulation Experiments}\label{sec:simulation}

\subsection{Consistency, Convergence Rate and Hypothesis Testing}\label{subsec:estimator-consistency}
In this section, we validate the consistency and convergence rate of the MARAC estimators. We consider a simple setup with $P=Q=1$ and $D=3$ and simulate the autoregressive coefficients $\matA_1^*,\matB_1^*$ such that they satisfy the stationarity condition in Theorem~\ref{thm:Joint-Stationarity}. We specify both $\matA_1^*,\matB_1^*$ and $\bSigma_r^*,\bSigma_c^*$ to have symmetric banded structures. To simulate $g_{1},g_{2},g_{3}$ (we drop the lag subscript) from the RKHS $\RKHS$, we choose $k(\cdot,\cdot)$ to be the Lebedev kernel~\citep{kennedy2013classification} and generate $g_1,g_2,g_3$ randomly from Gaussian processes with the Lebedev kernel as the covariance kernel. Finally, we simulate the auxiliary vector time series $\vecz_t\in\mathbb{R}^{3}$ from a VAR$(1)$ process. We include more details and visualizations of the simulation setups in the supplemental material.

\begin{figure}[!htb]
    \centering
    \includegraphics[width=0.98\textwidth]{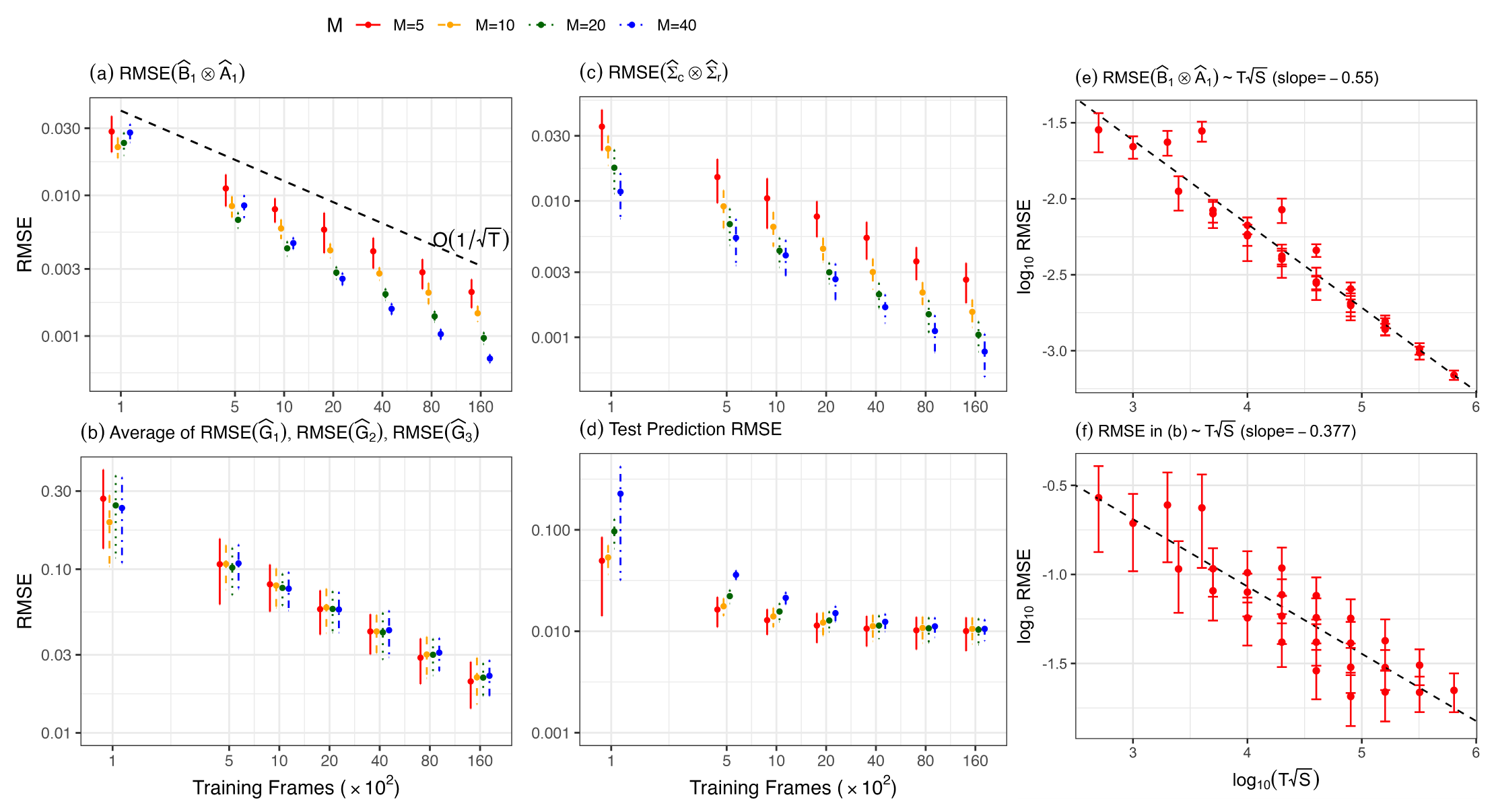}
    \caption{Panel (a), (b), (c) show the RMSE of the penalized MLE of the MARAC model. Panel (d) shows the testing set prediction RMSE subtracted by $1$, where $1$ is the noise variance of the simulated time series. Panels (a)-(d) have both axes plotted in $\log_{10}$ scale. (e) and (f) are the RMSE of the autoregressive parameters and auxiliary covariates parameters under different $T\sqrt{S}$, plotted with both axes in $\log_{10}$ scale together with a fitted linear regression line.}
    \label{fig:consistency-n-convergence}
\end{figure}

The evaluation metric is the rooted mean squared error (RMSE), defined as $\mathrm{RMSE}(\widehat{\boldsymbol{\Theta}}) = \twonorm{\widehat{\boldsymbol{\Theta}}-\boldsymbol{\Theta}^*}/\sqrt{d(\boldsymbol{\Theta}^*)}$, where $d(\boldsymbol{\Theta}^*)$ is the number of elements in $\boldsymbol{\Theta}^*$. We consider $\boldsymbol{\Theta}\in\{\matB_1\otimes\matA_1,\bSigmac\otimes\bSigma_r,\tenG_1,\tenG_2,\tenG_3\}$ and we report the average RMSE for $\tenG_1,\tenG_2,\tenG_3$. The dataset is configured with $M\in\{5,10,20,40\}$ and $N=M$. For each $M$, we train the MARAC model with $P=Q=1$ over $T_{\text{train}}\in\{1,5,10,20,40,80,160\}\times 10^2$ frames of the matrix time series and choose the tuning parameter $\lambda$ based on the prediction RMSE over a held-out validation set with $T_{\text{val}} = T_{\text{train}}/2$. We validate the prediction performance over a $5,000$-frame testing set. 
All results are reported with $20$ repetitions in Figure~\ref{fig:consistency-n-convergence}. 

The result shows that all model estimators are consistent. The convergence rate, under a fixed spatial dimensionality, is close to $1/\sqrt{T}$ (the black line in panel (a) shows a reference line of $O(1/\sqrt{T})$), echoing the result in Theorem~\ref{thm:PMLE-Asymp-Normal}. As the spatial dimensionality $S$ increases, the RMSE for $\widehat{\matB}_1\otimes\widehat{\matA}_{1}$ becomes even smaller, echoing the result in~\eqref{eq:high-dim-autoregressive-error} and Table~\ref{tab:conv-rate-summary}.  The RMSE of the nonparametric estimators $\widehat{g}_{1},\widehat{g}_{2},\widehat{g}_{3}$, under a fixed spatial dimensionality, also decay at a rate of $1/\sqrt{T}$, echoing the result in Theorem~\ref{thm:PMLE-Asymp-Normal} as well. The RMSE of the covariance matrix estimator $\widehat{\bSigma}_c\otimes\widehat{\bSigma}_r$ suggests that it is consistent, confirming the result of Proposition~\ref{thm:error-cov-consistency} and showing a convergence rate similar to $\widehat{\matB}_1\otimes\widehat{\matA}_{1}$, though we did not provide the exact convergence rate theoretically.

In this simulation, we fix the variance of each element of $\vect{\matE_t}$ to be unity. Therefore, the optimal testing set prediction RMSE should be unity. When plotting the test prediction RMSE in (d), we subtract $1$ from all RMSE results, and thus, the RMSE should be interpreted as the RMSE for the \textit{signal} part of the matrix time series. The test prediction RMSE for all cases converges to zero, and for matrices of higher dimensionality, we typically require more training frames to reach the same prediction performance. 

To validate the theoretical result of the high-dimensional MARAC in Theorem~\ref{thm:high_dimensional_MARAC}, we also plot the RMSE of $\widehat{\matB}_1\otimes\widehat{\matA}_{1}$ and $\widehat{g}_{1},\widehat{g}_{2},\widehat{g}_{3}$ against $T\sqrt{S}$ in panel (e) and (f) of Figure~\ref{fig:consistency-n-convergence}. The trend line is fitted by linear regression, and it shows that $\widehat{\matB}_1\otimes\widehat{\matA}_{1}$ converges roughly at the rate of $1/\sqrt{T}\sqrt[4]{S}$, which indicates that $c_{0,S}\asymp 1/\sqrt{S}$ under this specific setup. It also shows that the functional parameter's convergence rate is around $(T\sqrt{S})^{-3/8}$, which coincides with our simulation setup where $r_0\approx 3/4$ and the theoretical result in the last row of Table~\ref{tab:conv-rate-summary}. 

\begin{figure}[!htb]
    \centering
    \includegraphics[width=\textwidth]{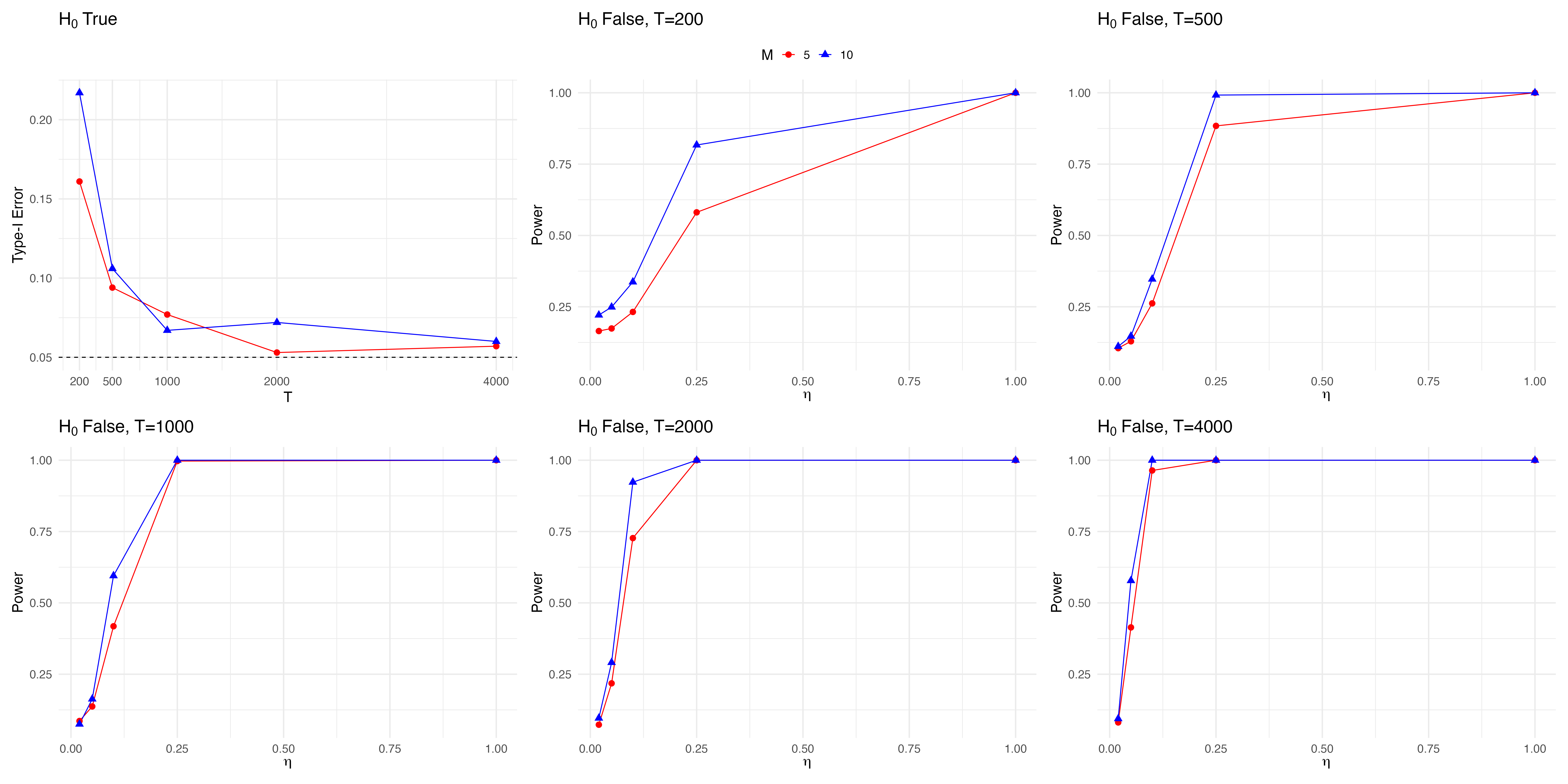}
    \caption{Specification test simulation results. The top left panel shows the Type-I Error rate out of 1000 repetitions across different sample sizes $T$, when $H_0$ is True ($Q=0$). The remaining panels show the power of the test, under different $T$, when the alternative hypothesis is true ($Q=1$) but with varying scaling factor $\eta$ (smaller $\eta$ means the norm of $\tenG^*_1$ is smaller).}
    \label{fig:Specification-Test-Simulation}
\end{figure}

We also conduct finite-sample simulations for the hypothesis testing discussed in Corollary~\ref{cor:SpecTest}. We set $(M,N)\in\{(5,5), (10,10)\}$, $P=1$ and test for both the scenarios of $Q=0$ ($H_0$ is true) and $Q=1$ ($H_0$ is False). For $Q=1$, we further introduce a scaling factor $\eta$ that controls the scale of $\tenG^*_1$, and thus a smaller $\eta$ makes the alternative hypothesis less distinguishable from the null hypothesis. We run the simulation for different sample sizes $T$ with $1000$ repetitions and report the Type I Error rate and power of the test in Figure~\ref{fig:Specification-Test-Simulation}.

In Corollary~\ref{cor:SpecTest}, we lower bound the degrees of freedom of the chi-square distribution by $MNQD-1$ and use it to determine the critical value, and thus our tests could lead to higher Type-I Error and power. However, this lower bound makes very little difference, even given the smaller matrix setup here, and we do see that the Type-I error rate reaches the specified level ($\alpha=0.05$) with relatively larger sample size $T$. The power of the test approaches unity as the sample size $T$ grows or the alternative hypothesis becomes more distinguishable (larger $\eta$). These results justify the applications of the proposed test in practice, and we will be using it in our data application in Section~\ref{sec:application}.

\subsection{Lag Selection Consistency}
In Section~\ref{subsec:model_selection}, we propose to select the lag parameters $P$ and $Q$ of the MARAC model using information criteria such as AIC and BIC. To validate the consistency of these model selection criteria, we simulate data from a MARAC$(2,2)$ model with $5\times 5$ matrix dimensionality. We consider a candidate model class with $1\le P, Q\le 4$, and each model is fitted with $T\in\{1,2,4,8\}\times 10^3$ frames with $\lambda$ being chosen from a held-out validation set. In Table~\ref{tab:lag_select_prob}, we report the proportion of times that AIC and BIC select the correct $P$, $Q$ individually (first two numbers in each parenthesis), and $(P,Q)$ jointly (last number in each parenthesis) from $100$ repetitions. 

\begin{table}[!htb]
    \centering
    \begin{tabular}{|c|c|c|c|c|}
    \hline
         & $T=1000$ & $T=2000$ & $T=4000$ & $T=8000$ \\\hline
     AIC & $(.54,.99,.53)$ & $(.55,.97,.53)$ & $(.59,.96,.55)$ & $(.65,.94,.59)$ \\\hline
     BIC & $(1.00,.09,.09)$ & $(.99,.56,.56)$ & $(.97,.97,.94)$ & $(.96,.99,.95)$ \\\hline
    \end{tabular}
    \caption{Probability that AIC and BIC select the correct $P$ (first number), $Q$ (second number) and $(P,Q)$ (third number) from 100 repetitions.} 
    \label{tab:lag_select_prob}
\end{table}

From Table~\ref{tab:lag_select_prob}, we find that AIC tends to select the model with more autoregressive lags, but BIC performs consistently better under large sample sizes. This coincides with the findings in \citet{hsu2021matrix} for the matrix autoregression model. 

\subsection{Comparison with Alternative Methods}\label{subsec;compare}
We compare our MARAC model against other competing methods for the matrix autoregression task. We simulate the matrix time series $\matX_t$ from a MARAC$(P,Q)$ model, with $P=Q\in\{1,2,3\}$, and the vector time series $\vecz_{t} \in \mathbb{R}^{3}$ from VAR$(1)$. The dataset is generated with $T_{\text{train}} = T_{\text{val}} = T_{\text{test}} = 2000$. Under each $(P,Q)$, we simulate with varying matrix dimensionality with $M=N\in\{5,10,20,40\}$. We evaluate the performance of each method via the testing set prediction RMSE. Each simulation scenario is repeated 20 times. 

\begin{figure}[!htb]
    \centering
    \includegraphics[width=0.98\textwidth]{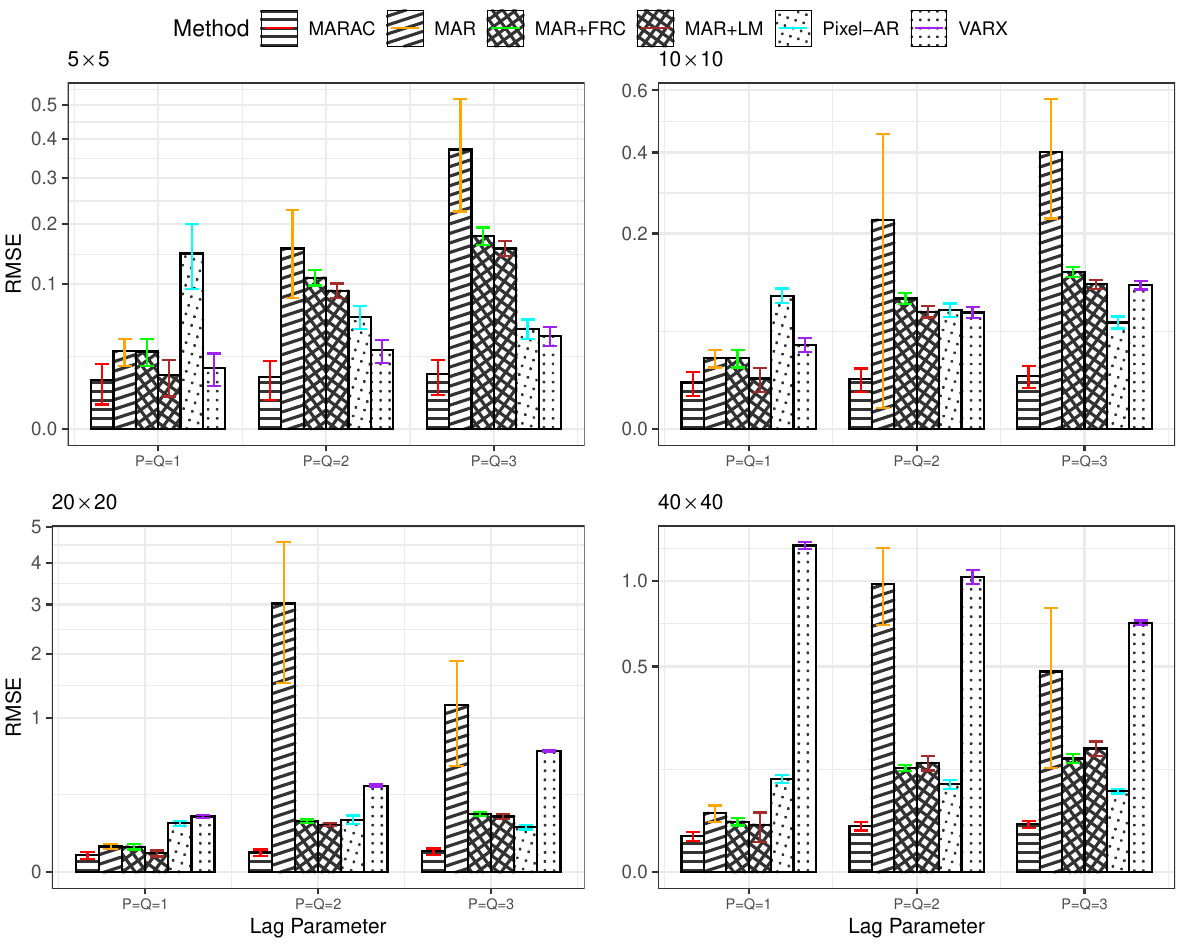}
    \caption{Testing set prediction RMSE comparison across six competing methods on the matrix autoregression task. Four panels correspond to four different matrix dimensionality (labeled on the top-left corner of each panel). Test prediction RMSE is subtracted by $1$ for better visualization, where $1$ is the noise variance of the simulated data. Error bar shows $95\%$ CI of the 20 repeated runs. For better visualization, we rearrange the spacing between ticks along the y-axis using a square root transformation.}
    \label{fig:method-compare-result}
\end{figure}

Under each $P,Q,M,N$ specification, we consider the following five competing methods besides our own MARAC$(P,Q)$ model. 
\begin{enumerate}
    \item MAR~\citep{chen2021autoregressive}: 
    \begin{equation*}
        \matX_{t} = \sum_{p=1}^{P} \matA_p\matX_{t-p}\matB_p^{\top} + \matE_t, \vect{\matE_t}\sim\mathcal{N}(\mathbf{0},\bSigmac\otimes\bSigmar).
    \end{equation*}
    \item MAR with fixed-rank co-kriging (MAR+FRC)~\citep{hsu2021matrix}:
    \begin{equation*}
        \matX_{t} = \sum_{p=1}^{P} \matA_p\matX_{t-p}\matB_p^{\top} + \matE_t, \vect{\matE_t}\sim\mathcal{N}(\mathbf{0},\sigma_{\eta}^{2}\matI + \mathbf{F}\mathbf{M}\mathbf{F}^{\top}),
    \end{equation*}
    where $\mathbf{F}\in\mathbb{R}^{MN\times QD}$ is the multi-resolution spline basis~\citep{tzeng2018resolution}.
    \item MAR followed by a tensor-on-scalar linear model (MAR+LM)~\citep{li2017parsimonious}:
    \begin{equation}\label{eq:MAR-LM-model}
        \matX_{t} - \sum_{p=1}^{P} \widehat{\matA}_p\matX_{t-p}\widehat{\matB}_p^{\top} = \sum_{q=1}^{Q}\tenG_q\tvprod\vecz_{t-q} + \matE_t, \vect{\matE_t}\sim\mathcal{N}(\mathbf{0},\sigma_{\eta}^{2}\matI),
    \end{equation}
    where $\widehat{\matA}_p,\widehat{\matB}_p$ come from a pre-trained MAR model and $\tenG_q$ can be a low-rank tensor. The MAR+LM model can be considered as a two-step procedure for fitting the MARAC model.
    \item Pixel-wise autoregression (Pixel-AR): for each $i\in[M], j\in[N]$, we have:
    \begin{equation*}
        [\matX_t]_{ij} = \alpha_{ij} + \sum_{p=1}^{P} \beta_{ijp}[\mathbf{X}_{t-p}]_{ij} + \sum_{q=1}^{Q} \boldsymbol{\gamma}_{ijq}^{\top}\mathbf{z}_{t-q} + [\matE_t]_{ij}, \quad [\matE_t]_{ij} \sim \mathcal{N}(0,\sigma_{ij}^{2}).
    \end{equation*}
    \item Vector Autoregression with Exogenous Predictor (VARX), which vectorizes the matrix time series and stacks them up with the vector time series as predictors.
\end{enumerate}
The results of the average prediction RMSE obtained from the 20 repeated runs are plotted in Figure~\ref{fig:method-compare-result}. Overall, our MARAC model outperforms the other competing methods under varying matrix dimensionality and lags. We make two additional remarks. First, when the matrix size is small (e.g., $5\times 5$), the vector autoregression model (VARX) performs almost as well as the MARAC model and is better than other methods. However, the performance of the VARX model gets worse quickly as the matrix becomes larger, indicating that sufficient dimension reduction is needed to deal with large matrix time series. The MARAC model is a parsimonious version of VARX for such purposes. Secondly, the MAR, MAR with fixed-rank co-kriging (MAR+FRC), and two-step MARAC (MAR+LM) all perform worse than MARAC. This shows that when the auxiliary time series predictors are present, it is sub-optimal to remove them from the model (MAR), incorporate them implicitly in the covariance structure (MAR+FRC), or fit them separately in a tensor-on-scalar regression model (MAR+LM). Putting both matrix and vector predictors in a unified framework like MARAC can be beneficial for improving prediction performances.

\section{Application to Global Total Electron Content Forecast}\label{sec:application}
For real data applications, we consider the problem of predicting the global total electron content (TEC) distribution, which we briefly introduce in Section~\ref{sec:intro}. The TEC data we use is the IGS (International GNSS Service) TEC data, which are freely available from the  National Aeronautics and  Space  Administration (NASA)  Crustal  Dynamics  Data  Information  System~\citep{hernandez2009igs}. 
The spatial-temporal resolution of the data is $2.5^{\circ} (\text{latitude}) \times 5^{\circ} (\text{longitude}) \times 15 (\text{minutes})$. We use whole-month data for September 2017, a matrix time series with $T = 2880$ and $M=71, N=73$. We use the 15-minute resolution IMF Bz and Sym-H time series for the auxiliary covariates, which are parameters related to the near-Earth magnetic field and plasma~\citep{papitashvili2014omni}. 
These covariates measure the solar wind strengths. Strong solar wind might lead to geomagnetic storms that could increase the global TEC significantly.

We formulate our MARAC model for the TEC prediction problem as:
\begin{equation}\label{eq:TEC-prediction-formula}
    \Delta\text{TEC}_{t+h} = \sum_{p=1}^{P} \matA_p \Delta\text{TEC}_{t-p}\matB_p^{\top} + \sum_{q=1}^{Q} \tenG_q \tvprod \Delta\vecz_{t-q} + \matE_t,
\end{equation}
where $h$ is the forecast latency time and $\Delta\vecz_{t} = \vecz_{t}-\vecz_{t-1}$ includes the change of IMF Bz, and Sym-H indices from time $t-1$ to $t$. We chose to forecast the $\Delta\text{TEC}_{t} = \text{TEC}_{t} - \text{TEC}_{t-1}$ series and use $\Delta\vecz_{t}$ instead of the raw $\vecz_{t}$ series as the auxiliary covariates to satisfy the joint stationarity condition in Theorem~\ref{thm:Joint-Stationarity}. To ensure we have better estimator convergence and valid inference, we downsample each matrix from $(71,73)$ to $(12,12)$ via averaging each local $6\times 6$ patch. 

We consider the forecasting scenario with $h\in \{4, 8, 12, \ldots, 72\}$, corresponding to making forecasts from 1 hour to 18 hours ahead. For each $h$, we fit our MARAC$(P,Q)$ model following~\eqref{eq:TEC-prediction-formula} with $1\le P \le 5$ and $1\le Q\le 3$. As a comparison, we also fit the MAR model with $1\le P\le 5$ and the MAR+LM model with $1\le P \le 5$ and $1\le Q\le 3$, see the definition of MAR+LM model in~\eqref{eq:MAR-LM-model}. 
The $2,880$ frames of matrix data are split into a $70\%$ training set, $15\%$ validation set, and a $15\%$ testing set following the chronological order. We choose the tuning parameter $\lambda$ for MARAC based on the validation set prediction RMSE. The lag parameters $P,Q$ are selected for all models based on the BIC. 

\begin{figure}[!htb]
    \centering
    \includegraphics[width=\textwidth]{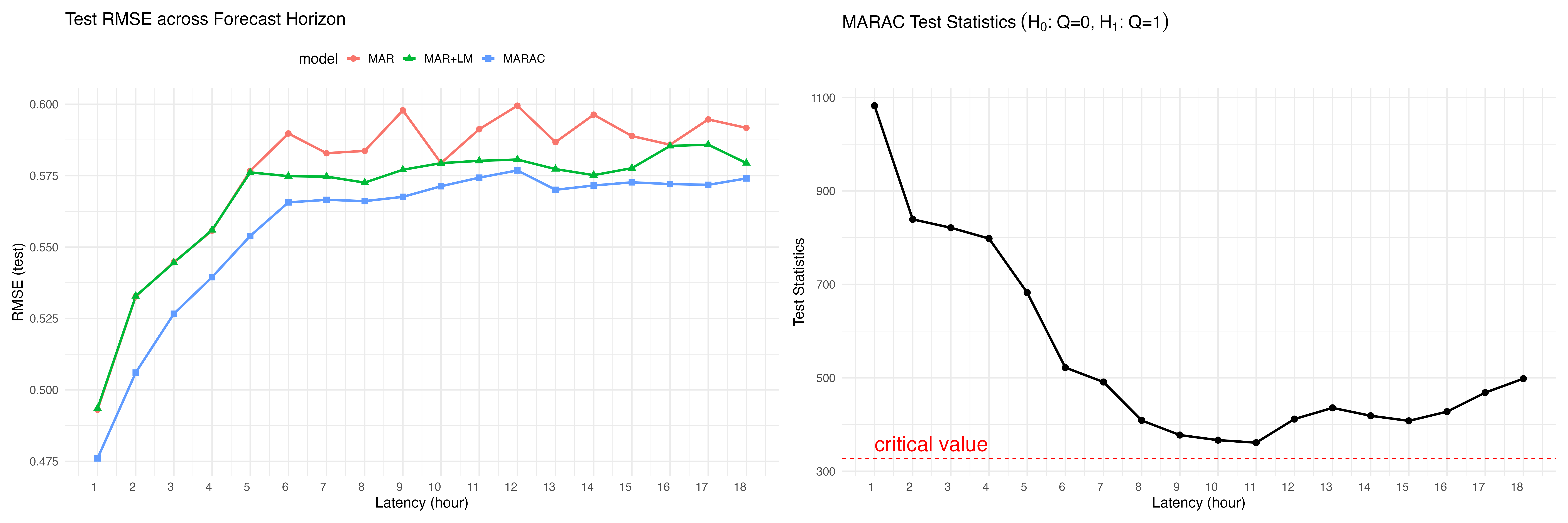}
    \caption{TEC prediction results. Left: test set prediction RMSE across three methods. Right: MARAC model test statistics, with model fitted with $Q=1$ and $P$ selected by BIC.}
    \label{fig:IGS-Downsampled-Prediction}
\end{figure}

We report the results in Figure~\ref{fig:IGS-Downsampled-Prediction}. From the left panel, it is clear that the MARAC model is consistently outperforming the other two competing methods across all forecast horizons. The addition of the auxiliary covariates improves the prediction accuracy, and this is also confirmed in the right panel, where all specification tests reject the null of $Q=0$.

To visualize the difference that the auxiliary covariates can make, in Figure~\ref{fig:IGS-Prediction-Example}, we fit a MARAC$(1,1)$ model to predict the $\Delta\text{TEC}_{t+12}$, namely the $\Delta\text{TEC}$ 3 hours later. For better visibility, we only downsample the data to $24\times 24$. To further distinguish the predictions made by different models, we take the sum of $90$ consecutive predictions/ground truth, and plot the results. It is clear from the results that the MARAC prediction tracks the target better than the competing methods, and the auxiliary covariates contribute to better predictions near the equatorial region (middle band of the plot), which is also the primary region of scientific interest. We believe that our method can help domain scientists to determine if scalar time series can predict spatial responses in similar application contexts.

\begin{figure}[!htb]
    \centering
    \includegraphics[width=\textwidth]{Figures/IGS_Delta_TEC_Prediction_Example.png}
    \caption{Example visualization of $\Delta\text{TEC}$ prediction with a forecast horizon of 3 hours. Left to right: original $\Delta\text{TEC}$ of size $71\times 73$; downsampled $\Delta\text{TEC}$ of size $24\times 24$; MARAC$(1,1)$ prediction result; Auxiliary covariate term prediction result from MARAC$(1,1)$; MAR$(1)$ prediction result; MAR+LM$(1,1)$ prediction result. The results are the sum across 90 consecutive frames.}
    \label{fig:IGS-Prediction-Example}
\end{figure}

\section{Summary}\label{sec:summary}
This paper proposes a new methodology for spatial-temporal matrix autoregression with non-spatial exogenous vector covariates. The model has an autoregressive component with bilinear transformations on the lagged matrix predictors and an additive auxiliary covariate component with a tensor-vector product between a tensor coefficient and the lagged vector covariates. We propose a penalized MLE estimation approach with a squared RKHS norm penalty and establish the estimator asymptotics under fixed and high matrix dimensionality. The model efficacy has been validated using both numerical experiments and an application to the global TEC forecast. 

The application of our model can be extended to other spatial data with exogenous, non-spatial predictors and is not restricted to matrix-valued data but can be generalized to the tensor setting and potentially data without a grid structure or containing missing data. Furthermore, our model nests a simpler model that does not include the autoregressive term, i.e., $P=0$, and thus can be applied to matrix-on-scalar regression with spatial data. Also, it is a natural extension of our paper to consider the case where $D$, the dimension of the auxiliary covariates, grows together with $M$ and $N$, and thus enables the modeling of high-dimensional auxiliary covariates for matrix/tensor response. We leave the discussions for these setups to future research.

\section*{Supplementary Materials}
The supplemental material contains details of the alternating minimization algorithm, technical proofs of all theorems and propositions of the paper, additional details of the simulation experiments, and the approximated estimating algorithm based on kernel truncation. Our code is available at \url{https://github.com/husun0822/MARAC}.
\par
\section*{Acknowledgements}
The authors thank Shasha Zou, Zihan Wang, and Yizhou Zhang for helpful discussions on the TEC data. YC acknowledges support from NSF DMS 2113397, NSF PHY 2027555, NASA Federal Award No. 80NSSC23M0192, and No. 80NSSC23M0191.
\par


\bibliography{references}

\begin{thebibliography}{66}
\providecommand{\natexlab}[1]{#1}
\providecommand{\url}[1]{\texttt{#1}}
\expandafter\ifx\csname urlstyle\endcsname\relax
  \providecommand{\doi}[1]{doi: #1}\else
  \providecommand{\doi}{doi: \begingroup \urlstyle{rm}\Url}\fi

\bibitem[Akaike(1998)]{akaike1998information}
Hirotogu Akaike.
\newblock Information {T}heory and an {E}xtension of the {M}aximum {L}ikelihood
  {P}rinciple.
\newblock In \emph{Selected papers of Hirotugu Akaike}, pages 199--213.
  Springer, 1998.

\bibitem[Attouch et~al.(2013)Attouch, Bolte, and
  Svaiter]{attouch2013convergence}
Hedy Attouch, J{\'e}r{\^o}me Bolte, and Benar~Fux Svaiter.
\newblock Convergence of {D}escent {M}ethods for {S}emi-{A}lgebraic and {T}ame
  {P}roblems: {P}roximal {A}lgorithms, {F}orward--{B}ackward {S}plitting, and
  {R}egularized {G}auss--{S}eidel {M}ethods.
\newblock \emph{Mathematical Programming}, 137\penalty0 (1-2):\penalty0
  91--129, 2013.

\bibitem[Banerjee et~al.(2005)Banerjee, Dhillon, Ghosh, Sra, and
  Ridgeway]{banerjee2005clustering}
Arindam Banerjee, Inderjit~S Dhillon, Joydeep Ghosh, Suvrit Sra, and Greg
  Ridgeway.
\newblock Clustering on the {U}nit {H}ypersphere using von {M}ises-{F}isher
  {D}istributions.
\newblock \emph{Journal of Machine Learning Research}, 6\penalty0 (9):\penalty0
  1345--1382, 2005.

\bibitem[Braun(2006)]{braun2006accurate}
Mikio~L Braun.
\newblock Accurate {E}rror {B}ounds for the {E}igenvalues of the {K}ernel
  {M}atrix.
\newblock \emph{The Journal of Machine Learning Research}, 7:\penalty0
  2303--2328, 2006.

\bibitem[Cai and Yuan(2012)]{cai2012}
T~Tony Cai and Ming Yuan.
\newblock Minimax and {A}daptive {P}rediction for {F}unctional {L}inear
  {R}egression.
\newblock \emph{Journal of the American Statistical Association}, 107\penalty0
  (499):\penalty0 1201--1216, 2012.

\bibitem[Chen and Fan(2023)]{chen2023statistical}
Elynn~Y Chen and Jianqing Fan.
\newblock Statistical {I}nference for {H}igh-{D}imensional {M}atrix-{V}ariate
  {F}actor {M}odels.
\newblock \emph{Journal of the American Statistical Association}, 118\penalty0
  (542):\penalty0 1038--1055, 2023.

\bibitem[Chen et~al.(2021)Chen, Xiao, and Yang]{chen2021autoregressive}
Rong Chen, Han Xiao, and Dan Yang.
\newblock Autoregressive {M}odels for {M}atrix-valued {T}ime {S}eries.
\newblock \emph{Journal of Econometrics}, 222\penalty0 (1):\penalty0 539--560,
  2021.

\bibitem[Cheng and Shang(2015)]{shang2015aos:b}
Guang Cheng and Zuofeng Shang.
\newblock Joint {A}symptotics for {S}emi-nonparametric {R}egression {M}odels
  with {P}artially {L}inear {S}tructure.
\newblock \emph{The Annals of Statistics}, 43:\penalty0 1351--1390, 2015.

\bibitem[Cressie(1986)]{cressie1986kriging}
Noel Cressie.
\newblock Kriging {N}onstationary {D}ata.
\newblock \emph{Journal of the American Statistical Association}, 81\penalty0
  (395):\penalty0 625--634, 1986.

\bibitem[Cressie and Johannesson(2008)]{cressie2008fixed}
Noel Cressie and Gardar Johannesson.
\newblock Fixed {R}ank {K}riging for {V}ery {L}arge {S}patial {D}ata {S}ets.
\newblock \emph{Journal of the Royal Statistical Society: Series B (Statistical
  Methodology)}, 70\penalty0 (1):\penalty0 209--226, 2008.

\bibitem[Cressie and Wikle(2015)]{cressie2015statistics}
Noel Cressie and Christopher~K Wikle.
\newblock \emph{Statistics for {S}patio-{T}emporal {D}ata}.
\newblock John Wiley \& Sons, 2015.

\bibitem[Cui et~al.(2018)Cui, Cheng, and Sun]{cui2018rkhs}
Wenquan Cui, Haoyang Cheng, and Jiajing Sun.
\newblock An {RKHS}-based {A}pproach to {D}ouble-{P}enalized {R}egression in
  {H}igh-dimensional {P}artially {L}inear {M}odels.
\newblock \emph{Journal of Multivariate Analysis}, 168:\penalty0 201--210,
  2018.

\bibitem[Dong et~al.(2020)Dong, Huang, Wu, and Zeng]{dong2020envsci}
Mingwang Dong, Linfu Huang, Xueqin Wu, and Qingguang Zeng.
\newblock Application of {L}east-{S}quares {M}ethod to {T}ime {S}eries
  {A}nalysis for 4dpm {M}atrix.
\newblock \emph{IOP Conference Series: Earth and Environmental Science},
  455\penalty0 (1):\penalty0 012200, feb 2020.
\newblock \doi{10.1088/1755-1315/455/1/012200}.
\newblock URL \url{https://dx.doi.org/10.1088/1755-1315/455/1/012200}.

\bibitem[Fosdick and Hoff(2014)]{fosdick2014separable}
BK~Fosdick and PD~Hoff.
\newblock Separable {F}actor {A}nalysis with {A}pplications to {M}ortality
  {D}ata.
\newblock \emph{The Annals of Applied Statistics}, 8\penalty0 (1):\penalty0
  120--147, 2014.

\bibitem[Gao and Tsay(2023)]{gao2023two}
Zhaoxing Gao and Ruey~S Tsay.
\newblock A {T}wo-way {T}ransformed {F}actor {M}odel for {M}atrix-{V}ariate
  {T}ime {S}eries.
\newblock \emph{Econometrics and Statistics}, 27:\penalty0 83--101, 2023.

\bibitem[Gao and Tsay(2025)]{gao2023denoising}
Zhaoxing Gao and Ruey~S Tsay.
\newblock Denoising and {M}ultilinear {P}rojected-{E}stimation of
  {H}igh-{D}imensional {M}atrix-{V}ariate {F}actor {T}ime {S}eries.
\newblock \emph{IEEE Transactions on Information Theory}, page in press, 2025.

\bibitem[Gu(2013)]{gu2013}
Chong Gu.
\newblock \emph{Smoothing {S}pline {ANOVA} models, 2nd edition}.
\newblock Springer, New York, 2013.

\bibitem[Guha and Guhaniyogi(2021)]{guha2021bayesian}
Sharmistha Guha and Rajarshi Guhaniyogi.
\newblock Bayesian {G}eneralized {S}parse {S}ymmetric {T}ensor-on-{V}ector
  {R}egression.
\newblock \emph{Technometrics}, 63\penalty0 (2):\penalty0 160--170, 2021.

\bibitem[Guhaniyogi et~al.(2017)Guhaniyogi, Qamar, and
  Dunson]{guhaniyogi2017bayesian}
Rajarshi Guhaniyogi, Shaan Qamar, and David~B Dunson.
\newblock Bayesian {T}ensor {R}egression.
\newblock \emph{The Journal of Machine Learning Research}, 18\penalty0
  (1):\penalty0 2733--2763, 2017.

\bibitem[Guo et~al.(2016)Guo, Wang, and Yao]{guo2016banded}
Shaojun Guo, Yazhen Wang, and Qiwei Yao.
\newblock High-dimensional and {B}anded {V}ector {A}utoregressions.
\newblock \emph{Biometrika}, 103\penalty0 (4):\penalty0 889--903, 10 2016.
\newblock ISSN 0006-3444.

\bibitem[Hall and Heyde(2014)]{hall2014martingale}
Peter Hall and Christopher~C Heyde.
\newblock \emph{Martingale {L}imit {T}heory and its {A}pplication}.
\newblock Academic press, 2014.

\bibitem[Hamilton(2020)]{hamilton2020time}
James~D Hamilton.
\newblock \emph{Time {S}eries {A}nalysis}.
\newblock Princeton University Press, 2020.

\bibitem[Hastie et~al.(2009)Hastie, Tibshirani, Friedman, and
  Friedman]{hastie2009}
Trevor Hastie, Robert Tibshirani, Jerome~H Friedman, and Jerome~H Friedman.
\newblock \emph{The {E}lements of {S}tatistical {L}earning: {D}ata {M}ining,
  {I}nference, and {P}rediction, 2nd edition}.
\newblock Springer, New York, 2009.

\bibitem[Hern{\'a}ndez-Pajares et~al.(2009)Hern{\'a}ndez-Pajares, Juan, Sanz,
  Orus, Garcia-Rigo, Feltens, Komjathy, Schaer, and
  Krankowski]{hernandez2009igs}
Manuel Hern{\'a}ndez-Pajares, JM~Juan, J~Sanz, R~Orus, A~Garcia-Rigo,
  J~Feltens, A~Komjathy, SC~Schaer, and A~Krankowski.
\newblock The {IGS} {VTEC} {M}aps: a {R}eliable {S}ource of {I}onospheric
  {I}nformation since 1998.
\newblock \emph{Journal of Geodesy}, 83:\penalty0 263--275, 2009.

\bibitem[Hoff(2011)]{hoff2011separable}
Peter~D Hoff.
\newblock Separable {C}ovariance {A}rrays via the {T}ucker {P}roduct, with
  {A}pplications to {M}ultivariate {R}elational {D}ata.
\newblock \emph{Bayesian Analysis}, 6\penalty0 (2):\penalty0 179--196, 2011.

\bibitem[Hsu et~al.(2021)Hsu, Huang, and Tsay]{hsu2021matrix}
Nan-Jung Hsu, Hsin-Cheng Huang, and Ruey~S Tsay.
\newblock Matrix {A}utoregressive {S}patio-{T}emporal {M}odels.
\newblock \emph{Journal of Computational and Graphical Statistics}, 30\penalty0
  (4):\penalty0 1143--1155, 2021.

\bibitem[Kang et~al.(2018)Kang, Reich, and Staicu]{kang2018scalar}
Jian Kang, Brian~J Reich, and Ana-Maria Staicu.
\newblock Scalar-on-{I}mage {R}egression via the {S}oft-{T}hresholded
  {G}aussian {P}rocess.
\newblock \emph{Biometrika}, 105\penalty0 (1):\penalty0 165--184, 2018.

\bibitem[Kennedy et~al.(2013)Kennedy, Sadeghi, Khalid, and
  McEwen]{kennedy2013classification}
Rodney~A Kennedy, Parastoo Sadeghi, Zubair Khalid, and Jason~D McEwen.
\newblock Classification and {C}onstruction of {C}losed-form {K}ernels for
  {S}ignal {R}epresentation on the 2-sphere.
\newblock In \emph{Wavelets and Sparsity XV}, volume 8858, pages 169--183.
  SPIE, 2013.

\bibitem[Kolda and Bader(2009)]{kolda2009tensor}
Tamara~G Kolda and Brett~W Bader.
\newblock Tensor {D}ecompositions and {A}pplications.
\newblock \emph{SIAM review}, 51\penalty0 (3):\penalty0 455--500, 2009.

\bibitem[Koltchinskii and Gin{\'e}(2000)]{koltchinskii2000random}
Vladimir Koltchinskii and Evarist Gin{\'e}.
\newblock Random {M}atrix {A}pproximation of {S}pectra of {I}ntegral
  {O}perators.
\newblock \emph{Bernoulli}, 6\penalty0 (1):\penalty0 113--167, 2000.

\bibitem[Li and Zhang(2017)]{li2017parsimonious}
Lexin Li and Xin Zhang.
\newblock Parsimonious {T}ensor {R}esponse {R}egression.
\newblock \emph{Journal of the American Statistical Association}, 112\penalty0
  (519):\penalty0 1131--1146, 2017.

\bibitem[Li et~al.(2018)Li, Xu, Zhou, and Li]{li2018tucker}
Xiaoshan Li, Da~Xu, Hua Zhou, and Lexin Li.
\newblock Tucker {T}ensor {R}egression and {N}euroimaging {A}nalysis.
\newblock \emph{Statistics in Biosciences}, 10\penalty0 (3):\penalty0 520--545,
  2018.

\bibitem[Li and Xiao(2021)]{li2021multi}
Zebang Li and Han Xiao.
\newblock Multi-linear {T}ensor {A}utoregressive {M}odels.
\newblock \emph{arXiv preprint arXiv:2110.00928}, 2021.

\bibitem[Liu et~al.(2020)Liu, Liu, and Zhu]{liu2020low}
Yipeng Liu, Jiani Liu, and Ce~Zhu.
\newblock Low-rank {T}ensor {T}rain {C}oefficient {A}rray {E}stimation for
  {T}ensor-on-{T}ensor {R}egression.
\newblock \emph{IEEE Transactions on Neural Networks and Learning Systems},
  31\penalty0 (12):\penalty0 5402--5411, 2020.

\bibitem[Lock(2018)]{lock2018tensor}
Eric~F Lock.
\newblock Tensor-on-{T}ensor {R}egression.
\newblock \emph{Journal of Computational and Graphical Statistics}, 27\penalty0
  (3):\penalty0 638--647, 2018.

\bibitem[Luo and Zhang(2024)]{luo2022tensor}
Yuetian Luo and Anru~R Zhang.
\newblock Tensor-on-tensor {R}egression: {R}iemannian {O}ptimization,
  {O}ver-parameterization, {S}tatistical-{C}omputational {G}ap and {T}heir
  {I}nterplay.
\newblock \emph{The Annals of Statistics}, 52\penalty0 (6):\penalty0
  2583--2612, 2024.

\bibitem[Lyu et~al.(2019)Lyu, Sun, Wang, Liu, Yang, and Cheng]{lyu2019tensor}
Xiang Lyu, Will~Wei Sun, Zhaoran Wang, Han Liu, Jian Yang, and Guang Cheng.
\newblock Tensor {G}raphical {M}odel: {N}on-convex {O}ptimization and
  {S}tatistical {I}nference.
\newblock \emph{IEEE Transactions on Pattern Analysis and Machine
  Intelligence}, 42\penalty0 (8):\penalty0 2024--2037, 2019.

\bibitem[Papadogeorgou et~al.(2021)Papadogeorgou, Zhang, and
  Dunson]{papadogeorgou2021soft}
Georgia Papadogeorgou, Zhengwu Zhang, and David~B Dunson.
\newblock Soft {T}ensor {R}egression.
\newblock \emph{The Journal of Machine Learning Research}, 22:\penalty0 219--1,
  2021.

\bibitem[Papitashvili et~al.(2014)Papitashvili, Bilitza, and
  King]{papitashvili2014omni}
Natasha Papitashvili, Dieter Bilitza, and Joseph King.
\newblock {OMNI}: a {D}escription of {N}ear-{E}arth {S}olar {W}ind
  {E}nvironment.
\newblock \emph{40th COSPAR Scientific Assembly}, 40:\penalty0 C0--1, 2014.

\bibitem[Rabusseau and Kadri(2016)]{rabusseau2016low}
Guillaume Rabusseau and Hachem Kadri.
\newblock Low-rank {R}egression with {T}ensor {R}esponses.
\newblock \emph{Advances in Neural Information Processing Systems}, 29, 2016.

\bibitem[Rudelson and Vershynin(2013)]{rudelson2013hanson}
Mark Rudelson and Roman Vershynin.
\newblock Hanson-{W}right {I}nequality and {S}ub-{G}aussian {C}oncentration.
\newblock \emph{Electronic Communications in Probability}, 18:\penalty0 1--9,
  2013.

\bibitem[Sch{\"o}lkopf et~al.(2001)Sch{\"o}lkopf, Herbrich, and
  Smola]{scholkopf2001generalized}
Bernhard Sch{\"o}lkopf, Ralf Herbrich, and Alex~J Smola.
\newblock A {G}eneralized {R}epresenter {T}heorem.
\newblock In \emph{International Conference on Computational Learning Theory},
  pages 416--426. Springer, 2001.

\bibitem[Schwarz(1978)]{schwarz1978estimating}
Gideon Schwarz.
\newblock Estimating the {D}imension of a {M}odel.
\newblock \emph{The Annals of Statistics}, 6\penalty0 (2):\penalty0 461--464,
  1978.

\bibitem[Shang and Cheng(2013)]{shang2013aos}
Zuofeng Shang and Guang Cheng.
\newblock Local and {G}lobal {A}symptotic {I}nference in {S}moothing {S}pline
  {M}odels.
\newblock \emph{The Annals of Statistics}, 41:\penalty0 2608--2638, 2013.

\bibitem[Shang and Cheng(2015)]{shang2015aos}
Zuofeng Shang and Guang Cheng.
\newblock Nonparametric {I}nference in {G}eneralized {F}unctional {L}inear
  {M}odels.
\newblock \emph{The Annals of Statistics}, 43:\penalty0 1742--1773, 2015.

\bibitem[Shen et~al.(2022)Shen, Xie, and Kong]{shen2022smooth}
Bo~Shen, Weijun Xie, and Zhenyu Kong.
\newblock Smooth {R}obust {T}ensor {C}ompletion for {B}ackground/{F}oreground
  {S}eparation with {M}issing {P}ixels: {N}ovel {A}lgorithm with {C}onvergence
  {G}uarantee.
\newblock \emph{The Journal of Machine Learning Research}, 23\penalty0
  (1):\penalty0 9757--9796, 2022.

\bibitem[Stock and Watson(2001)]{stock2001vector}
James~H Stock and Mark~W Watson.
\newblock Vector {A}utoregressions.
\newblock \emph{Journal of Economic perspectives}, 15\penalty0 (4):\penalty0
  101--115, 2001.

\bibitem[Sun et~al.(2022)Sun, Hua, Ren, Zou, Sun, and Chen]{sun2022matrix}
Hu~Sun, Zhijun Hua, Jiaen Ren, Shasha Zou, Yuekai Sun, and Yang Chen.
\newblock Matrix {C}ompletion {M}ethods for the {T}otal {E}lectron {C}ontent
  {V}ideo {R}econstruction.
\newblock \emph{The Annals of Applied Statistics}, 16\penalty0 (3):\penalty0
  1333--1358, 2022.

\bibitem[Sun et~al.(2023)Sun, Manchester, Jin, Liu, and Chen]{sun2023tensorGP}
Hu~Sun, Ward Manchester, Meng Jin, Yang Liu, and Yang Chen.
\newblock Tensor {G}aussian {P}rocess with {C}ontraction for {M}ulti-{C}hannel
  {I}maging {A}nalysis.
\newblock In \emph{International Conference on Machine Learning}, pages
  32913--32935. PMLR, 2023.

\bibitem[Sun and Li(2017)]{sun2017store}
Will~Wei Sun and Lexin Li.
\newblock {STORE}: {S}parse {T}ensor {R}esponse {R}egression and {N}euroimaging
  {A}nalysis.
\newblock \emph{The Journal of Machine Learning Research}, 18\penalty0
  (1):\penalty0 4908--4944, 2017.

\bibitem[Tsiligkaridis et~al.(2013)Tsiligkaridis, Hero~III, and
  Zhou]{tsiligkaridis2013convergence}
Theodoros Tsiligkaridis, Alfred~O Hero~III, and Shuheng Zhou.
\newblock On {C}onvergence of {K}ronecker {G}raphical {L}asso {A}lgorithms.
\newblock \emph{IEEE Transactions on Signal Processing}, 61\penalty0
  (7):\penalty0 1743--1755, 2013.

\bibitem[Tzeng and Huang(2018)]{tzeng2018resolution}
ShengLi Tzeng and Hsin-Cheng Huang.
\newblock Resolution {A}daptive {F}ixed {R}ank {K}riging.
\newblock \emph{Technometrics}, 60\penalty0 (2):\penalty0 198--208, 2018.

\bibitem[van Zanten and van~der Vaart(2008)]{van2008reproducing}
JH~van Zanten and Aad~W van~der Vaart.
\newblock Reproducing {K}ernel {H}ilbert {S}paces of {G}aussian {P}riors.
\newblock In \emph{Pushing the Limits of Contemporary Statistics: Contributions
  in honor of Jayanta K. Ghosh}, pages 200--222. Institute of Mathematical
  Statistics, 2008.

\bibitem[Wang et~al.(2022)Wang, Zheng, Lian, and Li]{wang2022high}
Di~Wang, Yao Zheng, Heng Lian, and Guodong Li.
\newblock High-dimensional {V}ector {A}utoregressive {T}ime {S}eries {M}odeling
  via {T}ensor {D}ecomposition.
\newblock \emph{Journal of the American Statistical Association}, 117\penalty0
  (539):\penalty0 1338--1356, 2022.

\bibitem[Wang et~al.(2024)Wang, Zheng, and Li]{wang2024high}
Di~Wang, Yao Zheng, and Guodong Li.
\newblock High-{D}imensional {L}ow-rank {T}ensor {A}utoregressive {T}ime
  {S}eries {M}odeling.
\newblock \emph{Journal of Econometrics}, 238\penalty0 (1):\penalty0 105544,
  2024.

\bibitem[Wang et~al.(2019)Wang, Liu, and Chen]{wang2019factor}
Dong Wang, Xialu Liu, and Rong Chen.
\newblock Factor {M}odels for {M}atrix-valued {H}igh-dimensional {T}ime
  {S}eries.
\newblock \emph{Journal of Econometrics}, 208\penalty0 (1):\penalty0 231--248,
  2019.

\bibitem[Wang and Li(2020)]{wang2020learning}
Miaoyan Wang and Lexin Li.
\newblock Learning from {B}inary {M}ultiway {D}ata: {P}robabilistic {T}ensor
  {D}ecomposition and its {S}tatistical {O}ptimality.
\newblock \emph{The Journal of Machine Learning Research}, 21\penalty0
  (154):\penalty0 1--38, 2020.

\bibitem[Wang et~al.(2017)Wang, Zhu, and Initiative]{wang2017generalized}
Xiao Wang, Hongtu Zhu, and Alzheimer’s Disease~Neuroimaging Initiative.
\newblock Generalized {S}calar-on-{I}mage {R}egression {M}odels via {T}otal
  {V}ariation.
\newblock \emph{Journal of the American Statistical Association}, 112\penalty0
  (519):\penalty0 1156--1168, 2017.

\bibitem[Wang et~al.(2021)Wang, Zou, Liu, Ren, and Aa]{wang2021}
Zihan Wang, Shasha Zou, Lei Liu, Jiaen Ren, and Ercha Aa.
\newblock Hemispheric {A}symmetries in the {M}id-latitude {I}onosphere {D}uring
  the {S}eptember 7–8, 2017 {S}torm: {M}ulti-instrument {O}bservations.
\newblock \emph{Journal of Geophysical Research: Space Physics}, 126:\penalty0
  e2020JA028829, 2021.
\newblock ISSN 2169-9402.
\newblock \doi{10.1029/2020JA028829}.

\bibitem[Williams and Rasmussen(2006)]{williams2006gaussian}
Christopher~K Williams and Carl~Edward Rasmussen.
\newblock \emph{Gaussian {P}rocesses for {M}achine {L}earning}, volume~2.
\newblock MIT press Cambridge, MA, 2006.

\bibitem[Xiao et~al.(2022)Xiao, Han, Chen, and Liu]{xiao2022reduced}
H~Xiao, Y~Han, R~Chen, and C~Liu.
\newblock Reduced {R}ank {A}utoregressive {M}odels for {M}atrix {T}ime
  {S}eries.
\newblock \emph{Journal of Business and Economic Statistics}, 2022.

\bibitem[Yang et~al.(2020)Yang, Shang, and Cheng]{shang:colt:2020}
Yun Yang, Zuofeng Shang, and Guang Cheng.
\newblock Non-asymptotic {A}nalysis for {N}onparametric {T}esting.
\newblock In \emph{33rd Annual Conference on Learning Theory}, pages 1--47.
  ACM, 2020.

\bibitem[Younas et~al.(2022)Younas, Khan, Amory-Mazaudier, Amaechi, and
  Fleury]{Younas2022}
Waqar Younas, Majid Khan, C.~Amory-Mazaudier, Paul~O. Amaechi, and R.~Fleury.
\newblock Middle and {L}ow {L}atitudes {H}emispheric {A}symmetries in {$\Sigma
  O/N2$} and {TEC} during {I}ntense {M}agnetic {S}torms of {S}olar {C}ycle 24.
\newblock \emph{Advances in Space Research}, 69:\penalty0 220--235, 2022.

\bibitem[Yuan and Cai(2010)]{yuan2010}
Ming Yuan and T~Tony Cai.
\newblock A {R}eproducing {K}ernel {Hilbert} {S}pace {A}pproach to {F}unctional
  {L}inear {R}egression.
\newblock \emph{The Annals of Statistics}, 38\penalty0 (6):\penalty0
  3412--3444, 2010.

\bibitem[Zhou et~al.(2013)Zhou, Li, and Zhu]{zhou2013tensor}
Hua Zhou, Lexin Li, and Hongtu Zhu.
\newblock Tensor {R}egression with {A}pplications in {N}euroimaging {D}ata
  {A}nalysis.
\newblock \emph{Journal of the American Statistical Association}, 108\penalty0
  (502):\penalty0 540--552, 2013.

\bibitem[Zhou(2014)]{zhou2014gemini}
Shuheng Zhou.
\newblock {GEMINI}: {G}raph {E}stimation with {M}atrix {V}ariate {N}ormal
  {I}nstances.
\newblock \emph{The Annals of Statistics}, 42\penalty0 (2):\penalty0 532--562,
  2014.

\end{thebibliography}


\newpage
\appendix
\begin{center}
{\large\bf SUPPLEMENTARY MATERIAL}
\end{center}

This supplemental material is organized as follows. Section~\ref{sec:algorithm_detail} presents the algorithmic details of the penalized MLE using alternating minimization outlined in Section~\ref{subsec:PMLE}. In Section~\ref{subsec:proof_KRR_equivalence}, we prove Proposition~\ref{thm:KRR-equivalence} on the equivalence of the estimation problem of MARAC to a kernel ridge regression problem. In Section~\ref{subsec:proof_joint_stationarity}, we prove Theorem~\ref{thm:Joint-Stationarity} on the joint stationarity condition of the matrix and auxiliary vector time series. Then in Section~\ref{sec:fix-dim}, we provide proofs of the theoretical results under fixed spatial dimensionality, including Proposition~\ref{thm:error-cov-consistency}, Theorem~\ref{thm:PMLE-Asymp-Normal} and Corollary~\ref{cor:SpecTest}. In Section~\ref{sec:grow-dim}, we present proofs of the theoretical results under high spatial dimensionality, namely Theorem~\ref{thm:high_dimensional_MARAC}. All essential lemmas used throughout the proofs are presented and proved in Section~\ref{app:proof-lemmas}. Finally, we include additional details of the simulation in Section~\ref{app:sim} as well as an approximated estimating algorithm for obtaining the penalized MLE via kernel truncation.

In this supplemental material, we use $\maxeigen{\cdot}$, $\rho_i(\cdot)$, $\mineigen{\cdot}$, and $\specnorm{\cdot}$ to denote the maximum, $i^{\rm{th}}$ largest, minimum eigenvalue, and spectral norm of a matrix. We use $a\vee b, a\wedge b$ to denote the maximum and minimum of $a$ and $b$, respectively. For two sequences of random variables, say $X_n, Y_n$, we use $X_n \lesssim Y_n$ to denote the case where $X_n/Y_n = O_P(1)$, and $X_n \gtrsim Y_n$ to denote the case where $Y_n/X_n = O_P(1)$. We then use $X_n\asymp Y_n$ to denote the case where both $X_n \lesssim Y_n$ and $X_n \gtrsim Y_n$ hold.

\section{Alternating Minimization Algorithm for PMLE}\label{sec:algorithm_detail}
To solve the optimization problem in~\eqref{eq:MARAC-PMLE-KRR} for $\matA_p$ at the $(l+1)^{\rm{th}}$ iteration, it suffices to solve the following least-square problem:
\begin{equation}\label{eq:Update-Ap-Subproblem}
    \min_{\matA_p} \left\{\sum_{t\in[T]} \text{tr}\left(\widetilde{\matX}_t(\matA_p)^{\top}\left(\bSigmar^{(l)}\right)^{-1}\widetilde{\matX}_t(\matA_p)\left(\bSigmac^{(l)}\right)^{-1}\right)\right\},
\end{equation}
where $\widetilde{\matX}_t(\matA_p)$ is the residual matrix when predicting $\matX_t$:
\begin{align*}
    \widetilde{\matX}_t(\matA_p) & = \matX_t - \sum_{p^{'}<p} \matA_{p^{'}}^{(l+1)}\matX_{t-p^{'}}\left(\matB_{p^{'}}^{(l+1)}\right)^{\top} - \sum_{p^{'}>p} \matA_{p^{'}}^{(l)}\matX_{t-p^{'}}\left(\matB_{p^{'}}^{(l)}\right)^{\top} \\
    & - \sum_{q\in[Q]}\tenG_q^{(l)}\tvprod\vecz_{t-q} - \matA_p\matX_{t-p}\left(\matB_p^{(l)}\right)^{\top} = \widetilde{\matX}_{t,-p} - \matA_p\matX_{t-p}\left(\matB_p^{(l)}\right)^{\top}
\end{align*}
and we use $\widetilde{\matX}_{t,-p}$ to denote the partial residual excluding the term involving $\matX_{t-p}$ and use $\tenG_q^{(l)}$ to denote the tensor coefficient satisfying $[\tenG_q^{(l)}]_{ijd}=\inner{[\matK]_{u:}^{\top}}{[\matgamma_q^{(l)}]_{:d}}$, with $u=i+(j-1)M$. The superscript $l$ represents the value at the $l^{\rm{th}}$ iteration. To simplify the notation, we define $\matPhi(\matA_t,\matB_t,\bSigma) = \sum_{t} \matA_t^{\top}\bSigma^{-1}\matB_t$, where $\bSigma,\matA_t,\matB_t$ are arbitrary matrices/vectors with conformal matrix sizes and we simply write $\matPhi(\matA_t,\bSigma)$ if $\matA_t=\matB_t$. Solving~\eqref{eq:Update-Ap-Subproblem} yields the following updating formula for $\matA_{p}^{(l+1)}$:
\begin{equation}\label{eq:Update-Ap}
    \matA_{p}^{(l+1)}\leftarrow \matPhi\left(\widetilde{\matX}_{t,-p}^{\top},\matB_{p}^{(l)}\matX_{t-p}^{\top},\bSigmac^{(l)}\right)\matPhi\left(\matB_{p}^{(l)}\matX_{t-p}^{\top},\bSigmac^{(l)}\right)^{-1}
\end{equation}
Similarly, we have the following updating formula for $\matB_{p}^{(l+1)}$:
\begin{equation}\label{eq:Update-Bp}
    \matB_{p}^{(l+1)} \leftarrow \matPhi\left(\widetilde{\matX}_{t,-p},\matA_{p}^{(l+1)}\matX_{t-p},\bSigmar^{(l)}\right)\matPhi\left(\matA_{p}^{(l+1)}\matX_{t-p},\bSigmar^{(l)}\right)^{-1}
\end{equation}

For updating $\matgamma_{q}$, or its vectorized version $\vecgamma_{q} = \vect{\matgamma_{q}}$, it is required to solve the following kernel ridge regression problem:
\begin{equation*}
    \min_{\vecgamma_q}  \left\{\frac{1}{2T} \matPhi\left(\widetilde{\vecx}_{t,-q} - \left(\vecz_{t-q}^{\top}\otimes\matK\right)\vecgamma_{q},\bSigma^{(l)}\right) + \frac{\lambda}{2}\vecgamma_{q}^{\top}\left(\matI_{D}\otimes\matK\right)\vecgamma_{q}\right\},
\end{equation*}
where $\bSigma^{(l)}=\bSigmac^{(l)}\otimes\bSigmar^{(l)}$ and $\widetilde{\vecx}_{t,-q}$ is the vectorized partial residual of $\matX_{t}$ by leaving out the lag-$q$ auxiliary predictor, defined in a similar way as $\widetilde{\matX}_{t,-p}$. Solving the kernel ridge regression leads to the following updating formula for $\vecgamma_{q}^{(l+1)}$:
\begin{equation}\label{eq:Update-Gammaq}
    \vecgamma_{q}^{(l+1)} \leftarrow \left[\left(\sum_{t\in[T]} \vecz_{t-q}\vecz_{t-q}^{\top}\right)\otimes \matK + \lambda T\left(\matI_{D}\otimes\bSigma^{(l)}\right)\right]^{-1}\left[\sum_{t\in[T]} \left(\vecz_{t-q}\otimes\widetilde{\vecx}_{t,-q}\right)\right].
\end{equation}
The step in~\eqref{eq:Update-Gammaq} can be slow since one needs to invert a square matrix of size $MND\times MND$. In the supplemental material, we propose an approximation to~\eqref{eq:Update-Gammaq} to avoid inverting large matrices.

The updating rule of $\bSigmar^{(l+1)}$ and $\bSigmac^{(l+1)}$ can be easily derived by taking their derivative in~\eqref{eq:MARAC-PMLE-KRR} and setting it to zero. Specifically, we have:
\begin{align}
    \bSigmar^{(l+1)} & \leftarrow \frac{1}{NT}\matPhi\left(\widetilde{\matX}_t^{\top},\bSigmac^{(l)}\right)\label{eq:Update-Sigmar} \\
    \bSigmac^{(l+1)} & \leftarrow \frac{1}{MT}\matPhi\left(\widetilde{\matX}_t,\bSigmar^{(l+1)}\right). \label{eq:Update-Sigmac}
\end{align}
where $\widetilde{\matX}_t$ is the full residual when predicting $\matX_t$.

The algorithm cycles through~\eqref{eq:Update-Ap},~\eqref{eq:Update-Bp},~\eqref{eq:Update-Gammaq},~\eqref{eq:Update-Sigmar} and~\eqref{eq:Update-Sigmac} and terminates when $\matB_{p}^{(l)}\otimes\matA_{p}^{(l)}$, $\tenG_{q}^{(l)}$, $\bSigmac^{(l)}\otimes\bSigmar^{(l)}$ have their relative changes between iterations fall under a pre-specified threshold. We summarize the algorithm in pseudo-code in Algorithm~\ref{alg:ALS_algorithm}.
\begin{algorithm}
\caption{Alternating Minimization Algorithm for PMLE}
\label{alg:ALS_algorithm}
\begin{algorithmic}
\State Randomly initialize parameters $\matTheta^{(0)} = \{\mathbf{A}_1^{(0)},\mathbf{B}_1^{(0)},\ldots,\mathbf{A}_P^{(0)},\mathbf{B}_P^{(0)},\matgamma_1^{(0)},\ldots,\matgamma_Q^{(0)},\bSigmar^{(0)},\bSigmac^{(0)}\}$.
\State $k\gets 0$.
\While{not converge}
\For{$\veceta^{(k)}$ in $[\mathbf{A}_1^{(k)},\mathbf{B}_1^{(k)},\ldots,\mathbf{A}_P^{(k)},\mathbf{B}_P^{(k)},\matgamma_1^{(k)},\ldots,\matgamma_Q^{(k)},\bSigmar^{(k)},\bSigmac^{(k)}]$}
\State $\veceta^{(k+1)} \gets \argmin_{\veceta} \mathfrak{L}_{\lambda}\left(\veceta; \matTheta^{(k)}\setminus \{\veceta^{(k)}\}\right)$. \Comment{Details in~\eqref{eq:Update-Ap},~\eqref{eq:Update-Bp},~\eqref{eq:Update-Gammaq},~\eqref{eq:Update-Sigmar} and~\eqref{eq:Update-Sigmac}.}
\State Replace $\veceta^{(k)}$ with $\veceta^{(k+1)}$ in $\matTheta^{(k)}$.
\EndFor
\State $\matTheta^{(k+1)} \gets \matTheta^{(k)}$. 
\State $k\gets k+1$.
\EndWhile
\For{$p=1,2,\ldots,P$}
\State $c\gets \text{sign}(\text{tr}(\mathbf{A}_p^{(k)}))\cdot \twonorm{\mathbf{A}_p^{(k)}}$.
\State $\matA_p^{(k)} \gets c^{-1}\cdot\matA_p^{(k)}, \matB_p^{(k)} \gets c\cdot\matB_p^{(k)}$.
\EndFor
\State \textbf{return} $\matTheta^{(k)}$.
\end{algorithmic}
\end{algorithm}

\begin{remark}{(Convergence of Kronecker Product)}
    When dealing with high-dimensional matrices, it is cumbersome to compute the change between $\matB_{p}^{(l)}\otimes\matA_{p}^{(l)}$ and $\matB_{p}^{(l+1)}\otimes\matA_{p}^{(l+1)}$ under the Frobenius norm. An upper bound of $\twonorm{\matB_{p}^{(l+1)}\otimes\matA_{p}^{(l+1)} - \matB_{p}^{(l)}\otimes\matA_{p}^{(l)}} $ can be used instead:
    \begin{equation}\label{eq:conv_check_kron}
         \twonorm{\matB_{p}^{(l+1)}-\matB_{p}^{(l)}}\cdot\twonorm{\matA_{p}^{(l+1)}} + \twonorm{\matB_{p}^{(l)}}\cdot\twonorm{\matA_{p}^{(l+1)}-\matA_{p}^{(l)}},
    \end{equation}
    and a similar bound can be used for the convergence check of $\bSigmac^{(l)}\otimes\bSigmar^{(l)}$.
\end{remark}

\section{Proof of Proposition~\ref{thm:KRR-equivalence}}\label{subsec:proof_KRR_equivalence}
\begin{proof}
For each function $g_{q,d}(\cdot)\in\RKHS$, we can decompose it as follows:
    \begin{equation*}
        g_{q,d}(\cdot) = \sum_{s\in\setS}\gamma_{q,d,s}k(\cdot,s) + \sum_{j=1}^{J} \alpha_{q,d,j}\phi_j(\cdot) + h_{q,d}(\cdot),
    \end{equation*}
where $h_{q,d}(\cdot)$ does not belong to the null space of $\RKHS$ nor the span of $\{k(\cdot,s)|s\in\setS\}$. Here we assume that the null space of $\RKHS$ contains only the zero function, so $\phi_j(\cdot) = 0, \text{ for all } j$. 

By the reproducing property of the kernel $k(\cdot,\cdot)$, we have $\inner{g_{q,d}}{k(\cdot,s^{\prime})}_{\RKHS} = g_{q,d}(s^{\prime}) = \sum_{s\in\setS} \gamma_{q,d,s}k(s,s^{\prime})$, which is independent of $h_{q,d}(\cdot)$, and therefore $h_{q,d}(\cdot)$ is independent of the prediction for $\vecx_t$ in the MARAC model. In addition, for any $h_{q,d}(\cdot)\notin \text{span}(\{k(\cdot,s)|s\in\setS\})$, we have:
\begin{equation*}
    \|g_{q,d}\|_{\RKHS}^{2} = \gamma_{q,d}^{\top}\matK\gamma_{q,d} + \|h_{q,d}\|_{\RKHS}^{2} \ge \|\sum_{s\in\setS}\gamma_{q,d,s}k(\cdot,s)\|_{\RKHS}^{2},
\end{equation*}
and the equality holds only if $h_{q,d}(\cdot)=0$. Consequently, the global minimizer for the constrained optimization problem~\eqref{eq:MARAC-PMLE-Problem} must have $h_{q,d}(\cdot) = 0$. It then follows that the squared RKHS functional norm penalty for $g_{q,d}$ can be written as $\gamma_{q,d}^{\top}\matK\gamma_{q,d}$ and the tensor coefficient $\tenG_q$ satisfies $\vect{[\tenG]_{::d}} = \matK\gamma_{q,d}$. The remainder of the proof is straightforward by simple linear algebra, and thus we omit it here.
\end{proof}

\section{Proof of Theorem 1}\label{subsec:proof_joint_stationarity}
\begin{proof}
Under Assumption~\ref{assump:vector_VAR} that the vector time series $\vecz_t$ follows a VAR$(\Qtilde)$ process, we can derive that the vectorized matrix time series $\matX_t$ and the vector time series $\vecz_t$ jointly follow a VAR$(\max(P,Q,\Qtilde))$ process, namely,
\begin{equation}\label{eq:matrix-vector-joint-VAR-replicate}
    \begin{bmatrix}
        \vecx_t \\
        \vecz_t
    \end{bmatrix} = \sum_{l=1}^{\max(P,Q,\Qtilde)} \begin{bmatrix}
        \left(\matB_l\otimes\matA_l\right)\odot\mathbf{1}_{\{l\le P\}} & \matG_l^{\top}\odot\mathbf{1}_{\{l\le Q\}} \\
        \matO_{D\times S} & \matC_l\odot\mathbf{1}_{\{l\le \Qtilde\}}
    \end{bmatrix}\begin{bmatrix}
        \vecx_{t-l} \\
        \vecz_{t-l}
    \end{bmatrix} + \begin{bmatrix}
        \vece_{t} \\
        \vecnu_{t}
    \end{bmatrix}.
\end{equation}
Let $L = \max(P,Q,\Qtilde)$ and $\vecy_t = [\vecx_t^{\top}, \vecz_t^{\top}]$. Denote the transition matrix in~\eqref{eq:matrix-vector-joint-VAR-replicate} at lag-$l$ as $\matJ_l \in \mathbb{R}^{(S+D)\times(S+D)}$ and the error term as $\vecu_t^{\top} = [\vece_{t}^{\top}, \vecnu_t^{\top}]$, then we can rewrite the VAR$(L)$ process in~\eqref{eq:matrix-vector-joint-VAR-replicate} as a VAR$(1)$ process as:
\begin{equation}\label{eq:matrix-vector-VAR1}
    \begin{bmatrix}
        \vecy_t \\
        \vecy_{t-1} \\
        \vdots \\
        \vecy_{t-L+1}
    \end{bmatrix} = \begin{bmatrix}
        \matJ_1 & \matJ_2 & \cdots & \matJ_{L-1} & \matJ_L \\
        \matI_{S+D} & \matO_{S+D} & \cdots & \cdots & \matO_{S+D} \\
        \matO_{S+D} & \matI_{S+D} & \matO_{S+D} & \cdots & \matO_{S+D} \\
        \vdots & \vdots & \ddots & \ddots & \vdots \\
        \matO_{S+D} & \matO_{S+D} & \cdots & \matI_{S+D} & \matO_{S+D}
    \end{bmatrix}\begin{bmatrix}
        \vecy_{t-1} \\
        \vecy_{t-2} \\
        \vdots \\
        \vecy_{t-L} 
    \end{bmatrix} + \begin{bmatrix}
            \vecu_t \\
            \mathbf{0}_{S+D} \\
            \vdots \\
            \mathbf{0}_{S+D}
        \end{bmatrix},
\end{equation}
where we use $\matO_{S+D}$ to denote a zero matrix of size $(S+D)\times (S+D)$. For this VAR$(1)$ process to be stationary, we require that $\mathrm{det}\left(\lambda\matI - \matJ\right) \neq 0$ for all $|\lambda| \ge 1, \lambda\in\mathbb{C}$, where $\matJ$ is the transition matrix in~\eqref{eq:matrix-vector-VAR1}. The determinant $\mathrm{det}\left(\lambda\matI - \matJ\right)$ can be simplified by column operations as:
\begin{align*}
    &\mathrm{det}\left(\lambda\matI - \matJ\right) \\
    &= \mathrm{det}
    \begin{bmatrix}
        \lambda^{L}\matI_{S} - \sum\limits_{l=1}^{L}\lambda^{L-l}\left(\matB_l\otimes\matA_l\right)\odot\mathbf{1}_{\{l\le P\}} & -\sum\limits_{l=1}^{L}\lambda^{L-l} \matG_l^{\top}\odot\mathbf{1}_{\{l\le Q\}} \\
        \matO & \lambda^{L}\matI_{D} - \sum\limits_{l=1}^{L}\lambda^{L-l} \matC_l\odot\mathbf{1}_{\{l\le \Qtilde\}}
    \end{bmatrix}\\
    & = \lambda^{2L} \mathrm{det}\left[\boldsymbol{\Phi}_1(\lambda)\right]\mathrm{det}\left[\boldsymbol{\Phi}_2(\lambda)\right],
\end{align*}
where $\matPhi_1(\lambda) = \matI_{S} - \sum_{p=1}^{P} \lambda^{-p}\left(\matB_p\otimes\matA_p\right)$ and $\matPhi_2(\lambda)=\matI_D - \sum_{\qtilde=1}^{\Qtilde}\lambda^{-\qtilde}\matC_{\qtilde}$, and setting $y=1/\lambda$ completes the proof. 
\end{proof}

\section{Theory under Fixed Spatial Dimension}\label{sec:fix-dim}
\subsection{Proof of Proposition~\ref{thm:error-cov-consistency}}\label{subsec:proof-error-cov-consistency}
\begin{proof}
For the brevity of the presentation, we fix $P,Q$ as $1$, but the proofs presented below can be easily extended to an arbitrary $P,Q$. For the vectorized MARAC$(1,1)$ model~\eqref{eq:MARAC-model-vec}, we can equivalently write it as:
\begin{equation}\label{eq:MARAC_compact_form}
    \vecx_t = \vecy_t\vectheta + \vece_t,
\end{equation}
where $\vecy_t = [\vecx_{t-1}^{\top}\otimes\matI_{S}; \vecz_{t-1}^{\top}\otimes\matK]$ and $\vectheta = [\vect{\matB_1\otimes\matA_1}^{\top}, \vecgamma_1^{\top}]^{\top}$. Using $\matOmega = \bSigma^{-1}$ to denote the precision matrix for $\vece_t$, we can rewrite the penalized likelihood in~\eqref{eq:MARAC-PMLE-KRR} for $(\vectheta,\matOmega)$ as:
\begin{equation}\label{eq:MARAC_PMLE_compact_form}
    h(\vectheta,\matOmega) = -\frac12\log\left|\matOmega\right| + \frac12\tr{\matOmega\matS(\vectheta)} + \frac{\lambda}{2}\vectheta^{\top}\widetilde{\matK}\vectheta,
\end{equation}
where $\matS(\vectheta) = T^{-1}\sum_{t=1}^{T}(\vecx_t-\vecy_t\vectheta)(\vecx_t-\vecy_t\vectheta)^{\top}$, $\widetilde{\matK}$ is defined as:
\begin{equation*}
    \widetilde{\matK} = \begin{bmatrix}
        \matO_{S\times S}\otimes \matK & \matO_{S\times D}\otimes\matK \\
        \matO_{D\times S}\otimes\matK & \matI_{D}\otimes \matK
    \end{bmatrix}.
\end{equation*}
We use $\vectheta^{*},\matOmega^{*}$ to denote the ground truth of $\vectheta,\matOmega$, respectively. We define $\setF_{\vectheta}$ and $\setF_{\matOmega}$ as:
\begin{align*}
    \setF_{\vectheta} & = \{[\vect{\matB_1\otimes\matA_1}^{\top}, \vecgamma_1^{\top}]^{\top}|\twonorm{\matA_1}=1,\text{sign}(\tr{\matA_1})=1\} \\
    \setF_{\matOmega} & = \{\bSigmac^{-1}\otimes\bSigmar^{-1}|\bSigmar\in\mathbb{R}^{M\times M},\bSigmac\in\mathbb{R}^{N\times N}, \mineigen{\bSigmar},\mineigen{\bSigmac} > 0\}.
\end{align*}
The estimators of MARAC, denoted as $\widehat{\vectheta},\widehat{\matOmega}$, is the minimizer of $h(\vectheta,\matOmega)$ with $\vectheta\in\setF_{\vectheta},\matOmega\in\setF_{\matOmega}$.

In order to establish the consistency of $\widehat{\bSigma}=\widehat{\matOmega}^{-1}$, it suffices to show that for any constant $c > 0$:
\begin{equation}\label{eq:consistency_equal_condition}
    \mathrm{P}\left(\inf_{\twonorm{\bar{\matOmega} - \matOmega^{*}}\ge c}\inf_{\bar{\vectheta}} h(\bar{\vectheta},\bar{\matOmega}) \le h(\vectheta^{*},\matOmega^{*})\right) \rightarrow 0, \text{ as }T\rightarrow \infty.
\end{equation}
This is because if~\eqref{eq:consistency_equal_condition} is established, then as $T\rightarrow\infty$ we have:
\begin{equation*}
    \mathrm{P}\left(\inf_{\twonorm{\bar{\matOmega} - \matOmega^{*}}\ge c}\inf_{\bar{\vectheta}\in\setF_{\vectheta}} h(\bar{\vectheta},\bar{\matOmega}) \ge \inf_{\twonorm{\bar{\matOmega} - \matOmega^{*}}\ge c}\inf_{\bar{\vectheta}} h(\bar{\vectheta},\bar{\matOmega}) > h(\vectheta^{*},\matOmega^{*})\ge h(\widehat{\vectheta},\widehat{\matOmega})\right)
\end{equation*}
approaching $1$ and thus we must have $\twonorm{\widehat{\matOmega} - \matOmega^{*}} < c$ with probability approaching $1$ as $T\rightarrow\infty$, and the consistency is established since $c$ is arbitrary.

To prove~\eqref{eq:consistency_equal_condition}, we first fix $\matOmega = \bar{\matOmega}$ and let $\widetilde{\vectheta}(\bar{\matOmega}) = \argmin_{\vectheta} h(\vectheta,\bar{\matOmega})$, thus we have: 
\begin{equation}\label{eq:unconstrained_GLS_solution}
    \widetilde{\vectheta}(\bar{\matOmega}) = \left(\frac{\sum_{t}\vecy_t^{\top}\bar{\matOmega}\vecy_t}{T} + \lambda\widetilde{\matK}\right)^{-1}\left(\frac{\sum_{t}\vecy_t^{\top}\bar{\matOmega}\vecx_t}{T}\right),
\end{equation}
which is a consistent estimator of $\vectheta^{*}$ for any $\bar{\matOmega}$ given that $\lambda\rightarrow 0$ and the matrix and vector time series are covariance-stationary. To see that $\widetilde{\vectheta}(\bar{\matOmega})\overset{p.}{\rightarrow}\vectheta^*$, notice that:
\begin{equation}\label{eq:theta-tilde-transformation}
    \widetilde{\vectheta}(\bar{\matOmega}) = (\matI-\lambda\widetilde{\matK})\vectheta^* + \left(\frac{\sum_{t}\vecy_t^{\top}\bar{\matOmega}\vecy_t}{T} + \lambda\widetilde{\matK}\right)^{-1}\left(\frac{\sum_{t}\vecy_t^{\top}\bar{\matOmega}\vece_t}{T}\right),
\end{equation}
and the first term converges to $\vectheta^*$ since $\lambda = o(1)$. In the second term of~\eqref{eq:theta-tilde-transformation}, we have:
\begin{equation}\label{eq:limit-yOmegey}
    \frac{\sum_{t}\vecy_t^{\top}\bar{\matOmega}\vecy_t}{T} + \lambda\widetilde{\matK} \overset{p.}{\rightarrow} \begin{bmatrix}
        \bSigma_{\vecx,\vecx}^*\otimes \bar{\matOmega} & \bSigma_{\vecx,\vecz}^*\otimes \bar{\matOmega}\matK \\
        \bSigma_{\vecz,\vecx}^*\otimes \matK\bar{\matOmega} & \bSigma_{\vecz,\vecz}^*\otimes \matK\bar{\matOmega}\matK
    \end{bmatrix},
\end{equation}
where $\bSigma_{\vecx,\vecx}^* = \mathrm{Var}(\vecx_t)$, $\bSigma_{\vecx,\vecz}^* = \mathrm{Cov}(\vecx_t,\vecz_t)$ and $\bSigma_{\vecz,\vecz}^* = \mathrm{Var}(\vecz_t)$. The convergence in probability in~\eqref{eq:limit-yOmegey} holds due to the joint stationarity of $\vecx_t$ and $\vecz_t$ and the assumption that $\lambda = o(1)$. We further note that the sequence $\{\vecy_t^\top\bar{\matOmega}\vece_t\}_{t=1}^T$ is a martingale difference sequence (MDS), and we have $\sum_{t=1}^T\vecy_t^\top\bar{\matOmega}\vece_t/T = O_P(T^{-1/2})$ by the central limit theorem (CLT) of MDS (see proposition 7.9 of~\citet{hamilton2020time} for the central limit theorem of martingale difference sequence). Combining this result together with~\eqref{eq:limit-yOmegey}, we conclude that the second term in~\eqref{eq:theta-tilde-transformation} is $o_P(1)$ and thus $\widetilde{\vectheta}(\bar{\Omega})$ is consistent for $\vectheta^*$.

Plugging $\widetilde{\vectheta}(\bar{\matOmega})$ into $h(\vectheta,\bar{\matOmega})$ yields the profile likelihood of $\bar{\matOmega}$:
\begin{equation*}
    \ell(\bar{\matOmega}) = -\frac12 \log|\bar{\matOmega}| + \frac12 \tr{\bar{\matOmega}\frac{\sum_t \vecx_t[\vecx_t - \vecy_t\widetilde{\vectheta}(\bar{\matOmega})]^{\top}}{T}}.
\end{equation*}
To prove~\eqref{eq:consistency_equal_condition}, it suffices to show that:
\begin{equation}\label{eq:consistency_equal_condition2}
    \mathrm{P}\left(\inf_{\twonorm{\bar{\matOmega} - \matOmega^{*}}\ge c} \ell(\bar{\matOmega}) \le \ell(\matOmega^{*})\right) \rightarrow 0, \text{ as }T\rightarrow\infty,
\end{equation}
since $\ell(\matOmega^{*}) \le h(\vectheta^{*},\matOmega^{*})$. Now, since $\widetilde{\vectheta}(\bar{\matOmega})\overset{p.}{\rightarrow}\vectheta^{*}$, we can write $\widetilde{\vectheta}(\bar{\matOmega}) = \vectheta^{*} + \veczeta$, with $\twonorm{\veczeta} = o_P(1)$. Using this new notation, we can rewrite $\ell(\bar{\matOmega})$ as:
\begin{align}
    \ell(\bar{\matOmega}) & = -\frac12 \log|\bar{\matOmega}| + \frac12 \tr{\bar{\matOmega}\frac{\sum_t \vecx_t\vece_t^{\top}}{T}} -\frac12\tr{\left(\frac{\sum_t \vecx_t^{\top}\bar{\matOmega}\vecy_t}{T}\right)\veczeta} \label{eq:profile-likelihood-full}\\
    & = \widetilde{\ell}(\bar{\matOmega}) - \frac12\tr{\left(\frac{\sum_t \vecx_t^{\top}\bar{\matOmega}\vecy_t}{T}\right)\veczeta}, \notag
\end{align}
where we define the first two terms in~\eqref{eq:profile-likelihood-full} as $\widetilde{\ell}(\bar{\matOmega})$.

By the Cauchy-Schwartz inequality, we have:
\begin{equation}\label{eq:profile-likelihood-op1-term}
    \left|\frac12\tr{\left(\frac{\sum_t \vecx_t^{\top}\bar{\matOmega}\vecy_t}{T}\right)\veczeta}\right| \le \frac12 \left\|\frac{\sum_t \vecx_t^{\top}\bar{\matOmega}\vecy_t}{T}\right\|_{\mathrm{F}}\cdot\twonorm{\veczeta}.
\end{equation}
By the definition of $\vecy_t$, we have:
\begin{equation*}
    \frac{\sum_t \vecx_t^{\top}\bar{\matOmega}\vecy_t}{T} = \left[\left(\frac{\sum_t \vecx_{t-1}\otimes \vecx_t}{T}\right)^\top\left(\matI_{S}\otimes\bar{\matOmega}\right);\left(\frac{\sum_t \vecz_{t-1}\otimes \vecx_t}{T}\right)^\top\left(\matI_{D}\otimes\bar{\matOmega}\matK\right)\right],
\end{equation*}
and notice that $\vecx_{t-1}\otimes\vecx_t$ and $\vecz_{t-1}\otimes\vecx_t$ are just rearranged versions of $\vecx_{t}\vecx_{t-1}^\top$ and $\vecx_t\vecz_{t-1}^{\top}$, respectively. Therefore, by the joint stationarity of $\vecx_t$ and $\vecz_t$, we have the time average of $\vecx_{t-1}\otimes\vecx_t$ and $\vecz_{t-1}\otimes\vecx_t$ converging to the rearranged version of some constant auto-covariance matrices and therefore we have the term on the right-hand side of~\eqref{eq:profile-likelihood-op1-term} being $o_P(1)$.

Given this argument, proving~\eqref{eq:consistency_equal_condition2} is now equivalent to proving:
\begin{equation}\label{eq:consistency_equal_condition3}
    \mathrm{P}\left(\inf_{\twonorm{\bar{\matOmega} - \matOmega^{*}}\ge c} \widetilde{\ell}(\bar{\matOmega}) \le \widetilde{\ell}(\matOmega^{*})\right) \rightarrow 0, \text{ as }T\rightarrow\infty.
\end{equation}

Define $\widetilde{\matOmega}$ as the unconstrained minimizer of $\widetilde{\ell}(\matOmega)$, then explicitly, we have:
\begin{align*}
    \widetilde{\matOmega} = \argmin_{\matOmega} \widetilde{\ell}(\matOmega) & = \left(\frac{\sum_t \vece_t\vecx_t^\top}{T}\right)^{-1} \\
    & = \left(\frac{\sum_t \vece_t\vece_t^\top}{T} + \frac{\sum_t \vece_t\left(\vecy_t\vectheta^*\right)^\top}{T}\right)^{-1} \overset{p.}{\rightarrow} \matOmega^*,
\end{align*}
where the final argument on the convergence in probability to $\matOmega^*$ is based on the fact that $\sum_{t=1}^{T} \vece_t\left(\vecy_t\vectheta^*\right)^\top/T = O_P(T^{-1/2})$ by the CLT of MDS. By the second-order Taylor expansion of $\widetilde{\ell}(\bar{\matOmega})$ at $\widetilde{\matOmega}$, we have:
\begin{equation}\label{eq:sigma-hat-taylor-expansion}
    \widetilde{\ell}(\bar{\matOmega}) = \widetilde{\ell}(\widetilde{\matOmega}) + \frac{1}{4} \vect{\bar{\matOmega} - \widetilde{\matOmega}}^{\top} \left[\check{\matOmega}^{-1}\otimes\check{\matOmega}^{-1}\right]\vect{\bar{\matOmega} - \widetilde{\matOmega}},
\end{equation}
where $\check{\matOmega} = \widetilde{\matOmega} + \eta (\bar{\matOmega} - \widetilde{\matOmega})$, for some $\eta\in[0,1]$. For any constant $c > 0$ such that $\twonorm{\bar{\matOmega} - \matOmega^*} = c$, let $c=\kappa \maxeigen{\matOmega^*}$, where $\kappa > 0$ is also a constant that relates to $c$ only. Consequently, we have:
\begin{equation*}
    \left|\maxeigen{\bar{\matOmega}} - \maxeigen{\matOmega^*}\right| \le \specnorm{\bar{\matOmega} - \matOmega^*} \le \twonorm{\bar{\matOmega} - \matOmega^*} = \kappa \maxeigen{\matOmega^*},
\end{equation*}
and thus $\maxeigen{\bar{\matOmega}} \le (1+\kappa)\maxeigen{\matOmega^*}$. Conditioning on the event that $\twonorm{\bar{\matOmega} - \matOmega^*} = c$, we first have $\twonorm{\bar{\matOmega} - \widetilde{\matOmega}} \ge c/2$ to hold with probability approaching one, due to the consistency of $\widetilde{\matOmega}$. Furthermore, we also have:
\begin{align*}
    \mineigen{\check{\matOmega}^{-1}\otimes\check{\matOmega}^{-1}} = \mineigen{\check{\matOmega}^{-1}}^2 & = \frac{1}{\maxeigen{\check{\matOmega}}^2} \\
    & \ge \left[\frac{1}{\maxeigen{\widetilde{\matOmega}} + \maxeigen{\bar{\matOmega}}}\right]^2 \\
    & \ge \left[\frac{1}{2\maxeigen{\matOmega^*} + (c + \maxeigen{\matOmega^*})}\right]^2 = \frac{1}{(3+\kappa)^2}\cdot\frac{1}{\maxeigen{\matOmega^*}^2},
\end{align*}
where the last inequality holds with probability approaching one since $\mathrm{P}\left[\maxeigen{\widetilde{\matOmega}} \le 2\maxeigen{\matOmega^*}\right]\rightarrow 1$. Utilizing these facts together with~\eqref{eq:sigma-hat-taylor-expansion}, we end up having:
\begin{equation}
    \mathrm{P}\left[\widetilde{\ell}(\bar{\matOmega}) \ge \widetilde{\ell}(\widetilde{\matOmega}) + \frac{1}{16}\cdot\left(\frac{\kappa}{3+\kappa}\right)^2\right] \rightarrow 1, \text{ as } T\rightarrow\infty,
\end{equation}
for any $\bar{\matOmega}$ such that $\twonorm{\bar{\matOmega}-\matOmega^*}=c=\kappa\maxeigen{\matOmega^*}$. Since $\kappa$ is an arbitrary positive constant and $\widetilde{\ell}(\widetilde{\matOmega})\overset{p.}{\rightarrow}\widetilde{\ell}(\matOmega^*)$, we establish~\eqref{eq:consistency_equal_condition3} and thereby completes the proof.
\end{proof}

\subsection{Proof of Theorem~\ref{thm:PMLE-Asymp-Normal}}\label{subsec:proof_asymptotic_normal}
To prove Theorem~\ref{thm:PMLE-Asymp-Normal}, we first establish the consistency and the convergence rate of the estimators in Lemma~\ref{lemma:convergence_rate} below. 

\begin{lemma}\label{lemma:convergence_rate}
    Under the same assumption as Theorem~\ref{thm:PMLE-Asymp-Normal}, all model estimators for MARAC are $\sqrt{T}$-consistent, namely:
    \begin{equation*}
        \twonorm{\widehat{\matA}_p-\matA_p^*} = O_P\left(\frac{1}{\sqrt{T}}\right), \twonorm{\widehat{\matB}_p-\matB_p^*} = O_P\left(\frac{1}{\sqrt{T}}\right), \twonorm{\widehat{\vecgamma}_{q}-\vecgamma_q^*} = O_P\left(\frac{1}{\sqrt{T}}\right),
    \end{equation*}
    for $p\in[P], q\in[Q]$. As a direct result, we also have:
    \begin{equation*}
        \twonorm{\widehat{\matB}_p\otimes\widehat{\matA}_p - \matB_p^*\otimes\matA^*_p} = O_P\left(\frac{1}{\sqrt{T}}\right), \text{ for }p\in[P].
    \end{equation*}
\end{lemma}
We delay the proof of Lemma~\ref{lemma:convergence_rate} to Section~\ref{subsec:proof-fixdim-conv-rate}. With this lemma, we are now ready to present the proof of Theorem~\ref{thm:PMLE-Asymp-Normal}.
\begin{proof}
For the simplicity of notation and presentation, we fix $P,Q$ as $1$, but the proving technique can be generalized to arbitrary $P,Q$. To start with, we revisit the updating rule for $\matA_p^{(l+1)}$ in~\eqref{eq:Update-Ap}. By plugging in the data-generating model for $\matX_t$ according to MARAC$(1,1)$ model, we can transform~\eqref{eq:Update-Ap} into:
\begin{equation*}
    \sum_{t\in[T]}\left[\Delta\matA_1\matX_{t-1}\widehat{\matB}_1^{\top} + \matA_1^*\matX_{t-1}\Delta\matB_1^{\top}+\Delta\tenG_1\tvprod\vecz_{t-1}-\matE_t\right]\widehat{\bSigma}_c^{-1}\widehat{\matB}_1\matX_{t-1}^{\top} = \matO_{M\times M},
\end{equation*}
where for any arbitrary matrix/tensor $\mathbf{M}$, we define $\Delta\mathbf{M}$ as $\Delta\mathbf{M} = \widehat{\mathbf{M}}-\mathbf{M}^*$. One can simplify the estimating equation above by left multiplying $\widehat{\bSigma}_r^{-1}$ and then vectorize both sides to obtain:
\begin{align*}
    & \sum_{t\in[T]} \left[(\matB_{1}^{*}\matX_{t-1}^{\top})^{\top}(\bSigmac^*)^{-1}(\matB_{1}^{*}\matX_{t-1}^{\top})\otimes(\bSigmar^*)^{-1}\right]\vect{\widehat{\matA}_1-\matA_1^*} \\
    + & \sum_{t\in[T]} \left[(\matB_{1}^{*}\matX_{t-1}^{\top})^{\top}(\bSigmac^*)^{-1}\otimes(\bSigmar^*)^{-1}\matA_1^*\matX_{t-1}\right]\vect{\widehat{\matB}_1^{\top}-(\matB_1^*)^{\top}} \\
    + & \sum_{t\in[T]} \left\{\vecz_{t-1}^{\top}\otimes\left[(\matB_{1}^{*}\matX_{t-1}^{\top})^{\top}(\bSigmac^*)^{-1}\otimes(\bSigmar^*)^{-1}\matK\right]\right\}\vect{\widehat{\vecgamma}_1-\vecgamma_1^*} \\
    = & \sum_{t\in[T]} \left[(\matB_{1}^{*}\matX_{t-1}^{\top})^{\top}(\bSigmac^*)^{-1}\otimes(\bSigmar^*)^{-1}\right]\vect{\matE_t} + o_{P}(\sqrt{T}).
\end{align*}
On the left-hand side of the equation above, we replace $\widehat{\matB}_1,\widehat{\bSigma}_r,\widehat{\bSigma}_c$ with their true values $\matB_1^*,\bSigmar^*,\bSigmac^*$, since the discrepancies are of order $o_P(1)$ and can thus be incorporated into the $o_P(\sqrt{T})$ term given the $\sqrt{T}$-consistency of $\widehat{\matA}_1,\widehat{\matB}_1,\widehat{\vecgamma}_1$. On the right-hand side, we have:
\begin{align*}
& \sum_t\vect{\widehat{\bSigma}_r^{-1}\matE_t\widehat{\bSigma}_c^{-1}\widehat{\matB}_1\matX_{t-1}^\top} \\
& = \sum_t\left[\vece_t^\top\otimes\left(\matX_{t-1}\otimes\matI_M\right)\right]\mathbf{vec}\left[\left(\widehat{\matB}_1^\top\otimes\matI_M\right)\widehat{\bSigma}^{-1}\right],
\end{align*}
where the process $\{\vece_t^\top\otimes\left(\matX_{t-1}\otimes\matI_M\right)\}_{t=1}^T$ is a martingale difference sequence and the martingale central limit theorem~\citep{hall2014martingale} implies that $\sum_t\left[\vece_t^\top\otimes\left(\matX_{t-1}\otimes\matI_M\right)\right] = O_P(\sqrt{T})$, and thus by the consistency of $\widehat{\bSigma}$ and $\widehat{\matB}_1$, we can replace $\widehat{\bSigma}$ and $\widehat{\matB}_1$ with their true values and incorporate the remainders into $o_P(\sqrt{T})$.

Similar transformations can be applied to~\eqref{eq:Update-Bp} and~\eqref{eq:Update-Gammaq}, where the penalty term is incorporated into $o_P(\sqrt{T})$ due to the assumption that $\lambda = o(T^{-\frac12})$. With the notation that $\matU_t = \matI_N\otimes \matA_1^{*}\matX_{t-1}$, $\matV_t=\matB_1^*\matX_{t-1}^{\top}\otimes\matI_M$, $\matY_t=\vecz_{t-1}^{\top}\otimes\matK$ and $\matW_t = [\matV_t;\matU_t;\matY_t]$, these transformed estimating equations can be converted altogether into:
\begin{align}
    \left(\frac1T\sum_{t\in[T]} \matW_t^{\top}(\bSigma^{*})^{-1}\matW_t\right)\vect{\widehat{\matTheta}-\matTheta^*} & = \frac1T\sum_{t\in[T]} \matW_t^{\top}(\bSigma^{*})^{-1}\vect{\matE_t} \nonumber \\
    & + o_{P}(T^{-1/2}),\label{eq:asymp-dist-equation}
\end{align}
where $\vect{\widehat{\matTheta}-\matTheta^*} = [\vect{\widehat{\tenA} - \tenA^{*}}^{\top}, \vect{\widehat{\tenB} - \tenB^{*}}^{\top}, \vect{\widehat{\tenR} - \tenR^{*}}^{\top}]^{\top}$, and $\widehat{\tenA},\widehat{\tenB},\widehat{\tenR}$ are defined as $[\widehat{\tenA}]_{::p} = \widehat{\matA}_p, [\widehat{\tenB}]_{::p} = \widehat{\matB}_p^{\top}$, $[\widehat{\tenR}]_{:dq} = \widehat{\vecgamma}_{q,d}$ and $\tenA^{*},\tenB^{*},\tenR^{*}$ are the corresponding true coefficients.

In~\eqref{eq:asymp-dist-equation}, we first establish that:
\begin{equation}\label{eq:asymp-dist-equation-1}
    (1/T)\sum_{t\in[T]} \matW_t^{\top}(\bSigma^{*})^{-1}\matW_t \overset{p.}{\rightarrow}\mathrm{E}\left[\matW_t^{\top}(\bSigma^{*})^{-1}\matW_t\right].
\end{equation}
To prove~\ref{eq:asymp-dist-equation-1}, by the assumption that $\matX_t$ and $\vecz_t$ are zero-meaned and jointly stationary, we have $T^{-1}\sum_{t\in [T]}\widetilde{\vecx}_t\widetilde{\vecx}_t^{\top} \overset{p.}{\rightarrow} \mathrm{E}[\widetilde{\vecx}_t\widetilde{\vecx}_t^{\top}]$ by Lemma~\ref{lemma:covariance-consistency} and Corollary~\ref{corollary:MARAC-covariance-consistency}, where $\widetilde{\vecx}_t = [\vecx_t^{\top},\vecz_t^{\top}]^{\top}$. See details of Lemma~\ref{lemma:covariance-consistency} and Corollary~\ref{corollary:MARAC-covariance-consistency} in Section~\ref{subsec:covariance-consistency}. Then since each element of $\matW_t^{\top}(\bSigma^{*})^{-1}\matW_t$ is a linear combination of terms in $\widetilde{\vecx}_t\widetilde{\vecx}_t^{\top}$ (thus a continuous mapping), it is straightforward that~\eqref{eq:asymp-dist-equation-1} holds elementwise.

Given~\eqref{eq:asymp-dist-equation-1} and the fact that $\widehat{\matTheta}$ is $\sqrt{T}$-consistent, we can rewrite~\eqref{eq:asymp-dist-equation} as:
\begin{align}
    \mathrm{E}\left[\matW_t^{\top}(\bSigma^{*})^{-1}\matW_t\right]\vect{\widehat{\matTheta}-\matTheta^*} & = \frac1T\sum_{t\in[T]} \matW_t^{\top}(\bSigma^{*})^{-1}\vect{\matE_t} \nonumber \\
    & + o_{P}(T^{-1/2}),\label{eq:asymp-dist-equation-2}
\end{align}

For the term on the right-hand side of~\eqref{eq:asymp-dist-equation-2}, first notice that the sequence $\{\veceta_t\}_{t=1}^{T}$, where $\veceta_t = \matW_t^{\top}(\bSigma^{*})^{-1}\vect{\matE_t}$,  is a zero-meaned, stationary vector martingale difference sequence (MDS), thanks to the independence of $\matE_t$ from the jointly stationary $\matX_{t-1}$ and $\vecz_{t-1}$. By the martingale central limit theorem~\citep{hall2014martingale}, we have:
\begin{equation}\label{eq:asymp-dist-equation-3}
    \frac{1}{\sqrt{T}}\sum_{t\in[T]} \matW_t^{\top}(\bSigma^{*})^{-1}\vect{\matE_t} \overset{d.}{\rightarrow} \mathcal{N}(\mathbf{0},\mathrm{E}\left[\matW_t^{\top}(\bSigma^{*})^{-1}\matW_t\right]).
\end{equation}

Combining~\eqref{eq:asymp-dist-equation-2} and~\eqref{eq:asymp-dist-equation-3}, we end up having:
\begin{equation}\label{eq:asymp-dist-equation4}
    \mathrm{E}\left[\matW_t^{\top}(\bSigma^{*})^{-1}\matW_t\right]\sqrt{T}\vect{\widehat{\matTheta}-\matTheta^*} \overset{d.}{\rightarrow} \mathcal{N}(\mathbf{0},\mathrm{E}\left[\matW_t^{\top}(\bSigma^{*})^{-1}\matW_t\right]).
\end{equation}
The asymptotic distribution of $\sqrt{T}\vect{\widehat{\matTheta}-\matTheta^*}$ can thus be derived by multiplying both sides of~\eqref{eq:asymp-dist-equation4} by the inverse of $\matL = \mathrm{E}\left[\matW_t^{\top}(\bSigma^{*})^{-1}\matW_t\right]$. However, the matrix $\matL$ is not a full-rank matrix, because $\matL\boldsymbol{\mu}=\mathbf{0}$, where $\boldsymbol{\mu} = [\vect{\tenA^*}^{\top}, -\vect{\tenB^*}^{\top}, \mathbf{0}^{\top}]^{\top}$. As a remedy, let $\boldsymbol{\zeta}=[\vect{\matA_1^{*}}^{\top}\mathbf{0}^{\top}]^{\top}\in\mathbb{R}^{M^2+N^2+DMN}$, then given the identifiability constraint that $\twonorm{\matA^*_1}=\twonorm{\widehat{\matA}_1} = 1$ and the fact that $\widehat{\matA}_1$ is $\sqrt{T}$-consistent, we have $\vect{\matA_1^*}^{\top}\vect{\widehat{\matA}_1 - \matA^*_1} = o_P(T^{-1/2})$. Therefore, we have:
\begin{equation}\label{eq:asymp-dist-equation5}
    \sqrt{T}\boldsymbol{\zeta}^{\top}\vect{\widehat{\matTheta}-\matTheta^*} \overset{p.}{\rightarrow} 0.
\end{equation}
Combining~\eqref{eq:asymp-dist-equation4} and~\eqref{eq:asymp-dist-equation5} and using the Slutsky's theorem, we have $\matH\sqrt{T}\mathbf{vec}(\widehat{\matTheta}-\matTheta^*) \overset{d.}{\rightarrow} \mathcal{N}(\mathbf{0},\matL)$, where $\matH = \matL + \boldsymbol{\zeta}\boldsymbol{\zeta}^{\top}$ and thus:
\begin{equation}\label{eq:asymp-dist-result}
    \sqrt{T}\mathbf{vec}(\widehat{\matTheta}-\matTheta^*) \overset{d.}{\rightarrow} \mathcal{N}(\mathbf{0},\matH^{-1}\matL\matH^{-1}).
\end{equation}

The final asymptotic distribution of $\mathrm{vec}(\widehat{\matB}_1^{\top})\otimes\mathrm{vec}(\widehat{\matA}_1)$ and $\matK\widehat{\vecgamma}_{q,d}$ can be derived easily from~\eqref{eq:asymp-dist-result} with the multivariate delta method, and we omit the details here.
\end{proof}

\subsection{Proof of Corollary~\ref{cor:SpecTest}}\label{subsec:SpecTestProof}
\begin{proof}
Based on the asymptotic distribution of the MARAC model estimators in~\eqref{eq:PMLE-Asymp-Dist}, it is straightforward that the marginal asymptotic distribution of $\htenG_1,\ldots,\htenG_Q$ follows:
\begin{equation}\label{eq:limit-dist-G}
    \sqrt{T}\begin{bmatrix}
        \vect{\htenG_1 - \tenG^*_1} \\
        \cdots \\
        \vect{\htenG_Q - \tenG^*_Q} 
    \end{bmatrix} \overset{d.}{\longrightarrow} \mathcal{N} \left( \mathbf{0}, \begin{bmatrix}
        \matO: \matI_{QD}\otimes\matK
    \end{bmatrix}\matXi\begin{bmatrix}
        \matO \\
        \matI_{QD}\otimes\matK
    \end{bmatrix}\right).
\end{equation}
Unwrapping the matrix $\matXi$, one can simplify the asymptotic variance in~\eqref{eq:limit-dist-G} as:
\begin{equation*}
    \mathbf{\Psi}\coloneqq \begin{bmatrix}
        \matO: \matI_{QD}\otimes\matK
    \end{bmatrix}\matXi\begin{bmatrix}
        \matO \\
        \matI_{QD}\otimes\matK
    \end{bmatrix} = \left(\matI\otimes\matK\right)\left[\matD - \matC\vecgamma\left(\matC\vecgamma\right)^\top\right]\left(\matI\otimes\matK\right),
\end{equation*}
where $\matD$ is the lower-right $MNQD\times MNQD$ block of $\matH^{-1}$, and $\matC$ is the lower-left block under the same block partition. To estimate the rank of matrix $\matPsi$, it is sufficient to estimate the rank of $\matD$, as $\matI\otimes\matK$ is full-rank, and $\matC\vecgamma\left(\matC\vecgamma\right)^\top$ is rank-1. Note that matrix $\matH$ is full-rank, and the top-left block of $\matH$, denoted as $\matH_{11}$, is:
\begin{equation*}
\mathrm{E}\left\{\begin{bmatrix}
\ddots & \cdots & \iddots & \ddots & \cdots & \iddots \\
\cdot & \matX_{t-i}\matB_i^\top\bSigma_c^{-1}\matB_j\matX_{t-j}^\top \otimes \bSigma_r^{-1} & \cdots & \cdots & \matX_{t-i}\matB_i^\top\bSigma_c^{-1}\otimes \bSigma_r^{-1}\matA_j\matX_{t-j} & \cdots \\
\iddots & \cdots & \ddots & \iddots & \cdots & \ddots \\
\ddots & \cdots & \iddots & \ddots & \cdots & \iddots \\
\cdot & \bSigma_c^{-1}\matB_i\matX_{t-i}^\top \otimes \matX_{t-j}^\top\matA_j^\top\bSigma_r^{-1} & \cdots & \cdots & \bSigma_c^{-1}\otimes \matX_{t-i}^\top\matA_i^\top\bSigma_r^{-1}\matA_j\matX_{t-j} & \cdots \\
\iddots & \cdots & \ddots & \iddots & \cdots & \ddots \\
\end{bmatrix}\right\} + \boldsymbol{\alpha}\boldsymbol{\alpha}^\top,
\end{equation*}
where $1\le i,j\le P$, and $\boldsymbol{\alpha} = [\vect{\matA_1}^\top,\ldots,\vect{\matA_P}^\top,\mathbf{0}^\top]^\top$. Here, all model parameters are the ground truth values, and we omit the asterisk notation for simplicity. This matrix is the key component of the asymptotic variance of the MAR($P$) model, see Theorem 3 of~\citet{chen2021autoregressive}, and is thus invertible. Consequently, the Schur complement of $\matH$ is invertible and thus $\matD$ is a full-rank matrix. Therefore, we have $\text{rank}(\matPsi) \ge MNQD-1$.

Finally, based on~\eqref{eq:limit-dist-G}, we have $T \cdot \left(\widehat{\mathbf{g}}-\mathbf{g}^*\right)^\top \mathbf{\Psi}^{\dagger}\left(\widehat{\mathbf{g}}-\mathbf{g}^*\right) \overset{d.}{\longrightarrow} \chi^2_{r}$, where $r=\text{rank}(\matPsi) \ge MNQD-1$, and thus completes the proof. In practice, when we utilize this result to test the hypothesis of $\mathbf{g}^*=\mathbf{0}$, we will plug in the estimator of all parameters and compute the test statistics $T \cdot \widehat{\mathbf{g}}^\top \widehat{\mathbf{\Psi}}^{\dagger}\widehat{\mathbf{g}}$, and set the critical region based on $\chi^2_{MNQD-1}$.
\end{proof}

\section{Theory under High Spatial Dimension}\label{sec:grow-dim}
\subsection{Proof of Theorem~\ref{thm:high_dimensional_MARAC}}\label{subsec:proof_high_dimensional_MARAC}
\begin{proof}
In this proof, we will fix $P,Q$ as $1$ again for the ease of presentation, but the technical details can be generalized to arbitrary $P,Q$. Since we fix the lags to be $1$, we drop the subscript of the coefficients for convenience. 

Under the specification of the MARAC$(1,1)$ model, we restate the model as:
\begin{equation*}
    \vecx_t = \left(\vecx_{t-1}^{\top}\otimes\mathbf{I}_{S}\right)\vect{\mathbf{B}^*\otimes\mathbf{A}^*} + \left(\vecz_{t-1}^{\top}\otimes\matK\right)\vecgamma^* + \vece_t,
\end{equation*}
where $S=MN$ and we introduce the following additional notations:
\begin{equation*}
    \matY_T \coloneqq \begin{bmatrix}
        \vecx_1 \\ 
        \vdots \\
        \vecx_T
    \end{bmatrix}, \quad  \wmatX_T \coloneqq \begin{bmatrix}
        \vecx_0^{\top} \\
        \vdots \\
        \vecx_{T-1}^{\top}
    \end{bmatrix} \otimes\matI_S, \quad \wvecz_T \coloneqq \begin{bmatrix}
        \vecz_0^{\top} \\
        \vdots \\
        \vecz_{T-1}^{\top}
    \end{bmatrix}, \quad  \vecep_T = \begin{bmatrix}
        \vece_1 \\
        \vdots \\
        \vece_T
    \end{bmatrix}.
\end{equation*}
We will drop the subscript $T$ for convenience. Let $\vecphi^* = \vect{\mathbf{B}^*\otimes\mathbf{A}^*}$, and $g_1^*,\ldots,g_D^* \in\RKHS$ be the true autoregressive and functional parameters. Correspondingly, let $\vecgamma_1^*,\ldots,\vecgamma_D^*$ be the coefficients for the representers when evaluating $g_1^*,\ldots,g_D^*$ on a matrix grid, i.e., $\matK\vecgamma_d^*$ is a discrete evaluation of $g_d^*$ on the matrix grid. Let $\setF_{\vecphi} = \{\vect{\matB\otimes\matA}|\twonorm{\matA}=\text{sign}(\tr{\matA})=1,\matA\in\mathbb{R}^{M\times M}, \matB\in\mathbb{R}^{N\times N}\}$. Using these new notations, the MARAC estimator is obtained by solving the following penalized least squares problem:
\begin{equation}\label{eq:MARAC-PMLE-LS-Problem}
    \min_{\vecphi\in\setF_{\vecphi}, \vecgamma\in\mathbb{R}^{SD}} \mathfrak{L}_{\lambda}(\vecphi,\vecgamma) \coloneqq \left\{\frac{1}{2T} \|\matY - \wmatX\vecphi - \left(\wvecz\otimes\matK\right)\vecgamma\|_{\mathrm{F}}^{2} + \frac{\lambda}{2} \vecgamma^{\top}\left(\matI_D\otimes\matK\right)\vecgamma\right\}.
\end{equation}
By fixing $\vecphi$, the estimator for $\vecgamma$ is given by $\widehat{\vecgamma}(\vecphi) = \argmin_{\vecgamma}\mathfrak{L}_{\lambda}(\vecphi,\vecgamma)$, and can be explicitly written as:
\begin{equation}\label{eq:gamma_given_phi_solution}
    \widehat{\vecgamma}(\vecphi) = T^{-1} \left(\widehat{\bSigma}_{\vecz}\otimes\matK + \lambda\cdot\matI_{SD}\right)^{-1}\left(\wvecz^{\top}\otimes\matI_S\right)\left(\matY - \wmatX\vecphi\right). 
\end{equation}
Plugging~\eqref{eq:gamma_given_phi_solution} into~\eqref{eq:MARAC-PMLE-LS-Problem} yields the profile likelihood for $\vecphi$:
\begin{equation}\label{eq:profile-likelihood-phi}
    \ell_\lambda(\vecphi) = \mathfrak{L}_{\lambda}(\vecphi,\widehat{\vecgamma}(\vecphi)) = \frac{1}{2T}\left(\matY - \wmatX\vecphi\right)^{\top}\matW\left(\matY - \wmatX\vecphi\right),
\end{equation}
where $\matW$ is defined as:
\begin{equation}\label{eq:matrix-W-definition}
    \matW = \left\{\matI-\frac{\left(\wvecz\otimes\matK\right)\left[\widehat{\bSigma}_{\vecz}\otimes\matK + \lambda\cdot\matI_{SD}\right]^{-1}\left(\wvecz^{\top}\otimes\matI_S\right)}{T}\right\} = \left(\matI + \frac{\wvecz\wvecz^{\top}}{\lambda T}\otimes\matK\right)^{-1},
\end{equation}
and the second equality in~\eqref{eq:matrix-W-definition} is by the Woodbury matrix identity. It can be seen that $\matW$ is positive semi-definite and has all of its eigenvalues within $(0,1)$. To improve the clarity and organization of the proof, we break down the proof into several major steps. In the first step, we establish the following result on $\widehat{\vecphi}$:
\begin{proposition}\label{prop:AR-conv-rate}
    Under the assumptions of Theorem~\ref{thm:high_dimensional_MARAC}, we have:
    \begin{equation}\label{eq:phi-error-bound}
    \left(\widehat{\vecphi} - \vecphi^*\right)^{\top}\left(\frac{\wmatX^{\top}\matW\wmatX}{T}\right)\left(\widehat{\vecphi} - \vecphi^*\right) \lesssim  O_{P}(C_g\lambda) + O_{P}(c_{1,S}\cdot SD/T),
    \end{equation}
    where $C_g = \sum_{d=1}^{D}\|g_d^*\|_{\RKHS}^2$.
\end{proposition}
In order to derive the convergence rate of $\widehat{\vecphi}$, we still require one additional result:
\begin{lemma}\label{thm:RE-Condition}
    Under the assumptions of Theorem~\ref{thm:high_dimensional_MARAC} and the requirement that $S\log S/T\rightarrow 0$, it holds that:
    \begin{equation}
        \underaccent{\bar}{\rho}\left(\wmatX^{\top}\matW\wmatX/T\right) \ge \frac{c_{0,S}}{2} > 0,
    \end{equation}
    with probability approaching $1$ as $S,T\rightarrow\infty$, where $\mineigen{\cdot}$ is the minimum eigenvalue of a matrix and $c_{0,S} = \mineigen{\bSigma_{\vecx,\vecx}^*-\left(\bSigma_{\vecz,\vecx}^*\right)^{\top}\left(\bSigma_{\vecz,\vecz}^*\right)^{-1}\bSigma_{\vecz,\vecx}^*}$.
\end{lemma}
The proof of Proposition~\ref{prop:AR-conv-rate} and Lemma~\ref{thm:RE-Condition} are relegated to Section~\ref{proof:AR-conv-rate} and~\ref{subsec:proof-RE-Condition}, respectively. Combining Proposition~\ref{prop:AR-conv-rate} and Lemma~\ref{thm:RE-Condition}, we can derive the error bound of $\widehat{\vecphi}$ as:
\begin{equation}\label{eq:AR-conv-rate}
    \frac1S\twonorm{\widehat{\vecphi}-\vecphi^*} \lesssim O_P(\sqrt{\frac{C_g\gamma_S}{c_{0,S}S}}) + O_P(\sqrt{\frac{c_{1,S}D}{c_{0,S}TS}}).
\end{equation}

Now with this error bound of the autoregressive parameter $\widehat{\vecphi}$, we further derive the prediction error bound for the functional parameters. To start with, we have:
\begin{align*}
    \frac{1}{\sqrt{TS}} \twonorm{(\wvecz\otimes\matK)(\widehat{\vecgamma}-\vecgamma^*)} & = \frac{1}{\sqrt{TS}}\left\|(\matI - \matW)(\matY-\wmatX\widehat{\vecphi}) - (\wvecz\otimes\matK)\vecgamma^* \right\|_{\mathrm{F}} \notag \\
    & \le \frac{1}{\sqrt{TS}} \bigg[\underbrace{\twonorm{(\matI-\matW)\vecep}}_{\text{\makebox[0pt]{$J_1$}}} + \underbrace{\twonorm{(\matI-\matW)\wmatX(\widehat{\vecphi}-\vecphi^*)}}_{\text{\makebox[0pt]{$J_2$}}} \\
    & + \underbrace{\twonorm{\matW(\wvecz\otimes\matK)\vecgamma^*}}_{\text{\makebox[0pt]{$J_3$}}}\bigg], \notag
\end{align*}
and we will bound the terms $J_1,J_2,J_3$ separately.

To bound $J_1$, we first establish two lemmas.
\begin{lemma}\label{lemma:order-I-W}
    Given the definition of $\matW$ in~\eqref{eq:matrix-W-definition} and under the assumptions of Theorem~\ref{thm:high_dimensional_MARAC}, we have $O_P(\gamma_S^{-1/2r_0})\le \tr{\matI-\matW} \le O_P(\sqrt{S}\gamma_S^{-1/2r_0})$, 
    where $\gamma_S=\lambda/S$. Furthermore, we have $\tr{\matW} \le SD$.
\end{lemma}
\begin{lemma}\label{thm:boundedness-gaussian-quadratic-form}
Given the definition of $\matW$ in~\eqref{eq:matrix-W-definition} and under the assumptions of Theorem~\ref{thm:high_dimensional_MARAC}, we have that: 
\begin{equation*}
    \vecep^\top\matW\vecep/\tr{\matW} = O_P(c_{1,S}),
\end{equation*}
where $c_{1,S} = \specnorm{\bSigma}$. Furthermore, we have $\vecep^\top\left(\matI-\matW\right)^2\vecep/\tr{\left(\matI-\matW\right)^2} = O_P(c_{1,S})$.
\end{lemma}
We leave the proof of Lemma~\ref{lemma:order-I-W} and Lemma~\ref{thm:boundedness-gaussian-quadratic-form} to Section~\ref{subsec:proof-order-I-W} and~\ref{subsec:proof-boundedness-gaussian-quadratic-form}. By Lemma~\ref{thm:boundedness-gaussian-quadratic-form}, we have:
\begin{equation*}
    J_1^2 \asymp c_{1,S}\cdot\tr{(\matI-\matW)^2} \lesssim c_{1,S}\cdot\tr{\matI-\matW}.
\end{equation*}
And by Lemma~\ref{lemma:order-I-W}, we have $J_1 \le O_P(c_{1,S}^{1/2}\cdot S^{1/4}\gamma_{S}^{-1/4r_0})$.

For $J_2$, we have the following bound:
\begin{align}
    J_2 & = \twonorm{(\matI-\matW)\matW^{-1/2}\matW^{1/2}\wmatX(\widehat{\vecphi}-\vecphi^*)} \\
    & \le \specnorm{(\matI-\matW)\matW^{-1/2}}\cdot\twonorm{\matW^{1/2}\wmatX(\widehat{\vecphi}-\vecphi^*)} \notag \\
    & \le \specnorm{\matW^{-1/2}}\cdot\twonorm{\matW^{1/2}\wmatX(\widehat{\vecphi}-\vecphi^*)}. \label{eq:J2-intermediate-bound} 
\end{align}
To bound $\specnorm{\matW^{-1/2}}$, we can take advantage of the simpler form of $\matW$ using the Woodbury matrix identity in~\eqref{eq:matrix-W-definition} and obtain:
\begin{align*}
    \specnorm{\matW^{-1/2}} & = \maxeigen{\matW^{-1}}^{\frac12} = \bar{\rho}\left(\matI + (\lambda T)^{-1}\wvecz\wvecz^{\top}\otimes\matK\right)^{\frac12} \\
    & \le \left[1 + \lambda^{-1}\maxeigen{\matK}\maxeigen{T^{-1}\wvecz\wvecz^{\top}}\right]^{\frac12} \le \left[1 + \lambda^{-1}\maxeigen{\matK}\tr{\widehat{\bSigma}_{\vecz}}\right]^{\frac12}.
\end{align*}
In Lemma~\ref{lemma:covariance-consistency}, which we state later in Section~\ref{subsec:covariance-consistency}, we have shown that for $N$-dimensional stationary vector autoregressive process, the covariance estimator is consistent in the spectral norm as long as $N\log N/T\rightarrow 0$. Therefore, since $\{\vecz_t\}_{t=1}^{T}$ follows a stationary VAR$(\Qtilde)$ process and its dimensionality $D$ is fixed, we have $\|\widehat{\bSigma}_{\vecz} - \bSigma_{\vecz}^*\|_{s} \overset{p.}{\rightarrow} 0$ and thus with probability approaching $1$, we have $\mathrm{tr}(\widehat{\bSigma}_{\vecz}) \le 2\mathrm{tr}(\bSigma_{\vecz}^*)$. Therefore, we have $\specnorm{\matW^{-1/2}} \le O_P(\sqrt{1+c_0/\lambda})$, where $c_0$ is a constant related to $\tr{\bSigma_{\vecz}^*}$ and $\maxeigen{\matK}$. Combining this with the result in Proposition~\ref{prop:AR-conv-rate}, we can bound $J_2$ via its upper bound~\eqref{eq:J2-intermediate-bound} as:
\begin{equation}\label{eq:J2-bound}
    J_2 \le O_P\left(\sqrt{C_g\lambda T}\right) + O_P\left(\sqrt{C_g T}\right) + O_P(\sqrt{c_{1,S}S}) + O_P\left(\sqrt{c_{1,S}\gamma_S^{-1}}\right).
\end{equation}
Finally, for $J_3$, we first notice that: 
\begin{equation*}
    J_3 = \twonorm{\matW(\wvecz\otimes\matK)\vecgamma^*} \le \|\matW^{1/2}\|_{s}\cdot\twonorm{\matW^{1/2}(\wvecz\otimes\matK)\vecgamma^*} \le \twonorm{\matW^{1/2}(\wvecz\otimes\matK)\vecgamma^*}.
\end{equation*}
The upper bound of $J_3$ above can be further bounded by:
\begin{align}
    \twonorm{\matW^{1/2}(\wvecz\otimes\matK)\vecgamma^*}^2 & = (\lambda T) [(\matI_D\otimes\matK)\vecgamma^*]^{\top}\left\{\matI_{SD} - \left(\lambda^{-1}\widehat{\bSigma}_{\vecz}\otimes\matK+\matI_{SD}\right)^{-1}\right\}\vecgamma^* \notag \\
    & = (\lambda T)\left(\sum_{d=1}^{D} \|g_{d}^{*}\|_{\RKHS}^2\right) \notag \\
    & - (\lambda^2 T)\left(\vecgamma^*\right)^{\top}\left[\left(\matI_D\otimes\matK\right)\left(\widehat{\bSigma}_{\vecz}\otimes\matK+\lambda\matI_{SD}\right)^{-1}\right]\vecgamma^* \notag \\
    & \le C_{g}\lambda T, \label{eq:I1-bound}
\end{align}
where $C_{g} = \sum_{d=1}^{D} \|g_{d}^{*}\|_{\RKHS}^2$ is the norm of all the underlying functional parameters. The last inequality of~\eqref{eq:I1-bound} follows from the fact that the quadratic form led by $\lambda^2 T$ is non-negative. To see why, first note that:
\begin{equation*}    \left(\matI_D\otimes\matK\right)\left(\widehat{\bSigma}_{\vecz}\otimes\matK+\lambda\matI_{SD}\right)^{-1} = \left(\widehat{\bSigma}_{\vecz}\otimes\matI_S\right)^{-1} - \left[\widehat{\bSigma}_{\vecz}\otimes\matI_S + \lambda^{-1} \widehat{\bSigma}^2_{\vecz}\otimes\matK\right]^{-1}.
\end{equation*}
Then, we have the following lemma:
\begin{lemma}\label{thm:psd-matrix-lemma}
    If $\matA,\matB$ are symmetric, positive definite real matrices and $\matA-\matB$ is positive semi-definite, then $\matB^{-1}-\matA^{-1}$ is also positive semi-definite.
\end{lemma}
We leave the proof to Section~\ref{proof:psd-matrix-lemma}. Let $\mathbf{M}=\widehat{\bSigma}_{\vecz}\otimes\matI_S+\lambda^{-1} \widehat{\bSigma}^2_{\vecz}\otimes\matK$ and $\mathbf{N}=\widehat{\bSigma}_{\vecz}\otimes\matI_S$, then both $\mathbf{M}$ and $\mathbf{N}$ are positive definite and $\mathbf{M}-\mathbf{N}$ is positive semi-definite. By Lemma~\ref{thm:psd-matrix-lemma}, we have $\mathbf{N}^{-1}-\mathbf{M}^{-1}$ being positive semi-definite and thus~\eqref{eq:I1-bound} holds.

Using the result in~\eqref{eq:I1-bound}, we eventually have $J_3 \le O_P(\sqrt{C_g\lambda T})$. Combining all the bounds for $J_1,J_2,J_3$, we end up with:
\begin{align*}
    \frac{1}{\sqrt{TS}} \twonorm{(\wvecz\otimes\matK)(\widehat{\vecgamma}-\vecgamma^*)}  & \le O_P\left(\frac{\sqrt{c_{1,S}}\sqrt{\gamma_S^{-1/{2r_0}}}}{\sqrt{T}\sqrt[4]{S}}\right) + O_P(\sqrt{\gamma_S}) \\
    & + O_P\left(\frac{1}{\sqrt{S}}\right) + O_P\left(\sqrt{\frac{c_{1,S}}{T}}\right) +  O_P\left(\frac{\sqrt{c_{1,S}\gamma_S^{-1}}}{\sqrt{TS}}\right).
\end{align*}
\end{proof}

\subsection{Proof of Proposition~\ref{prop:AR-conv-rate}}\label{proof:AR-conv-rate}
\begin{proof}
    The MARAC estimator $\widehat{\vecphi}$ is the minimizer of $\ell_\lambda(\vecphi)$, defined in~\eqref{eq:profile-likelihood-phi}, for all $\vecphi\in\setF_\phi$ and thus $\ell_\lambda(\widehat{\vecphi}) \le \ell_{\lambda}(\vecphi^*)$. Equivalently, this means that:
\begin{equation*}
    \frac{1}{2}\left(\widehat{\vecphi} - \vecphi^*\right)^{\top}\left(\frac{\wmatX^{\top}\matW\wmatX}{T}\right)\left(\widehat{\vecphi} - \vecphi^*\right) \notag \le \frac1T\left[\left(\wvecz\otimes\matK\right)\vecgamma^*+\vecep\right]^{\top}\matW\wmatX\left(\widehat{\vecphi} - \vecphi^*\right). 
\end{equation*}
Let $\boldsymbol{\delta} = \matW^{1/2}\wmatX(\widehat{\vecphi}-\vecphi^*)/\sqrt{T}$ and $\boldsymbol{\omega}=\matW^{1/2}\left[\left(\wvecz\otimes\matK\right)\vecgamma^*+\vecep\right]/\sqrt{T}$, then the inequality can be simply written as $\boldsymbol{\delta}^{\top}\boldsymbol{\delta} \le 2\boldsymbol{\delta}^{\top}\boldsymbol{\omega}$, and we can upper bound our quantity of interest, namely $\boldsymbol{\delta}^{\top}\boldsymbol{\delta}$, as:
\begin{equation*}
    \boldsymbol{\delta}^{\top}\boldsymbol{\delta} \le 2(\boldsymbol{\delta}-\boldsymbol{\omega})^{\top}(\boldsymbol{\delta}-\boldsymbol{\omega}) + 2\boldsymbol{\omega}^{\top}\boldsymbol{\omega} \le 4\boldsymbol{\omega}^{\top}\boldsymbol{\omega}.
\end{equation*}
Therefore, the bound of $\twonorm{\boldsymbol{\delta}}^2$ can be obtained via the bound of $\twonorm{\boldsymbol{\omega}}^2$. We have the following upper bound for $\twonorm{\boldsymbol{\omega}}^2$:
\begin{align}
    \twonorm{\boldsymbol{\delta}}^2 \le 4\twonorm{\boldsymbol{\omega}}^2 & = \frac4T \left[\left(\wvecz\otimes\matK\right)\vecgamma^*+\vecep\right]^{\top}\matW\left[\left(\wvecz\otimes\matK\right)\vecgamma^*+\vecep\right] \notag\\
    & \le \frac8T \left[\underbrace{\twonorm{\matW^{1/2}\left(\wvecz\otimes\matK\right)\vecgamma^*}^2}_{\text{\makebox[0pt]{$I_1$}}} + \underbrace{\twonorm{\matW^{1/2}\vecep}^2}_{\text{\makebox[0pt]{$I_2$}}}\right], \label{eq:delta-norm-bound}
\end{align}
where the last inequality follows from the fact that $\matW$ is positive semi-definite.

For $I_1$, it can be bounded by~\eqref{eq:I1-bound} and thus $I_1\le C_{g}\lambda T$. To bound $I_2$, we utilize Lemma~\ref{thm:boundedness-gaussian-quadratic-form} and bound $I_2$ as $I_2 \asymp c_{1,S}\cdot\tr{\matW} \le c_{1,S}\cdot SD$. Combining the bounds for $I_1$ and $I_2$, we have:
\begin{equation*}
    \twonorm{\boldsymbol{\delta}}^2 =\left(\widehat{\vecphi} - \vecphi^*\right)^{\top}\left(\frac{\wmatX^{\top}\matW\wmatX}{T}\right)\left(\widehat{\vecphi} - \vecphi^*\right) \lesssim  O_{P}(C_g\lambda) + O_{P}(c_{1,S}\cdot SD/T),
\end{equation*}
which completes the proof.
\end{proof}

\section{Technical Lemmas \& Proofs}\label{app:proof-lemmas}
In this section, we first introduce Lemma~\ref{lemma:covariance-consistency} on the consistency of the covariance matrix estimator for any stationary vector autoregressive process and then Corollary~\ref{corollary:MARAC-covariance-consistency} on the consistency of the covariance estimator of our MARAC model, given the joint stationarity condition. Then we provide proof for Lemma~\ref{lemma:convergence_rate} used in Section~\ref{subsec:proof_asymptotic_normal} when proving Theorem~\ref{thm:PMLE-Asymp-Normal} on the asymptotic normality under fixed spatial dimension. Then we provide proofs for Lemma~\ref{thm:RE-Condition},~\ref{lemma:order-I-W},~\ref{thm:boundedness-gaussian-quadratic-form} and~\ref{thm:psd-matrix-lemma} used in Section~\ref{sec:grow-dim} when proving the error bounds with high spatial dimensionality.

\subsection{Statement of Lemma~\ref{lemma:covariance-consistency}}\label{subsec:covariance-consistency}
In Lemma~\ref{lemma:covariance-consistency}, we restate the result of Propositions 6 and 7 of~\citet{li2021multi}, which covers the general result of the consistency of the estimator for the lag-$0$ auto-covariance matrix of a stationary VAR$(p)$ process.
\begin{lemma}\label{lemma:covariance-consistency}
    Let $\vecx_t\in\mathbb{R}^{N}$ be a zero-meaned stationary VAR$(p)$ process: $\vecx_t = \sum_{l=1}^{p}\matPhi_p\vecx_{t-p}+\vecxi_{t}$, where $\vecxi_t$ have independent sub-Gaussian entries. Let $\widehat{\bSigma} = (1/T)\sum_{t=1}^{T} \vecx_t\vecx_t^{\top}$ and $\bSigma = \mathrm{E}[\widehat{\bSigma}]$, then we have:
    \begin{equation}\label{eq:tail-bound-covariance}
        \mathrm{E}\|\widehat{\bSigma}-\bSigma\|_{s} \le C\left(\sqrt{\frac{N\log N}{T}} + \frac{N\log N}{T}\right)\|\bSigma\|_{s},
    \end{equation}
    where $C$ is an absolute constant.
\end{lemma}
We refer our readers to Appendix C.3 of~\citet{li2021multi} for the proof. As a corollary of Lemma~\ref{lemma:covariance-consistency}, we have the following results:
\begin{corollary}\label{corollary:MARAC-covariance-consistency}
Assume that $\{\vecz_t\}_{t=1}^{T}$ is generated by a stationary VAR$(\Qtilde)$ process: $\vecz_t=\sum_{\qtilde=1}^{\Qtilde} \matC_{\qtilde}\vecz_{t-\qtilde} + \vecnu_t$, with $\vecnu_t$ having independent sub-Gaussian entries, then with $\widehat{\bSigma}_{\vecz} = (1/T)\sum_{t=1}^{T}\vecz_t\vecz_t^\top$ and $\bSigma_{\vecz}^*=\mathrm{E}[\widehat{\bSigma}_{\vecz}]$, we have:
\begin{equation}\label{eq:consistency-covariance-z}
    \mathrm{P}\left(\left\|\widehat{\bSigma}_{\vecz} - \bSigma_{\vecz}^*\right\|_{s}\ge \epsilon\right) \le C\epsilon^{-1}\left(\sqrt{\frac{D}{T}}+\frac{D}{T}\right),
\end{equation}
with $C$ being an absolute constant and $\epsilon$ being a fixed positive real number, and thus $\left\|\widehat{\bSigma}_{\vecz} - \bSigma_{\vecz}^*\right\|_{s}\overset{p.}{\rightarrow} 0$.

Let $\{\matX_t\}_{t=1}^{T}$ be a zero-meaned matrix time series generated by the MARAC model with lag $P,Q$ and $\{\vecz_t\}_{t=1}^{T}$ satisfies the assumption above and $\{\matX_t,\vecz_t\}_{t=1}^{T}$ are jointly stationary in the sense of Theorem~\ref{thm:Joint-Stationarity}. Assume further that $\matE_t$ has i.i.d. Gaussian entries with constant variance $\sigma^2$, then for $\vecy_t = [\vecx_t^{\top},\vecz_t^{\top}]^{\top}$, $\widehat{\bSigma}_0 = (1/T)\sum_{t=1}^{T}\vecy_t\vecy_t^\top$ and $\bSigma_0^*=\mathrm{E}[\vecy_t\vecy_t^\top]$, we have:
\begin{equation}\label{eq:consistency-covariance-MARAC}
    \mathrm{E}\left\|\widehat{\bSigma}_0 - \bSigma_0^*\right\|_{s} \le C\left(\sqrt{\frac{S\log S}{T}}+\frac{S\log S}{T}\right)\|\bSigma_0^*\|_{s},
\end{equation}
where $C$ is an absolute constant.
\end{corollary}
\begin{proof}
The proof of~\eqref{eq:consistency-covariance-z} is straightforward from Lemma~\ref{lemma:covariance-consistency} together with Markov inequality. The proof of~\eqref{eq:consistency-covariance-MARAC} also follows from Lemma~\ref{lemma:covariance-consistency} since $\{\vecy_t\}_{t=1}^T$ follows a stationary VAR$(\text{max}(P,Q,\Qtilde))$ process with i.i.d. sub-Gaussian noise (see~\eqref{eq:matrix-vector-joint-VAR-replicate}) and $\mathrm{E}[(1/T)\sum_{t=1}^{T}\vecy_t\vecy_t^\top] = \mathrm{E}[\vecy_t\vecy_t^\top]$ due to stationarity.
\end{proof}
Note that the convergence of the variance estimator in spectral norm also indicates that each element of the variance estimator converges in probability. Also, the assumption that $\matE_t$ has i.i.d. Gaussian entries can be relaxed to $\matE_t$ having independent sub-Gaussian entries.

\subsection{Proof of Lemma~\ref{lemma:convergence_rate}}\label{subsec:proof-fixdim-conv-rate}
\begin{proof}
    Without loss of generality, we fix $P,Q$ as 1 and use the same notation as~\eqref{eq:MARAC_compact_form} in Section~\ref{subsec:proof-error-cov-consistency}, so the MARAC model can be written as $\vecx_t = \vecy_t\vectheta^* + \vece_t$. Correspondingly, the penalized log-likelihood $h(\vectheta,\matOmega)$ is specified by~\eqref{eq:MARAC_PMLE_compact_form} and given any $\bar{\matOmega}$, we have $\widetilde{\vectheta}(\bar{\matOmega}) = \argmin_{\vectheta} h(\vectheta,\bar{\matOmega})$ as specified by~\eqref{eq:unconstrained_GLS_solution}. 
    Given the decomposition of $\widetilde{\vectheta}(\bar{\matOmega})$ in \eqref{eq:theta-tilde-transformation}, we have:
    \begin{equation*}
        \widetilde{\vectheta}(\bar{\matOmega}) - \vectheta^* = -\lambda\widetilde{\matK}\vectheta^* + \left(\frac{\sum_{t}\vecy_t^{\top}\bar{\matOmega}\vecy_t}{T} + \lambda\widetilde{\matK}\right)^{-1}\left(\frac{\sum_{t}\vecy_t^{\top}\bar{\matOmega}\vece_t}{T}\right),
    \end{equation*}
    where $\twonorm{\lambda\widetilde{\matK}\vectheta^*}=o(T^{-1/2})$ since $\lambda = o(T^{-1/2})$ and the norm of the second term is $O_P(T^{-1/2})$. To show that the norm of the second term is $O_P(T^{-1/2})$, we first observe that:
    \begin{align*}
        & \left\|\left(\frac{\sum_{t}\vecy_t^{\top}\bar{\matOmega}\vecy_t}{T} + \lambda\widetilde{\matK}\right)^{-1}\left(\frac{\sum_{t}\vecy_t^{\top}\bar{\matOmega}\vece_t}{T}\right)\right\|_{\mathrm{F}} \\
        & \le \left\|\underbrace{\left(\frac{\sum_{t}\vecy_t^{\top}\bar{\matOmega}\vecy_t}{T} + \lambda\widetilde{\matK}\right)^{-1}}_{\text{\makebox[0pt]{$\matL_T^{-1}$}}}\right\|_{\mathrm{F}}\cdot\left\|\underbrace{\left(\frac{\sum_{t}\vecy_t^{\top}\bar{\matOmega}\vece_t}{T}\right)}_{\text{\makebox[0pt]{$\matR_T$}}}\right\|_{\mathrm{F}}.
    \end{align*}
    For the sequence of random matrices $\{\matL_T\}_{T=1}^{\infty}$, we have:
    \begin{equation*}
        \matL_T = \frac{\sum_{t}\vecy_t^{\top}\bar{\matOmega}\vecy_t}{T} + \lambda\widetilde{\matK} \overset{p.}{\rightarrow} \begin{bmatrix}
            \mathrm{Cov}(\vecx_t,\vecx_t)\otimes\bar{\matOmega} & \mathrm{Cov}(\vecx_t,\vecz_t)\otimes\bar{\matOmega}\matK \\
            \mathrm{Cov}(\vecz_t,\vecx_t)\otimes\matK\bar{\matOmega} & \mathrm{Cov}(\vecz_t,\vecz_t)\otimes\matK\bar{\matOmega}\matK
        \end{bmatrix},
    \end{equation*}
    and we define the limiting matrix as $\matL$. To show this, first note that the covariance estimator $\widehat{\mathrm{Var}}([\vecx_t^\top,\vecz_t^\top]^\top) = T^{-1}\sum_{t} [\vecx_t^\top,\vecz_t^\top]^\top[\vecx_t^\top,\vecz_t^\top]$ converges in probability to the true covariance $\mathrm{Var}([\vecx_t^\top,\vecz_t^\top]^\top)$, which we prove separately in Corollary~\ref{corollary:MARAC-covariance-consistency}. Secondly, notice that $\lambda=o(T^{-1/2})$, thus we have $\lambda\widetilde{\matK}\rightarrow \matO$ and thus we have the convergence in probability of $\matL_T$ to $\matL$ holds.

    Notice that the limiting matrix $\matL$ is invertible because the matrix $\matL^{\prime}$, defined as:
    \begin{equation*}
        \matL^{\prime} = \begin{bmatrix}
            \matI\otimes\matK & \matO \\
            \matO & \matI
        \end{bmatrix}\matL \begin{bmatrix}
            \matI\otimes\matK & \matO \\
            \matO & \matI
        \end{bmatrix} = \mathrm{Var}([\vecx_t^\top,\vecz_t^\top]^\top)\otimes (\matK\bar{\matOmega}\matK),
    \end{equation*}
    is invertible. To see why, firstly note that $\mathrm{Var}([\vecx_t^\top,\vecz_t^\top]^\top)$ is invertible because we can express $[\vecx_t^\top,\vecz_t^\top]^\top$ as $\sum_{j=0}^{\infty}\matPhi_j[\vece_t^\top,\vecnu_t^\top]^\top$, where $\{\matPhi_j\}_{j=0}^{\infty}$ is a sequence of matrices whose elements are absolutely summable and $\matPhi_0=\matI$, therefore, we have $\mineigen{\mathrm{Var}([\vecx_t^\top,\vecz_t^\top]^\top)}\ge \mineigen{\mathrm{Var}([\vece_t^\top,\vecnu_t^\top]^\top)} > 0$. Secondly, by Assumption~\ref{assump:Gq-Gammaq}, we have $\mineigen{\matK} > 0$ and we also have $\mineigen{\bar{\matOmega}} > 0$ by definition, therefore we have $\matK\bar{\matOmega}\matK$ to be positive definite. The invertibility of $\matL$ and the fact that $\matL_T\overset{p.}{\rightarrow}\matL$ indicates that $\matL_T^{-1}\overset{p.}{\rightarrow}\matL^{-1}$, since matrix inversion is a continuous function of the input matrix and the convergence in probability carries over under continuous transformations. Eventually, this leads to the conclusion that $\|\matL_T^{-1}\|_{\mathrm{F}} = O_P(1)$.

    For the sequence of random matrices $\{\matR_T\}_{T=1}^{\infty}$, we note that the sequence $\{\vecy_t^{\top}\bar{\matOmega}\vece_t\}_{t=1}^{\infty}$ is a martingale difference sequence (MDS) such that $\twonorm{\matR_T} = O_P(T^{-1/2})$ (see proposition 7.9 of~\citet{hamilton2020time} for the central limit theorem of martingale difference sequence). Combining the result of $\twonorm{\matL_T}$ and $\twonorm{\matR_T}$, we conclude that $\twonorm{\widetilde{\vectheta}(\bar{\matOmega}) - \vectheta^*} = O_P(T^{-1/2})$.
    
    Fix $\matOmega = \bar{\matOmega}$, we can decompose $h(\vectheta,\bar{\matOmega})$ via the second-order Taylor expansion as follows:
    \begin{align}
        h(\vectheta,\bar{\matOmega}) & = h(\widetilde{\vectheta}(\bar{\matOmega}),\bar{\matOmega}) + \frac12 (\vectheta - \widetilde{\vectheta}(\bar{\matOmega}))^{\top}\left(\frac{\sum_t \vecy_t^{\top}\bar{\matOmega}\vecy_t}{T} + \lambda\widetilde{\matK}\right)(\vectheta - \widetilde{\vectheta}(\bar{\matOmega})) \nonumber \\ 
        & \ge h(\widetilde{\vectheta}(\bar{\matOmega}),\bar{\matOmega}) + \frac{1}{2} \mineigen{\matL_T}\twonorm{\vectheta-\widetilde{\vectheta}(\bar{\matOmega})}^2, \label{eq:lower_bound_PMLE_function}
    \end{align}
and recall that $\matL_T=T^{-1}\sum_t \vecy_t^\top\bar{\matOmega}\vecy_t+\lambda \widetilde{\matK}$. In the previous proof, we've shown that $\matL_T\overset{p.}{\rightarrow}\matL$, with $\matL$ being a positive definite matrix. Therefore, with probability approaching $1$, we have $\mineigen{\matL_T} \ge \mineigen{\matL}/2 > 0$.

With the lower bound on $\mineigen{\matL_T}$, we can claim that for some constant $C_1 > 0$:
\begin{align}
    & \inf_{\bar{\matOmega}\in\setF_{\matOmega}: \twonorm{\bar{\matOmega} - \matOmega^*}\le C_1}h(\vectheta,\bar{\matOmega}) \notag \\
    & \ge \inf_{\bar{\matOmega}\in\setF_{\matOmega}: \twonorm{\bar{\matOmega} - \matOmega^*}\le C_1} \left\{h(\widetilde{\vectheta}(\bar{\matOmega}),\bar{\matOmega}) + \frac14\mineigen{\matL}\cdot\twonorm{\vectheta-\widetilde{\vectheta}(\bar{\matOmega})}^2\right\},\label{eq:lower_bound_PMLE_function2}
\end{align}
with probability approaching $1$. Now consider $\vectheta$ belongs to the set $\{\vectheta\in\setF_{\vectheta}|\sqrt{T}\twonorm{\vectheta - \vectheta^*} \ge c_T\}$, where $c_T\rightarrow\infty$ is an arbitrary sequence that diverges to infinity. Within this set, we have:
\begin{equation}
    \twonorm{\vectheta - \widetilde{\vectheta}(\bar{\matOmega})} \ge \frac{c_T}{\sqrt{T}} - \twonorm{\vectheta^* - \widetilde{\vectheta}(\bar{\matOmega})},
\end{equation}
thus $\twonorm{\vectheta - \widetilde{\vectheta}(\bar{\matOmega})} \gtrsim O_P(c^{\prime}_T/\sqrt{T})$ for some sequence $c^{\prime}_T\rightarrow\infty$ since $\twonorm{\widetilde{\vectheta}(\bar{\matOmega}) - \vectheta^*} = O_P(T^{-1/2})$. By the Taylor expansion in~\eqref{eq:lower_bound_PMLE_function}, we can conclude that $h(\vectheta^*,\bar{\matOmega})=h(\widetilde{\vectheta}(\bar{\matOmega}),\bar{\matOmega}) + O_P(T^{-1})$, also using that $\twonorm{\widetilde{\vectheta}(\bar{\matOmega}) - \vectheta^*} = O_P(T^{-1/2})$. Combining this result together with the order of $\twonorm{\vectheta - \widetilde{\vectheta}(\bar{\matOmega})}$, we have the following hold according to~\eqref{eq:lower_bound_PMLE_function2}:
\begin{equation}\label{eq:lower_bound_PMLE_function3}
    \mathrm{P}\left(\inf_{\sqrt{T}\twonorm{\vectheta-\vectheta^*}\ge c_T}\inf_{\bar\matOmega\in\setF_{\matOmega}:\twonorm{\bar{\matOmega} - \matOmega^*}\le C_1}h(\vectheta,\bar{\matOmega}) > \inf_{\bar\matOmega\in\setF_{\matOmega}:\twonorm{\bar{\matOmega} - \matOmega^*}\le C_1}h(\vectheta^*,\bar{\matOmega})\right)\rightarrow 1.
\end{equation}
The result in~\eqref{eq:lower_bound_PMLE_function3} indicates that for any $\vectheta$ that lies outside of the set $\{\vectheta\in\setF_{\vectheta}|\sqrt{T}\twonorm{\vectheta-\vectheta^*} < c_T\}$, the penalized log-likelihood is no smaller than a sub-optimal solution with probability approaching $1$. Therefore, with probability approaching $1$, one must have $\sqrt{T}\twonorm{\vectheta-\vectheta^*}\le c_T$. And since the choice of $c_T$ is arbitrary, we can conclude that $\twonorm{\widehat{\vectheta} - \vectheta^*} = O_P(T^{-1/2})$ and thus each block of $\widehat{\vectheta}$, namely $\widehat{\matA}_p,\widehat{\matB}_p,\widehat{\vecgamma}_q$ converges to their ground truth value at the rate of $T^{-1/2}$.

The convergence rate of $\widehat{\matB}_p\otimes\widehat{\matA_p}$ can be derived from the following inequality:
\begin{equation*}
    \twonorm{\widehat{\matB}_p\otimes\widehat{\matA}_p - \matB_p^*\otimes\matA_p^*} \le \twonorm{\widehat{\matB}_p}\cdot\twonorm{\widehat{\matA}_p-\matA_p^*} + \twonorm{\widehat{\matB}_p-\matB_p^*}\cdot\twonorm{\matA_p^*},
\end{equation*}
as well as the convergence rate of $\widehat{\matA}_p$ and $\widehat{\matB}_p$.

\end{proof}

\subsection{Proof of Lemma~\ref{thm:RE-Condition}}\label{subsec:proof-RE-Condition}
\begin{proof}
Based on the definition of $\matW$ in equation~\eqref{eq:matrix-W-definition}, we have
\begin{align}
    \frac{\wmatX^{\top}\matW\wmatX}{T} & = \widehat{\bSigma}_{\vecx,\vecx}\otimes\matI_S - \left(\widehat{\bSigma}_{\vecz,\vecx}^{\top}\otimes\matK\right)\left(\widehat{\bSigma}_{\vecz,\vecz}\otimes\matK+\lambda\matI_{SD}\right)^{-1}\left(\widehat{\bSigma}_{\vecz,\vecx}\otimes\matI_S\right) \notag \\
    & = \left(\widehat{\bSigma}_{\vecx,\vecx} - \widehat{\bSigma}_{\vecz,\vecx}^{\top}\widehat{\bSigma}_{\vecz,\vecz}^{-1}\widehat{\bSigma}_{\vecz,\vecx}\right)\otimes\matI_S \notag \\
    & + \left(\widehat{\bSigma}_{\vecz,\vecx}\otimes\matI_S\right)^{\top}\left[\widehat{\bSigma}_{\vecz,\vecz}^2\otimes\lambda^{-1}\matK+\widehat{\bSigma}_{\vecz,\vecz}\otimes\matI_S\right]^{-1}\left(\widehat{\bSigma}_{\vecz,\vecx}\otimes\matI_S\right), \label{eq:RE-lower-bound}
\end{align}
where the second term in~\eqref{eq:RE-lower-bound} is positive semi-definite since both $\mineigen{\widehat{\bSigma}_{\vecz,\vecz}}$ and $\mineigen{\matK}$ are non-negative and the whole term is symmetric. Therefore, by Weyl's inequality, one can lower bound $\mineigen{\wmatX^{\top}\matW\wmatX/T}$ by $\mineigen{\widehat{\bSigma}_{\vecx,\vecx} - \widehat{\bSigma}_{\vecz,\vecx}^{\top}\widehat{\bSigma}_{\vecz,\vecz}^{-1}\widehat{\bSigma}_{\vecz,\vecx}}$. For simplicity, we will use $\matA,\matB,\matC$ to denote $\bSigma_{\vecx,\vecx}^*,\bSigma_{\vecz,\vecx}^*,(\bSigma_{\vecz,\vecz}^*)^{-1}$, and $\hmatA,\hmatB,\hmatC$ to denote $\widehat{\bSigma}_{\vecx,\vecx},\widehat{\bSigma}_{\vecz,\vecx},\widehat{\bSigma}_{\vecz,\vecz}^{-1}$, respectively. We will use $\widehat{\bSigma}$ and $\bSigma^*$ to denote the estimated and true covariance matrix of $[\vecx_t^{\top}, \vecz_t^{\top}]^{\top}$. It is evident that $\|\matA\|_s\le \|\bSigma^*\|_s$ and $\|\matB\|_{s}\le \|\bSigma^*\|_s$, since both $\matA$ and $\matB$ are blocks of $\bSigma^*$ and can thus be represented as $\matE_1^\top\bSigma^*\matE_2$ with $\matE_1,\matE_2$ being two block matrices with unity spectral norm.

The rest of the proof focuses on showing that with $S\log S/T\rightarrow 0$, $\mineigen{\widehat{\bSigma}_{\vecx,\vecx} - \widehat{\bSigma}_{\vecz,\vecx}^{\top}\widehat{\bSigma}_{\vecz,\vecz}^{-1}\widehat{\bSigma}_{\vecz,\vecx}}\overset{p.}{\rightarrow} \mineigen{\bSigma_{\vecx,\vecx}^* - \left(\bSigma_{\vecz,\vecx}^*\right)^{\top}\left(\bSigma_{\vecz,\vecz}^*\right)^{-1}\bSigma_{\vecz,\vecx}^*}$. For brevity, we omit the subscript $s$ for the spectral norm notation and simply use $\|\cdot\|$ in this proof.

To start with, we have:
\begin{align}
    & \|\hmatA-\hmatB^{\top}\hmatC\hmatB - (\matA-\matB^{\top}\matC\matB)\| \notag \\
    & \le \|\hmatA - \matA\| + \|\hmatB^{\top}\hmatC\hmatB - \matB^{\top}\hmatC\matB\| + \|\matB^{\top}\hmatC\matB - \matB^{\top}\matC\matB\| \notag\\
    & \le \|\widehat{\bSigma}-\bSigma\| + \|(\hmatB-\matB)^{\top}\hmatC\hmatB\| + \|\matB^{\top}\matC(\hmatB-\matB)\| + \|\matB^{\top}(\hmatC-\matC)\hmatB\| \notag\\
    & \le \|\widehat{\bSigma}-\bSigma\| + \|\hmatB-\matB\|\cdot\left(\|\hmatC\|\cdot\|\hmatB\| + \|\matC\|\cdot\|\matB\|\right) \notag\\
    & + \|\matB\|\cdot\|\hmatB\|\cdot\|\hmatC-\matC\|.\label{eq:min-eigen-bound1}
\end{align}
Based on Corollary~\ref{corollary:MARAC-covariance-consistency}, under the condition that $S\log S/T\rightarrow 0$ and the conditions that $\vecz_t$ follows a stationary VAR$(\Qtilde)$ process and is jointly stationary with $\vecx_t$, we have $\|\hmatC-\matC\|\overset{p.}{\rightarrow} 0$ and $\|\widehat{\bSigma}-\bSigma^*\|\overset{p.}{\rightarrow} 0$. Therefore, with probability approaching $1$, we have $\|\hmatC\|\le 2\|\matC\|$, $\|\hmatB-\matB\| \le \|\widehat{\bSigma}-\bSigma^*\| \le 2\|\bSigma^*\|$ and $\|\hmatB\|\le 3\|\bSigma^*\|$.

Combining these results and the upper bound in~\eqref{eq:min-eigen-bound1}, with probability approaching $1$, we have:
\begin{align}
    \|\hmatA-\hmatB^{\top}\hmatC\hmatB - (\matA-\matB^{\top}\matC\matB)\| & \le \left(1+7\|\matC\|\cdot\|\bSigma^*\|\right)\cdot\|\widehat{\bSigma}-\bSigma^*\| \notag \\
    & + 3\|\bSigma^*\|^2\cdot\|\hmatC-\matC\|.\label{eq:min-eigen-bound2}
\end{align}
The upper bound in~\eqref{eq:min-eigen-bound2} can be arbitrarily small as $S,T\rightarrow\infty$ since $\|\hmatC-\matC\|\overset{p.}{\rightarrow}0$ and $\|\widehat{\bSigma}-\bSigma^*\|\overset{p.}{\rightarrow}0$. 

Eventually, with probability approaching $1$, we have:
\begin{equation}\label{eq:min-eigen-bound3}
    \mineigen{\widehat{\bSigma}_{\vecx,\vecx} - \widehat{\bSigma}_{\vecz,\vecx}^{\top}\widehat{\bSigma}_{\vecz,\vecz}^{-1}\widehat{\bSigma}_{\vecz,\vecx}} \ge \frac 12 \ubar{\rho}\left(\bSigma_{\vecx,\vecx}^* - \left(\bSigma_{\vecz,\vecx}^*\right)^{\top}\left(\bSigma_{\vecz,\vecz}^*\right)^{-1}\bSigma_{\vecz,\vecx}^*\right) = \frac{c_{0,S}}{2}.
\end{equation}
This completes the proof.

\end{proof}

\subsection{Proof of Lemma~\ref{lemma:order-I-W}}\label{subsec:proof-order-I-W}
\begin{proof}
By the definition of $\matW$ in~\eqref{eq:matrix-W-definition}, we have:
\begin{align}
    \tr{\matI-\matW} & = \mathrm{tr}\left[\left(\widehat{\bSigma}_{\vecz}\otimes\matK + \lambda\matI_{SD}\right)^{-1}\left(\widehat{\bSigma}_{\vecz}\otimes\matK\right)\right] \notag \\
    & = \sum_{s=1}^{S}\sum_{d=1}^{D} \frac{\rho_d(\widehat{\bSigma}_{\vecz})\rho_s(\matK)}{\lambda + \rho_d(\widehat{\bSigma}_{\vecz})\rho_s(\matK)} \le D\cdot\sum_{s=1}^{S} \frac{1}{1+\lambda\maxeigen{\widehat{\bSigma}_{\vecz}}^{-1}\rho_s(\matK)^{-1}}. \label{eq:trace-I-W-upper-bound}
\end{align}
Using Lemma~\ref{lemma:covariance-consistency}, we can bound $\maxeigen{\widehat{\bSigma}_{\vecz}}$ by $2\maxeigen{\bSigma_{\vecz}^*}$ with probability approaching $1$ as $T\rightarrow\infty$. Conditioning on this high probability event and using the Assumption~\ref{assump:kernel-eigen-decay} that the kernel function is separable, the kernel Gram matrix $\matK$ can be written as $\matK_2\otimes \matK_1$ and thus~\eqref{eq:trace-I-W-upper-bound} can be bounded as:
\begin{equation}\label{eq:trace-I-W-upper-bound2}
    D\cdot\sum_{s=1}^{S} \frac{1}{1+\lambda\maxeigen{\widehat{\bSigma}_{\vecz}}^{-1}\rho_s(\matK)^{-1}} \le D\cdot \sum_{i=1}^{M}\sum_{j=1}^{N} \frac{1}{1+c_{\vecz}\lambda\rho_{i}(\matK_1)^{-1}\rho_{j}(\matK_2)^{-1}},
\end{equation}
where $c_{\vecz}=1/2\maxeigen{\bSigma_{\vecz}^*}$. As $M, N\rightarrow\infty$, based on Assumption~\ref{assump:location-uniform-sample}, we have $\rho_i(\matK_1)\rightarrow Mi^{-r_0}$ and $\rho_j(\matK_2)\rightarrow Nj^{-r_0}$. Consequently, we can find two constants $0 < c_1 < c_2$, with $c_1$ being sufficiently small and $c_2$ being sufficiently large, such that:
\begin{align}
    \sum_{i=1}^{M}\sum_{j=1}^{N} \frac{1}{1+c_2\lambda(ij)^{r_0}/S} & \le \sum_{i=1}^{M}\sum_{j=1}^{N} \frac{1}{1+c_{\vecz}\lambda\rho_{i}(\matK_1)^{-1}\rho_{j}(\matK_2)^{-1}} \notag \\
    & \le \sum_{i=1}^{M}\sum_{j=1}^{N} \frac{1}{1+c_1\lambda(ij)^{r_0}/S}, \label{eq:order-I-W-bound2}
\end{align}
where we, with a little abuse of notations, incorporate $c_{\vecz}$ into $c_1,c_2$. To estimate the order of the lower and upper bound in~\eqref{eq:order-I-W-bound2}, we first notice that for any constant $c > 0$, one has:
\begin{equation}\label{eq:sequence-sum}
    \sum_{i=1}^{M\wedge N}\frac{1}{1+c\lambda i^{2r_0}/S} \le \sum_{i=1}^{M}\sum_{j=1}^{N} \frac{1}{1 + c\lambda(ij)^{r_0}/S} \le 2(M\vee N)\sum_{i=1}^{M\vee N}\frac{1}{1+c\lambda i^{2r_0}/S}. 
\end{equation}
To approximate the sum in~\eqref{eq:sequence-sum}, notice that:
\begin{equation*}
    \sum_{i=1}^{M\vee N}\frac{1}{1+c\lambda i^{2r_0}/S} = (S/c\lambda)^{1/2r_0}\cdot\sum_{i=1}^{M\vee N} \frac{1}{1+[\frac{i}{(S/c\lambda)^{1/2r_0}}]^{2r_0}}\cdot \frac{1}{(S/c\lambda)^{1/2r_0}}, 
\end{equation*}
and furthermore, we have:
\begin{equation*}
    \lim_{S\rightarrow \infty} \sum_{i=1}^{M\vee N} \frac{1}{1+[\frac{i}{(S/c\lambda)^{1/2r_0}}]^{2r_0}}\cdot \frac{1}{(S/c\lambda)^{1/2r_0}} = \int_{0}^{C} \frac{1}{1+x^{2r_0}}\mathrm{d}x < \infty,
\end{equation*}
where $C = \lim_{S\rightarrow \infty} c(M\vee N)^{2r_0}\cdot\gamma_S$. In the assumptions of Theorem~\ref{thm:high_dimensional_MARAC}, we assume that $M\vee N = O(\sqrt{S})$ and $\lim_{S\rightarrow\infty} \gamma_S\cdot S^{r_0}\rightarrow C_1$ where $0 < C_1 \le \infty$. As a result, we have $C$ being either a finite value or infinity, thus we have:
\begin{equation}
    \lim_{S\rightarrow \infty}\sum_{i=1}^{M\vee N}\frac{1}{1+c\lambda i^{2r_0}/S} = \int_{0}^{C} \frac{1}{1+x^{2r_0}}\mathrm{d}x\cdot \lim_{S\rightarrow \infty}(S/c\lambda)^{1/2r_0} = O(\gamma_S^{-1/2r_0}).\label{eq:J1-bound}
\end{equation}

Combining~\eqref{eq:trace-I-W-upper-bound},~\eqref{eq:trace-I-W-upper-bound2},~\eqref{eq:order-I-W-bound2} and~\eqref{eq:J1-bound}, we have $ \tr{\matI-\matW} \lesssim O_P((M\vee N)\gamma_S^{-1/2r_0}) = O_P(\sqrt{S}\gamma_S^{-1/2r_0})$. To obtain the lower bound of $\tr{\matI-\matW}$, we have:
\begin{equation*}
    \tr{\matI-\matW} \ge D\cdot \sum_{s=1}^S \frac{1}{1+\lambda c_{\vecz}^{\prime} \rho_s(\matK)^{-1}} \ge D\cdot\sum_{i=1}^{M}\sum_{j=1}^{N} \frac{1}{1+c_3\lambda(ij)^{r_0}/S},
\end{equation*}
which holds with probability approaching $1$ and $c_{\vecz}^{\prime}=2/\mineigen{\bSigma_\vecz^*}$ and the second inequality follows from~\eqref{eq:order-I-W-bound2}. To further lower bound the double summation, we have:
\begin{equation*}
    \cdot\sum_{i=1}^{M}\sum_{j=1}^{N} \frac{1}{1+c_3\lambda(ij)^{r_0}/S} \ge \sum_{i=1}^{M\wedge N}\frac{1}{1+c_3\lambda(ij)^{r_0}/S}.
\end{equation*}
This new lower bound can be approximated with the same method as~\eqref{eq:J1-bound} under the assumption that $M\wedge N=O(\sqrt{S})$. We can obtain the lower bound of $\tr{\matI-\matW}$ as $O_P(\gamma_S^{-1/2r_0})$, which establishes the final result.

The upper bound of $\tr{\matW}$ is trivial since:
\begin{equation*}
    \tr{\matW} = \sum_{s=1}^{S}\sum_{d=1}^{D} \frac{\lambda}{\lambda + \rho_d(\widehat{\bSigma}_{\vecz})\rho_s(\matK)} \le SD.
\end{equation*}
\end{proof}

\subsection{Proof of Lemma~\ref{thm:boundedness-gaussian-quadratic-form}}\label{subsec:proof-boundedness-gaussian-quadratic-form}
\begin{proof}
Let $\matW^{\prime}=(\matI_T\otimes\bSigma^{1/2})\matW(\matI_T\otimes\bSigma^{1/2})$, then by the Hanson-Wright inequality~\citep{rudelson2013hanson}, for any fixed $\matW$, with $c, t > 0$ being constants and $K=\sqrt{8/3}$, we have:
\begin{equation}\label{eq:hanson-wright-inequality}
\mathrm{P}\left[\left|\vecep^\top\matW\vecep - \matE\left[\vecep^\top\matW\vecep\right]\right| > t\bigg|\matW\right] \le 2\exp\left[-c\cdot\min\left(\frac{t^2}{K^4\twonorm{\matW^\prime}^2},\frac{t}{K^2\specnorm{\matW^\prime}}\right)\right].
\end{equation}
We denote each of the $S\times S$ sub-matrix along the diagonal of $\matW$ as $\matW_1,\ldots,\matW_T$, then for $\matE\left[\vecep^\top\matW\vecep|\matW\right]$, we have:
\begin{equation*}
\matE\left[\vecep^\top\matW\vecep|\matW\right] \overset{(1)}{=} \sum_{t=1}^{T}\left\langle \matW_t, \bSigma\right\rangle \overset{(2)}{\le} \sum_{t=1}^{T} \specnorm{\bSigma}\cdot\left\|\matW_t\right\|_{*} \overset{(3)}{=} \specnorm{\bSigma}\cdot\tr{\matW},
\end{equation*}
where $\langle\cdot,\cdot\rangle$ denotes the matrix inner product and $\|\cdot\|_{*}$ denotes the matrix nuclear norm. For (1), this is because of the definition of $\vecep^\top\matW^\prime\vecep$ as well as the independence between $\matW$ and $\vecep$. For (2), this inequality holds for the matrix/tensor inner product, and we refer our reader to Lemma 1 of~\citet{wang2020learning}. Similarly, we also have $\matE\left[\vecep^\top\matW\vecep|\matW\right] \ge \mineigen{\bSigma}\cdot\tr{\matW}$.

For (3), we can prove it via the semi-definiteness of $\matW$:
\begin{equation*}
\sum_{t=1}^T \left\|\matW_t\right\|_{*} = \sum_{t=1}^T \left\|\matL_t^\top\matW\matL_t\right\|_{*} = \tr{\matW \cdot \left(\sum_{t=1}^T \matL_t\matL_t^\top \right)} = \tr{\matW} = \left\|\matW\right\|_{*},
\end{equation*}
where $\matL_t=[\overbrace{\matO,\ldots,\matO}^{\text{\makebox[0pt]{t-1 blocks}}},\matI,\overbrace{\matO,\ldots,\matO}^{\text{\makebox[0pt]{T-t blocks}}}]^\top$.

Letting $t=\matE\left[\vecep^\top\matW\vecep|\matW\right]/2$, then we have:
\begin{align}
2\exp\left[-c\min\left(\frac{t^2}{K^4\twonorm{\matW}^2},\frac{t}{K^2\specnorm{\matW}}\right)\right] & \le 2\exp\left[-c\cdot\min\left(\frac{\mineigen{\bSigma}^2\cdot\tr{\matW}^2}{K^4\twonorm{\matW^\prime}}^2, \frac{\mineigen{\bSigma}\cdot\tr{\matW}}{K^2\cdot\specnorm{\matW^\prime}}\right)\right] \nonumber \\
& \le 2\exp\left[-c\cdot\min\left(\frac{\mineigen{\bSigma}^2\cdot\tr{\matW}^2}{K^4\maxeigen{\bSigma}^2\cdot\twonorm{\matW}}^2, \frac{\mineigen{\bSigma}\cdot\tr{\matW}}{K^2\cdot\maxeigen{\bSigma}}\right)\right]\nonumber \\
& \le 2\exp\left[-\frac{c}{C_1^2K^4}\cdot\tr{\matW}\right] \label{eq:hanson-wright-inequality-bound}
\end{align}
We can lower bound the trace of $\matW$ as follows. First, note that:
\begin{equation*}
    \tr{\matW} = \sum_{s=1}^{S}\sum_{d=1}^{D} \frac{\lambda}{\lambda + \rho_d(\widehat{\bSigma}_{\vecz})\rho_s(\matK)} \ge SD\cdot\frac{\lambda}{\lambda + \maxeigen{\widehat{\bSigma}_\vecz}\maxeigen{\matK}}.
\end{equation*}
By the assumption that $\maxeigen{\matK}$ is bounded and that the fact that $\maxeigen{\widehat{\bSigma}_\vecz} \le 2\maxeigen{\bSigma_\vecz^*}$ with probability approaching $1$ as $T\rightarrow\infty$, we have:
\begin{equation}\label{eq:trace-W-tail-bound}
    \mathrm{P}\left[\tr{\matW}\ge \frac{SD\lambda}{\lambda+\bar{c}}\right] \rightarrow 1, \quad \text{ as }T\rightarrow\infty,
\end{equation}
where $\bar{c}=2\maxeigen{\bSigma_\vecz^*}\maxeigen{\matK}$. Since $r_0<2$ and $\gamma_S\cdot S^{r_0}\rightarrow C_1$ as $S\rightarrow\infty$, with $C_1$ being either a positive constant or infinity, we have $\gamma_S \cdot S^2 = \lambda \cdot S\rightarrow\infty$. Therefore, we have $\tr{\matW}\rightarrow\infty$ with probability approaching $1$, as $S,T\rightarrow\infty$. 

With these results, we can now upper bound the unconditional probability of the event $\{\left|\vecep^\top\matW\vecep - \matE\left[\vecep^\top\matW\vecep\right]\right| > \matE\left[\vecep^\top\matW\vecep\right]/2\}$ as follows:
\begin{align}
    & \mathrm{P}\left[\left|\vecep^\top\matW\vecep - \matE\left[\vecep^\top\matW\vecep\right]\right| > \matE\left[\vecep^\top\matW\vecep\right]/2\right] \notag \\
    & \le \mathrm{E}\left[2\exp\left[-\frac{c}{C_1^2K^4}\cdot\tr{\matW}\right]\right] \notag \\
    & \le 2\left\{1\cdot\mathrm{P}\left(\tr{W}<\frac{SD\lambda}{\lambda+\bar{c}}\right)+ \exp\left[-\frac{c}{C_1^2K^4}\cdot\frac{SD\lambda}{\lambda+\bar{c}}\right]\cdot\mathrm{P}\left(\tr{W}\ge\frac{SD\lambda}{\lambda+\bar{c}}\right)\right\}\rightarrow 0.\label{eq:quadratic-final-bound}
\end{align}
This indicates that $\vecep^\top\matW\vecep$ concentrates around its mean $\matE\left[\vecep^\top\matW\vecep\right]$ with high probability, and thus $\vecep^\top\matW\vecep/\tr{\matW} = O_P(\specnorm{\bSigma}) = O_P(c_{1,S})$. To establish $\vecep^\top(\matI-\matW)^2\vecep/\tr{(\matI-\matW)^2} = O_P(\specnorm{\bSigma}) = O_P(c_{1,S})$, we 
first note the unboundedness of $\tr{(\matI-\matW)^2}$ by following the same idea as the proof for Lemma~\ref{lemma:order-I-W}, where we have:
\begin{equation*}
    \tr{(\matI-\matW)^2} \ge (S/c\lambda)^{1/2r_0}\cdot\sum_{i=1}^{M\wedge N}\left\{\frac{1}{1+\left[\frac{i}{(S/c\lambda)^{1/2r_0}}\right]^{2r_0}}\right\}^2(S/c\lambda)^{-1/2r_0},
\end{equation*}
with probability approaching $1$ and $c$ is some constant. The remainder of the proof follows exactly the same steps, and we omit the rest of the details here.
\end{proof}

\subsection{Proof of Lemma~\ref{thm:psd-matrix-lemma}}\label{proof:psd-matrix-lemma}
\begin{proof}
For any two arbitrary symmetric matrices $\mathbf{M},\mathbf{N}$ with identical sizes, we use $\mathbf{M} \gtrsim\mathbf{N}$ to indicate that $\mathbf{M}-\mathbf{N}$ is positive semi-definite, and we use $\mathbf{M}^{1/2}$ to denote the symmetric, positive semi-definite square root matrix of $\mathbf{M}$. 

Since $\matA-\matB$ is positive semi-definite, multiplying it by $\matB^{-1/2}$ on both left and right sides of $\matA-\matB$, we have $\matB^{-1/2}\matA\matB^{-1/2} \gtrsim \matI$. Therefore, we have $\matB^{-1/2}\matA^{1/2}\matA^{1/2}\matB^{-1/2} \gtrsim \matI$. Notice that the matrix $\matA^{1/2}\matB^{-1/2}$ is invertible and thus has no zero eigenvalues. As a result, all eigenvalues of $\matB^{-1/2}\matA^{1/2}\matA^{1/2}\matB^{-1/2}$ are the same as the eigenvalues of $\matA^{1/2}\matB^{-1/2}\matB^{-1/2}\matA^{1/2}$ and thus $\matA^{1/2}\matB^{-1/2}\matB^{-1/2}\matA^{1/2} \gtrsim \matI$. Multiplying both sides by $\matA^{-1/2}$ on both the left and right sides yields $\matB^{-1}\gtrsim \matA^{-1}$, which completes the proof.
\end{proof}

\section{Additional Details on Simulation and Algorithm}\label{app:sim}
\subsection{Simulation Setup}
\begin{figure}[!htb]
    \centering
    \includegraphics[width=0.98\textwidth]{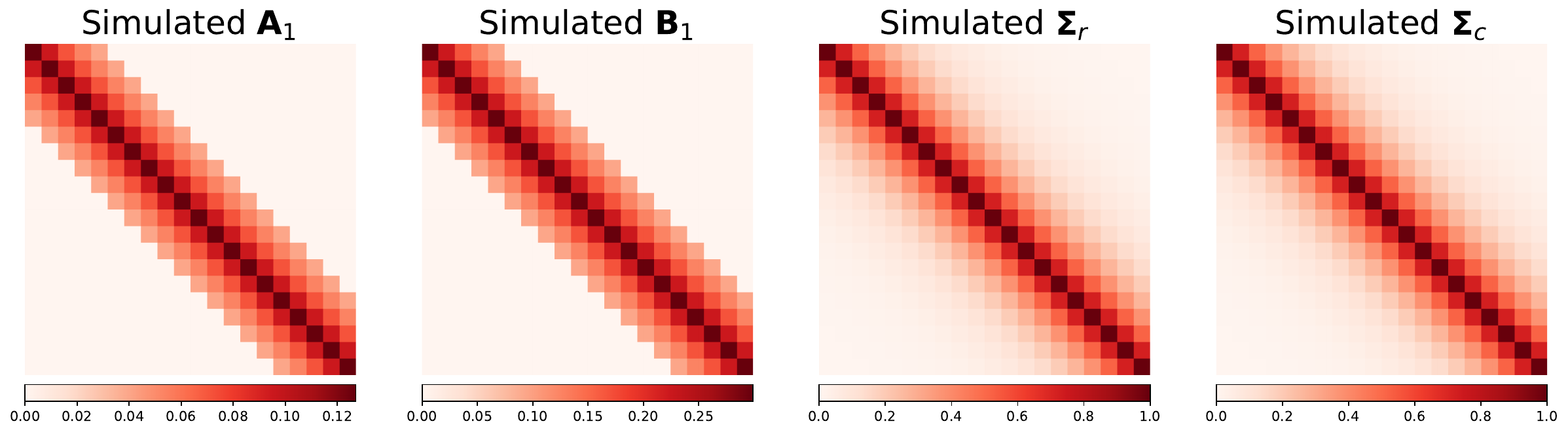}
    \caption{Visualization of the simulated $\mathbf{A}_{1},\mathbf{B}_{1}, \bSigma_{r},\bSigma_{c}$ with $M=N=20$.}
    \label{fig:Simulated_A_B_Sigma}
\end{figure}

We generate the simulated dataset according to the MARAC$(P,Q)$ model specified by~\eqref{eq:MARAC-model} and~\eqref{eq:Et_Covariance}. We simulate the autoregressive coefficients $\mathbf{A}_{p},\mathbf{B}_{p}$ such that they satisfy the stationarity condition specified in Theorem \ref{thm:Joint-Stationarity} and have a banded structure. We use a similar setup for generating $\bSigmar,\bSigmac$ with their diagonals fixed at unity. In Figure \ref{fig:Simulated_A_B_Sigma}, we plot the simulated $\mathbf{A}_{1},\mathbf{B}_{1}, \bSigma_{r},\bSigma_{c}$ when $(M,N)=(20,20)$.

To generate $g_1,g_2,g_3\in\RKHS$ and mimic the spatial grid in our real data application in Section~\ref{sec:application}, we specify the 2-D spatial grid with the two dimensions being latitude and longitude of points on a unit sphere $\mathbb{S}^{2}$. Each of the evenly spaced $M\times N$ grid points has its polar-azimuthal coordinate pair as $(\theta_{i},\phi_{j})\in [0^{\circ},180^{\circ}]\times [0^{\circ}, 360^{\circ}], i\in[M], j\in[N]$, and one projects the sampled grid points on the sphere onto a plane to form an $M\times N$ matrix. The polar $\theta$ (co-latitude) and azimuthal $\phi$ (longitude) angles are very commonly used in the spherical coordinate system, with the corresponding Euclidean coordinates being
$(x,y,z)=(\sin(\theta)\cos(\phi),\sin(\theta)\sin(\phi),\cos(\theta))$. 

As for the spatial kernel, we choose the Lebedev kernel:
\begin{equation}\label{eq:Lebedev_Kernel}
    k_{\eta}(s_{1}, s_{2}) = \left(\frac{1}{4\pi} + \frac{\eta}{12\pi}\right) - \frac{\eta}{8\pi}\sqrt{\frac{1 - \left<s_{1}, s_{2}\right>}{2}}, \quad s_{1}, s_{2} \in \mathbb{S}^{2},
\end{equation}
where $\left<\cdot,\cdot\right>$ denotes the angle between two points on the sphere $\mathbb{S}^{2}$ and $\eta$ is a hyperparameter of the kernel. In the simulation experiment as well as the real data application, we fix $\eta = 3$. The Lebedev kernel has the spherical harmonics functions as its eigenfunction:
\begin{equation*}
    k_{\eta}(s_{1}, s_{2}) = \frac{1}{4\pi} + \sum_{l=1}^{\infty} \frac{\eta}{(4l^{2}-1)(2l+3)} \sum_{m=-l}^{l} Y_{l}^{m}(s_{1})Y_{l}^{m}(s_{2}),
\end{equation*}
where $Y_{l}^{m}(\cdot)$ is a series of orthonormal real spherical harmonics bases defined on sphere $\mathbb{S}^{2}$:
\begin{equation*}
Y_{l}^{m}(s) = Y_{l}^{m}(\theta,\phi) = \begin{dcases}
      \sqrt{2}N_{lm}P_{l}^{m}(\cos(\theta))\cos(m\phi) & \text{if $m > 0$} \\
      N_{l0}P_{l}^{0}(\cos(\theta)) & \text{if $m = 0$} \\
      \sqrt{2}N_{l|m|}P_{l}^{|m|}(\cos(\theta))\sin(|m|\phi) & \text{if $m < 0$}
    \end{dcases}  ,
\end{equation*}
with $N_{lm} = \sqrt{(2l+1)(l-m)!/(4\pi(l+m)!)}$, and $P_{l}^{m}(\cdot)$ being the associated Legendre polynomials of order $l$. We refer our readers to \citet{kennedy2013classification} for detailed information about the spherical harmonics functions and the associated isotropic kernels. Under our 2-D grid setup and the choice of kernel, we have found that empirically, the kernel Gram matrix $\matK$ has its eigen spectrum decaying at a rate of $\rho_i(\matK) \approx i^{-r}$ with $r\in [1.3,1.5]$.

\begin{figure}[!htb]
    \centering
    \includegraphics[width=0.98\textwidth]{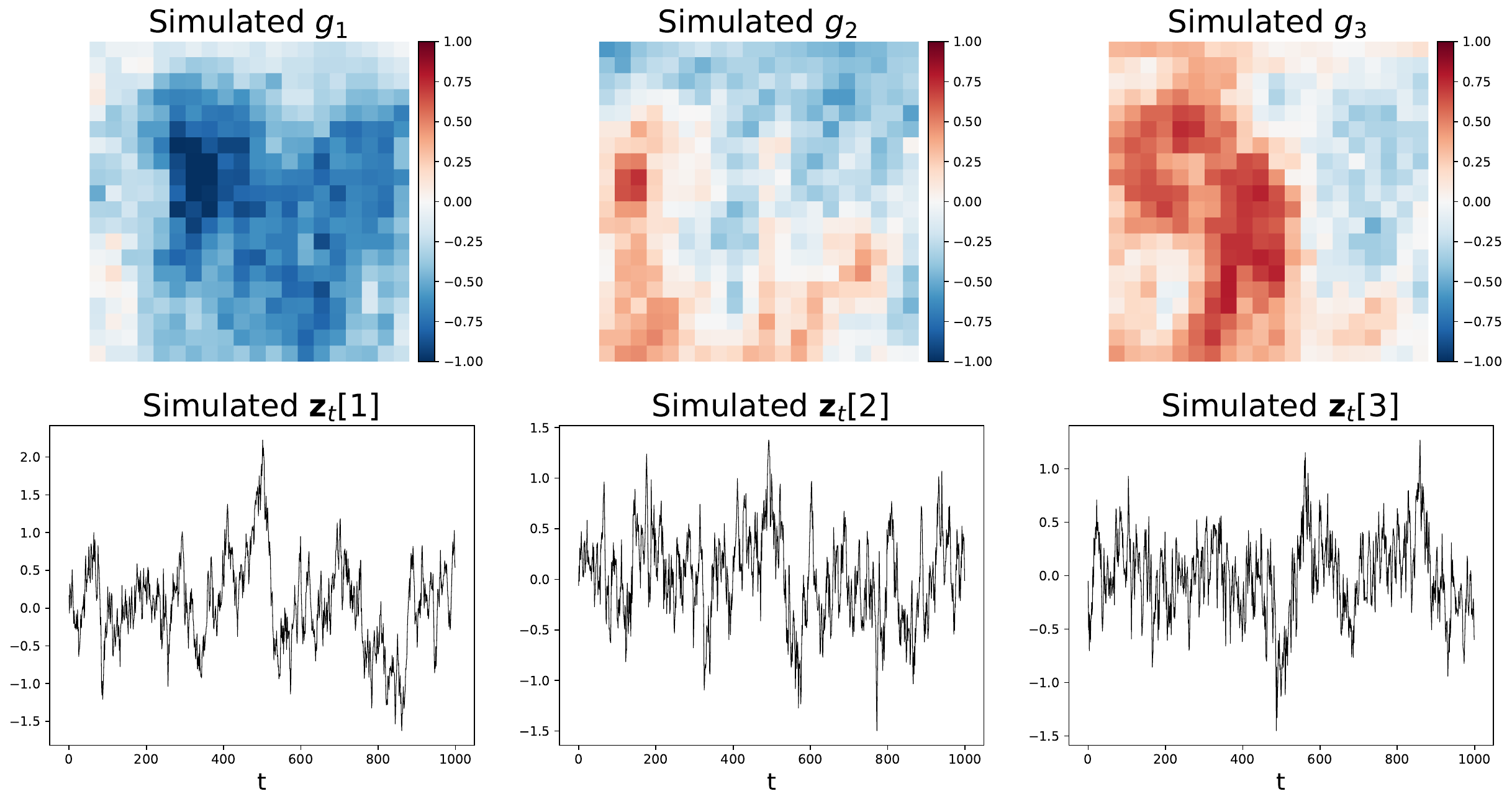}
    \caption{Simulated functional parameters $g_1,g_2,g_3$ evaluated on a $20\times 20$ spatial grid (top row) and the corresponding auxiliary vector time series (bottom row).}
    \label{fig:Simulated_g_Z}
\end{figure}

We randomly sample $g_1,g_2,g_3$ from Gaussian processes with a covariance kernel being the Lebedev kernel in~\eqref{eq:Lebedev_Kernel}. Finally, we simulate the vector time series $\vecz_t$ using a VAR$(1)$ process. In Figure~\ref{fig:Simulated_g_Z}, we visualize the simulated functional parameters as well as the vector time series from one random draw.

\subsection{Approximated Penalized MLE with Kernel Truncation}\label{subsec:truncation}
The iterative algorithm in Section~\ref{subsec:PMLE} requires inverting an $MND\times MND$ matrix in~\eqref{eq:Update-Gammaq} when updating $\vecgamma_{q}$, i.e., the coefficients of the representer functions $k(\cdot,s)$. One way to reduce the computational complexity without any approximation is to divide the step of updating $\vecgamma_{q} = [\vecgamma_{q,1}^{\top}:\cdots:\vecgamma_{q,D}^{\top}]^{\top}$ to updating one block of parameters at a time following the order of $\vecgamma_{q,1}\rightarrow\cdots\rightarrow\vecgamma_{q,D}$. However, such a procedure requires inverting a matrix of size $MN\times MN$, which could still be high-dimensional.


To circumvent the issue of inverting large matrices, we can approximate the linear combination of all $MN$ representers using a set of $R<<MN$ basis functions, i.e., $\matK\vecgamma_{q,d} \approx \matKR\vectheta_{q,d}$, where $\matKR\in\mathbb{R}^{MN\times R}, \vectheta_{q,d}\in\mathbb{R}^{R}$. For example, one can reduce the spatial resolution by subsampling a fraction of the rows and columns of the matrix and only use the representers at the subsampled ``knots" as the basis functions. In this subsection, we consider an alternative approach by truncating the Mercer decomposition in~\eqref{eq:Kernel-Decomposition}. A similar technique can be found in~\citet{kang2018scalar}.

Given the eigen-decomposition of $k(\cdot,\cdot)$ in~\eqref{eq:Kernel-Decomposition}, one can truncate the decomposition at the $R^{\rm{th}}$ largest eigenvalue $\lambda_{R}$ and get an approximation: $k(\cdot,\cdot)\approx\sum_{r\le R} \lambda_{r}\psi_{r}(\cdot)\psi_{r}(\cdot)$. We will use the set of eigen-functions $\{\psi_{1}(\cdot),\ldots,\psi_{R}(\cdot)\}$ for faster computation. The choice of $R$ depends on the decaying rate of the eigenvalue sequence $\{\lambda_{r}\}_{r=1}^{\infty}$ (thus the smoothness of the underlying functional parameters) and can be done via cross-validation in practice. Our simulation result shows that the estimation and prediction errors shrink monotonically as $R\rightarrow\infty$. Therefore, $R$ can be chosen based on the computational resources available. The kernel truncation speeds up the computation at the cost of providing an overly-smoothed estimator, as we demonstrate later in this section.


Given the kernel truncation, any functional parameter $g_{q,d}(\cdot)$ is now approximated as: $g_{q,d}(\cdot)\approx \sum_{r\in[R]} [\vectheta_{q,d}]_{r}\psi_{r}(\cdot)$. The parameter to be estimated now is $\matTheta_{q} = [\vectheta_{q,1};\cdots;\vectheta_{q,D}]\in\mathbb{R}^{R\times D}$, whose dimension is much lower than before ($\matgamma_q\in\mathbb{R}^{MN\times D}$). Estimating $\matTheta_{q}$ requires solving a ridge regression problem, and the updating formula for $\vect{\matTheta_q}=\vectheta_q$ can be written as:
\begin{equation*}
    \vectheta_q^{(l+1)} \leftarrow \left[\matPhi\left(\vecz_{t-q}^{\top}\otimes \matKR,\bSigma^{(l)}\right) + \lambda T\left(\matI_{D}\otimes\matLambda_{R}^{-1}\right)\right]^{-1}\matPhi\left(\vecz_{t-q}^{\top}\otimes \matKR,\widetilde{\vecx}_{t,-q},\bSigma^{(l)}\right),
\end{equation*}
where $\matKR\in \mathbb{R}^{MN\times R}$ satisfies $[\matKR]_{ur} = \psi_{r}(s_{ij}), u=i+(j-1)M$, and $\matLambda_{r} = \text{diag}(\lambda_1,\ldots,\lambda_R)$, with $\lambda_r$ being the $r^{\rm{th}}$ largest eigenvalue of the Mercer decomposition of $k(\cdot,\cdot)$. Now we only need to invert a matrix of size $RD \times RD$, which speeds up the computation.

\begin{figure}[!htb]
    \centering
    \includegraphics[width=0.98\textwidth]{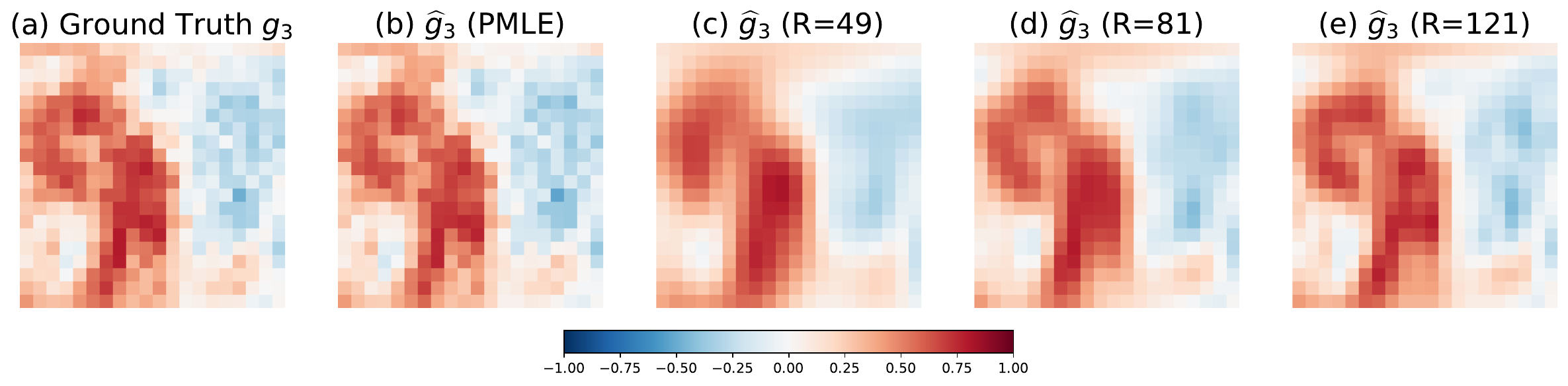}
    \caption{Ground truth $g_{3}$ (panel (a)), the penalized MLE estimator $\widehat{g}_{3}$ (panel (b)), and the truncated penalized MLE estimator $\widehat{g}_{3}$ using $R\in\{49,81,121\}$ basis functions. $M=20$.}
    \label{fig:PMLE-vs-Truncation}
\end{figure}

In Figure~\ref{fig:PMLE-vs-Truncation}, we visualize the ground truth of $g_3$ and both its penalized MLE and truncated penalized MLE estimators. It is evident that the truncated penalized MLE estimators give a smooth approximation to $g_{3}$, and the approximation gets better when $R$ gets larger. The choice of $R$ should be as large as possible for accuracy, so one can determine $R$ based on the computational resources available.

\end{document}